%% file: IAXO_Physics_Potential_v20.tex
\documentclass[a4paper,11pt]{article}
\pdfoutput=1 

\usepackage[utf8]{inputenc}
\usepackage{color}
\usepackage[dvipsnames]{xcolor}
\usepackage{jcappub} 
\usepackage{aas_macros}
\usepackage{tikz}
\usetikzlibrary{patterns}
\usetikzlibrary{plotmarks}
\usetikzlibrary{external}
\usepackage{soul}
\usepackage{slashed}
\usepackage{cancel}

\newcommand{\J}[1]{\color{orange} #1 \color{black}}
\newcommand{\JJ}[1]{\color{red} [-[#1]-] \color{black}}
\newcommand{\K}[1]{\color{blue} #1 \color{black}}

\newcommand{\JJJ}[1]{\color{Blue} #1 \color{black}}

\usepackage{epsfig}

\usepackage{graphicx}
\usepackage{amssymb}
\usepackage{epstopdf}

\renewcommand\({\left(}
\renewcommand\){\right)}

\newcommand{\be}{\begin{equation}}
\newcommand{\ee}{\end{equation}}
\newcommand{\bea}{\begin{eqnarray}}
\newcommand{\eea}{\end{eqnarray}}

\newcommand{\exclude}[1]{}

\newcommand{\gagamma}{g_{a\gamma}}
\newcommand{\ckcs}{counts~keV$^{-1}$~cm$^{-2}$~s$^{-1}$}
\newcommand{\calN}{ {\cal N}}
\newcommand{\calE}{ {\cal E}}

\title{Physics potential of the International Axion Observatory (IAXO)}

\input{IAXO_Physics_Potential_Authorlist.tex}

\emailAdd{Igor.Irastorza@cern.ch}

\abstract{
We review the physics potential of a next generation search for solar axions: the International Axion Observatory (IAXO).
Endowed with a sensitivity to discover axion-like particles (ALPs) with a coupling to photons as small as $\gagamma\sim 10^{-12}$~GeV$^{-1}$, or to electrons $g_{ae}\sim$10$^{-13}$, IAXO has the potential to find the QCD axion in the 1~meV$\sim$1~eV mass range where it solves the strong CP problem, can account for the cold dark matter of the Universe and be responsible for the anomalous cooling observed in a number of stellar systems.
At the same time, IAXO will have enough sensitivity to detect lower mass axions invoked to explain: 1) the origin of the anomalous ``transparency'' of the Universe to gamma-rays, 2) the observed soft X-ray excess from galaxy clusters or 3) some inflationary models. In addition, we review string theory axions with parameters accessible by IAXO and discuss their potential role in cosmology as Dark Matter and Dark Radiation as well as their connections to the above mentioned conundrums.
}

\begin{document}
\maketitle
\flushbottom

\section{Introduction}
\label{sec:introduction}

\input{sections/introduction.tex}

\section{Axions from high-energy physics}
\label{sec:StringAxions}
\subsection{Stringy axions}

\input{sections/stringyaxions.tex}

\subsection{The low-energy Lagrangian and field-theoretic models}

\input{sections/axionmodels.tex}

\section{A meV mass QCD axion as dark matter candidate}
\label{sec:axiondm}

\input{sections/darkmatter.tex}

\section{Axions as dark radiation}
\label{sec:AxionDR}

\input{sections/darkradiation.tex}

\section{Axions and inflation}
\label{sec:inflation}

\input{sections/inflation.tex}

\section{Hints for axions in the anomalous cooling of stellar objects}
\label{sec:cooling_hints}

\input{sections/cooling.tex}
\section{Axion-like particles in the propagation of photons over astronomical distances}
\label{sec:transparency}

\input{sections/transparency2.tex}

\section{Updated sensitivity prospects for IAXO}
\label{sec:iaxo}

\input{sections/iaxosensitivity.tex}

\section{Discussion and conclusions}
\label{sec:conclusions}

Axions and more generic ALPs are recently receiving an increasing attention from the experimental community. This is in part due to the lack of experimental confirmation of the WIMP dark matter paradigm in recent searches both at colliders and direct detection experiments. Axions are known to be attractive dark matter candidates, but despite continued experimental activity almost since their proposal more than 30 years ago, most of the allowed parameter space remains largely unexplored so far. Recently a plethora of ideas and new small-scale initiatives are being put forward~\cite{Irastorza:2018dyq}, under the assumption that dark matter is entirely made of axions, and mostly addressed to the to low-mass range $m_a\sim 10^{-6}-10^{-4}$~eV. However, as described in section \ref{sec:axiondm}, axion (or ALP) dark matter can be also realized in other mass ranges, in particular at higher mass values, for which dark matter experiments are increasingly difficult to realize. In addition, axions/ALPs may compose only a subdominant fraction of the galactic dark matter density, a possibility that may have a particular theoretical interest~\cite{Baer:2011uz}. In general, a signal in a dark matter experiment is proportional to $g_{a\gamma}^2\rho_a$, being the local galactic axion density $\rho_a$ degenerate with $g_{a\gamma}$. Experiments not relying on the axion being the dark matter are needed to break such degeneracy. In any case, axions are strongly motivated by theory, without relying on the dark matter question. They constitute our only compelling solution to the strong-CP problem in the Standard Model. As reviewed in section~\ref{sec:StringAxions}, axions and ALPs are also very generic low energy signatures of high-energy completions of the SM in extra dimensions.

Experiments that invoke both the production and detection of ALPs entirely in the laboratory are the least model-dependent. Among these experiments, only the ALPS-II project will reach a sensitivity beyond current astrophysical and experimental bounds on $g_{a\gamma}$, for ALP masses below $m_a<10^{-4}$~eV. In this context, the search for solar axions constitutes a good compromise between model-independency (as the emission of axions by the Sun is a robust prediction of any axion model), and intensity of the source axion flux. The IAXO project embodies the technical know-how accumulated in previous realizations of the axion helioscope concept, most in particular the CAST experiment at CERN, extended to a much larger scale. While CAST has been the first axion helioscope reaching a sensitivity to $g_{a\gamma}$ comparable to astrophysical bounds, IAXO will largely advance well beyond them. In particular, and as a summary of the material reviewed in previous sections:

\begin{itemize}
  \item IAXO will cover a large fraction of unexplored ALP parameter. In particular, it will probe a large range of QCD axion models in the mass range $m_a \sim$ 1~meV -- 1~eV.
  \item Most of this region is not attainable by any other experimental technique, stressing the complementarity of IAXO in the wider axion experimental landscape.
  \item IAXO will comfortably cover the range of $g_{a\gamma}$ invoked as ALP solutions to the possible anomalies observed in the propagation of high energy photon over astronomical distances, fully testing this hypothesis.
  \item  IAXO will cover a large fraction of the region of parameter space invoked as possible solutions to the anomalies observed in the cooling of several stellar systems, both involving $g_{a\gamma}$ and $g_{ae}$.
  \item The parameter space to be covered by IAXO could contain a viable dark matter candidate. In particular, post-inflation models with $N_{\rm DM}>1$ allows for QCD axion with $m_a$ in this range to account for the totality of the DM density. In addition, that very ($g_{a\gamma}$,$m_a$) region is also suggested by ALP-miracle models recently proposed in which ALPs can account for both DM and inflation.
  \item Combined with a positive detection in a haloscope, it will break the $g_{a\gamma}^2\rho_a$  degeneracy and determine the local density of the galactic axionic dark matter.
  \item Although not covered in this paper, the IAXO infrastructure could be used to host additional experimental setups, in particular microwave cavities or other devices able to directly search for DM axions.
\end{itemize}

IAXO will play a prominent role in the new generation of axion experiments currently under proposal or preparation. It is highly complementary with the other experimental frontiers (laboratory and relic axions). Collectively a relatively large fraction of the most motivated parameter space for axions will be explored, potentially leading to substantial progress in the low energy frontier in the coming years.

\acknowledgments


We acknowledge support from the the European Research Council (ERC)
under the European Union's Horizon 2020 research and innovation programme, grant agreement ERC-2017-AdG-788781 (IAXO+) as well as ERC-2018-StG-802836 (AxScale), and from the Spanish MINECO under grant FPA2016-76978-C3-1-P. 
Part of this work was performed under the auspices of the U.S. Department of Energy by Lawrence Livermore National Laboratory under Contract DE-AC52-07NA27344. JR is supported by the Ramon y Cajal Fellowship 2012-10597, the grant FPA2015-65745-P (MINECO/FEDER), the EU through the ITN ``Elusives'' H2020-MSCA-ITN-2015/674896 and the Deutsche Forschungsgemeinschaft under grant SFB-1258 as a Mercator Fellow. We also acknowledge partial support of RFBR (grants 17-02-00305A, 16-29-13014 ofi-m).

\bibliographystyle{bib/JHEP}
\bibliography{bib/updatedinspires,bib/iaxobib,bib/iaxobibdm,bib/iaxobibcool,bib/iaxobibastro,bib/iaxo-th}

\end{document}

%% file: IAXO_Physics_Potential_Authorlist.tex
\newcommand{\IRFU}{a}
\newcommand{\INAF}{b}
\newcommand{\CERN}{c}
\newcommand{\UNIZAR}{d}
\newcommand{\Bologna}{e}
\newcommand{\INFNBologna}{f}
\newcommand{\ICCUB}{g}
\newcommand{\RPCTPOxford}{h}
\newcommand{\Tohoku}{i}
\newcommand{\Petersburg}{j}
\newcommand{\LLNL}{k}
\newcommand{\UBonn}{l}
\newcommand{\USiegen}{m}
\newcommand{\UHEI}{n}
\newcommand{\Barry}{o}
\newcommand{\INR}{p}
\newcommand{\Zagreb}{q}
\newcommand{\DESY}{r}
\newcommand{\Stockholm}{s}
\newcommand{\Stanford}{t}
\newcommand{\ICREA}{u}
\newcommand{\MIPT}{v}
\newcommand{\MIT}{w}
\newcommand{\MPI}{x}
\newcommand{\INFNPavia}{y}
\newcommand{\Mainz}{z}
\newcommand{\KavliTokio}{aa}
\newcommand{\CapeTown}{ab}
\newcommand{\IHEP}{ac}
\newcommand{\KAIST}{ad}

\author[\IRFU]{E.~Armengaud,}
\author[\IRFU]{D.~Attie,}
\author[\INAF]{S.~Basso,}
\author[\IRFU]{P.~Brun,}
\author[\CERN]{N.~Bykovskiy,}
\author[\UNIZAR]{J.~M.~Carmona,}
\author[\UNIZAR]{J.~F.~Castel,}
\author[\UNIZAR]{S.~Cebri\'{a}n,}
\author[\Bologna,\INFNBologna]{M.~Cicoli,}
\author[\INAF]{M.~Civitani, }
\author[\ICCUB]{C.~Cogollos,}
\author[\RPCTPOxford]{J.~P.~Conlon,}
\author[\ICCUB]{D.~Costa,}
\author[\UNIZAR]{T.~Dafni,}
\author[\Tohoku]{R.~Daido,}
\author[\Petersburg]{A.V.~Derbin,}
\author[\LLNL]{M.~A.~Descalle,}
\author[\UBonn]{K.~Desch,}
\author[\Petersburg]{I.S.~Dratchnev,}
\author[\CERN]{B.~D\"{o}brich,}
\author[\CERN]{A.~Dudarev,}
\author[\IRFU]{E.~Ferrer-Ribas,}
\author[\USiegen]{I.~Fleck,}
\author[\UNIZAR]{J.~Gal\'{a}n,}
\author[\INAF]{G.~Galanti,}
\author[\ICCUB]{L.~Garrido,}
\author[\ICCUB]{D.~Gascon,}
\author[\UHEI]{L.~Gastaldo,}
\author[\ICCUB]{C.~Germani,}
\author[\INAF]{G.~Ghisellini,}
\author[\Barry]{M.~Giannotti,}
\author[\IRFU]{I.~Giomataris,}
\author[\INR]{S.~Gninenko,}
\author[\INR]{N.~Golubev,}
\author[\ICCUB]{R.~Graciani,}
\author[\UNIZAR,1]{I.~G.~Irastorza\note{Corresponding author},}
\author[\Zagreb]{K.~Jakov\v ci\' c,}
\author[\UBonn]{J.~Kaminski,}
\author[\Zagreb]{M.~Kr\v {c}mar,}
\author[\UBonn]{C.~Krieger,}
\author[\Zagreb]{B.~Laki\'{c},}
\author[\IRFU]{T.~Lasserre,}
\author[\IRFU]{P.~Laurent,}
\author[\IRFU]{O.~Limousin,}
\author[\DESY]{A.~Lindner,}
\author[\Petersburg]{I.~Lomskaya,}
\author[\INR]{B.~Lubsandorzhiev,} 
\author[\UNIZAR]{G.~Luz\'{o}n,}
\author[\Stockholm]{M.~C.~D.~Marsh,}
\author[\UNIZAR]{C.~Mergalejo,}
\author[\ICCUB]{F.~Mescia,}
\author[\Stanford]{M.~Meyer,}
\author[\ICCUB,\ICREA]{J.~Miralda-Escud\'{e},}
\author[\UNIZAR]{H.~Mirallas,}
\author[\Petersburg]{V.~N.~Muratova,}
\author[\IRFU]{X.~F.~Navick, } 
\author[\IRFU]{C.~Nones, }
\author[\ICCUB]{A.~Notari,}
\author[\INR,\MIPT]{A.~Nozik,}
\author[\UNIZAR]{A.~Ortiz de Sol\'{o}rzano,}
\author[\INR]{V.~Pantuev,}
\author[\IRFU]{T.~Papaevangelou,}
\author[\INAF]{G.~Pareschi,}
\author[\MIT]{K.~Perez,}
\author[\ICCUB]{E.~Picatoste,}
\author[\LLNL]{M.~J.~Pivovaroff,}
\author[\UNIZAR,\MPI]{J.~Redondo,}
\author[\DESY]{A.~Ringwald,}
\author[\INFNPavia]{M.~Roncadelli,}
\author[\UNIZAR]{E.~Ruiz-Ch\'{o}liz,}
\author[\LLNL]{J.~Ruz,}
\author[\MPI]{K.~Saikawa,}
\author[\ICCUB]{J.~Salvad\'{o},}
\author[\UNIZAR]{M.~P.~Samperiz,}
\author[\UBonn]{T.~Schiffer,}
\author[\UBonn]{S.~Schmidt,}
\author[\DESY]{U.~Schneekloth,}
\author[\Mainz]{M.~Schott,}
\author[\CERN]{H.~Silva,}
\author[\INAF]{G.~Tagliaferri,}
\author[\Tohoku,\KavliTokio]{F.~Takahashi,}
\author[\INAF]{F.~Tavecchio,}
\author[\CERN]{H.~ten~Kate,}
\author[\INR]{I.~Tkachev,}
\author[\INR]{S.~Troitsky,}
\author[\Petersburg]{E.~Unzhakov,}
\author[\IRFU]{P.~Vedrine,}
\author[\LLNL]{J.~K.~Vogel,}
\author[\Mainz]{C.~Weinsheimer,}
\author[\CapeTown]{A.~Weltman,}
\author[\IHEP,\KAIST]{W.~Yin.}

\newcommand{\ICREAName}{Instituci\'o Catalana de Recerca i Estudis Avan\c cats, Passeig Llu\'\i s Companys 23, 08010-Barcelona, Spain}
\newcommand{\MainzName}{Institute of Physics, University of Mainz,  Mainz, Germany}
\newcommand{\UNIZARName}{Grupo de F\'isica Nuclear y Astropart\'{\i}culas, Departamento de F\'isica Te\'orica, \\ Universidad de Zaragoza C/ P. Cerbuna 12 50009, Zaragoza, Spain}
\newcommand{\CERNName}{European Laboratory for Particle Physics (CERN), Geneva, Switzerland}
\newcommand{\PetersburgName}{NRC KI Petersburg Nuclear Physics Institute, 188300, Gatchina, Russia}
\newcommand{\TohokuName}{Department of Physics, Tohoku University, Sendai, Japan}
\newcommand{\KavliTokioName}{Kavli IPMU (WPI), UTIAS, The University of Tokyo, Kashiwa,  Japan}
\newcommand{\IHEPName}{IHEP, Chinese Academy of Sciences, Beijing, China}
\newcommand{\MITName}{Department of Physics, Massachusetts Institute of Technology, Cambridge, MA 02139, USA}
\newcommand{\BolognaName}{Dipartimento di Fisica e Astronomia, Universit\`a di Bologna, \\ via Irnerio 46, 40126 Bologna, Italy}
\newcommand{\INFNBolognaName}{INFN, Sezione di Bologna, viale Berti Pichat 6/2,
40127 Bologna, Italy}
\newcommand{\ICCUBName}{Institut de Ci\`encies del Cosmos, Universitat de Barcelona, Mart\'i i Franqu\`es 1, 08028 Barcelona, Spain}
\newcommand{\CapeTownName}{The Cosmology and Gravity Group, Department of Mathematics and Applied Mathematics,University of Cape Town, Private Bag, Rondebosch, 7700, South Africa}
\newcommand{\RPCTPOxfordName}{Rudolf Peierls Centre for Theoretical Physics, Clarendon Laboratory, Parks Road, Oxford OX1 3PU, UK}
\newcommand{\StockholmName}{Oskar Klein Centre for Cosmoparticle Physics, Stockholm University, Albanova, Stockholm SE-10691, Sweden}
\newcommand{\MPIName}{Max-Planck-Institut f\"ur Physik (Werner-Heisenberg-Institut), F\"ohringer Ring 6, D-80805 M\"unchen, Germany}
\newcommand{\DESYName}{Deutsches Elektronen-Synchrotron (DESY),Notkestr. 85, 22607 Hamburg, Germany}
\newcommand{\ZagrebName}{Rudjer Bo\v{s}kovi\'{c} Institute, Zagreb, Croatia}
\newcommand{\BarryName}{Physical Sciences, Barry University, 11300 NE 2nd Ave., Miami Shores, FL 33161, USA.}
\newcommand{\IRFUName}{IRFU, CEA, Universit\'{e} Paris-Saclay, F-91191 Gif-sur-Yvette, France}
\newcommand{\LLNLName}{Lawrence Livermore National Laboratory, Livermore, CA, U.S.A.}
\newcommand{\UBonnName}{Physikalisches Institut, Universit\"at Bonn, Nussallee 12, 53115 Bonn, Germany}
\newcommand{\StanfordName}{W. W. Hansen Experimental Physics Laboratory, Kavli  Institute  for  Particle  Astrophysics  and  Cosmology, Department  of  Physics  and  SLAC  National  Accelerator  Laboratory, Stanford  University,  Stanford,  California  94305,  USA}
\newcommand{\INAFName}{INAF - Osservatorio astronomico di Brera, Via E. Bianchi 46, Merate (LC), I-23807, Italy}
\newcommand{\UHEIName}{Kirchhoff Institute for Physics, Heidelberg University, INF 227 69120 Heidelberg Germany}
\newcommand{\INRName}{Institute for Nuclear Research (INR), Russian Academy of Sciences, Moscow, Russia}
\newcommand{\MIPTName}{Moscow Institute of Physics and Technology, 9 Institutskiy per., Dolgoprudny, Moscow Region, 141700, Russian Federation}
\newcommand{\KAISTName}{Department of Physics, KAIST, Daejeon, Korea}
\newcommand{\USiegenName}{Department Physik, Universit\"{a}t Siegen, Siegen, Germany}
\newcommand{\INFNPaviaName}{INFN, Sezione di Pavia, via A. Bassi 6, I-27100 Pavia, Italy}

\affiliation[\IRFU]{\IRFUName}
\affiliation[\INAF]{\INAFName}
\affiliation[\CERN]{\CERNName}
\affiliation[\UNIZAR]{\UNIZARName}
\affiliation[\Bologna]{\BolognaName}
\affiliation[\INFNBologna]{\INFNBolognaName}
\affiliation[\ICCUB]{\ICCUBName}
\affiliation[\RPCTPOxford]{\RPCTPOxfordName}
\affiliation[\Tohoku]{\TohokuName}
\affiliation[\Petersburg]{\PetersburgName}
\affiliation[\LLNL]{\LLNLName}
\affiliation[\UBonn]{\UBonnName}
\affiliation[\USiegen]{\USiegenName}
\affiliation[\UHEI]{\UHEIName}
\affiliation[\Barry]{\BarryName}
\affiliation[\INR]{\INRName}
\affiliation[\Zagreb]{\ZagrebName}
\affiliation[\DESY]{\DESYName}
\affiliation[\Stockholm]{\StockholmName}
\affiliation[\Stanford]{\StanfordName}
\affiliation[\ICREA]{\ICREAName}
\affiliation[\MIPT]{\MIPTName}
\affiliation[\MIT]{\MITName}
\affiliation[\MPI]{\MPIName}
\affiliation[\INFNPavia]{\INFNPaviaName}
\affiliation[\Mainz]{\MainzName}
\affiliation[\KavliTokio]{\KavliTokioName}
\affiliation[\CapeTown]{\CapeTownName}
\affiliation[\IHEP]{\IHEPName}
\affiliation[\KAIST]{\KAISTName}

%% file: sections/introduction.tex

Axions and axion-like particles (ALPs), as well as other more generic categories of particles (weakly interacting sub-eV particles, WISPs) at the low-mass frontier \cite{Jaeckel:2010ni,Ringwald:2012hr,Hewett:2012ns} are acquiring a strong interest as a portal for new physics, candidates to the dark Universe, or as solutions of poorly understood astrophysical phenomena. The detection of these particles in terrestrial experiments is currently pursued by a number of experimental techniques, see~\cite{Graham:2015ouw,Irastorza:2018dyq} for recent reviews. In this experimental landscape, the International Axion Observatory (IAXO)~\cite{Irastorza:1567109,Armengaud:2014gea} stands out as one of the most ambitious projects under consideration. It is our purpose here to review the theoretical, cosmological and astrophysical motivation to carry out the primary goal of IAXO, i.e. the search for solar axions with sensitivity much beyond previous similar searches and well into unexplored parameter space. We will focus on more recent developments affecting regions of parameter space at reach of IAXO, as well as those highlighting its novelty and complementarity within the larger set of axion experimental efforts.

The QCD axion is a hypothetical $0^-$ particle predicted in the Peccei-Quinn mechanism to solve the strong CP problem~\cite{Cheng:1987gp} of the standard model (SM) of particle physics. In the pure SM, the parity (P) and time-reversal (T) violation effects observed so far can be attributed to the phase of the CKM matrix, which is relatively large $\delta\sim 30^o$ and has its origin in the Yukawa couplings of the Higgs to fermionic fields. The problem is that theory predicts the existence of another P,T violating phase, $\bar\theta$, which appears in the Lagrangian density multiplying the topological-charge density of QCD, 
\be
\label{thetaaxion}
{\cal L}_{\bar\theta} = -\frac{\alpha_s}{8\pi} G_{\mu\nu}^a \widetilde G^{a\mu\nu}\bar \theta \, . 
\ee
This term produces CP violating effects like electric-dipole moments (EDMs) for hadrons, which have never been observed. 
The strongest upper limit on strong CP violation comes from the neutron EDM, $|d_n|<3.0\times 10^{-13}$ e fm~\cite{Afach:2015sja}. 
Given the calculation $d_n=(2.4\pm 1.0)\bar\theta \times 10^{-3}$ e fm~\cite{Pospelov:2005pr}, one finds an extremely strong upper bound,  
\be
\label{barthetaexp}
|\bar\theta| < 1.3\times 10^{-10}.
\ee 
Indeed, $\bar\theta$ arises in the SM as the sum of two contributions: the $\theta$-angle defining a gauge-invariant QCD vacuum and a common phase of the quark-mass matrix. The latter has an origin similar to the CKM phase and the former has no clear a-priori relation with them so it is extremely suspicious that these two will cancel so precisely as \eqref{barthetaexp}.  

The solution proposed by  Peccei and Quinn~\cite{Peccei:1977hh,Peccei:1977ur} to alleviate the ``strong CP issue" is based on the observation that 
the QCD vacuum energy, $V_{\rm QCD}(\bar\theta)$, has an absolute minimum\footnote{In a strict sense this is true if $\bar\theta$ is the only source of CP violation in the SM. 
One expects a tiny shift of the minimum due to the non-zero CKM and from any other source of CP violation that can be communicated radiatively to the gluonic sector.  } at $\bar\theta=0$. They proposed the existence of an extra global U(1) symmetry, spontaneously broken and colour anomalous. Due to the colour anomaly, the concomitant Goldstone-boson field, dubbed ``axion"~\cite{Wilczek:1977pj} by Wilczek and denoted here by $A$, develops in the effective low-energy theory an anomalous coupling to gluons,
\be
\label{axicoup}
{\cal L}_{Ag} = -\frac{\alpha_s}{8\pi} G_{\mu\nu}^a \widetilde G^{a\mu\nu}\frac{A}{f_A},  
\ee
where $f_A$ is the so-called axion decay constant, a new energy scale related to the scale of PQ symmetry breaking. 
Effectively, $\bar\theta$ becomes replaced by $\bar\theta+\langle A\rangle/f_A$, where $\langle A\rangle$ is the vacuum-expectation-value (VEV) of the axion. 
Since the axion has no other potential energy terms in the Lagrangian (as long as PQ is a symmetry at the classical level) it will adjust itself to minimise $V_{\rm QCD}(\bar\theta+A/f_A)$ by taking the VEV $\langle A\rangle = -\theta f_A$, which cancels all the CP violating effects of $\bar\theta$!. 
Note that the axion field effectively becomes a dynamical version (it is a space-time dependent field) of the $\bar\theta$ angle, which is a mere constant.
 
The QCD potential gives the axion field a mass,
\be
\label{mass}
m_A =\frac{\sqrt{\chi}}{f_A}\simeq 5.7\, {\rm meV} \frac{10^9\, {\rm GeV}}{f_A}, 
\ee
(where $\chi = \left.\partial_{\bar\theta}^2 V_{\rm QCD}\right|_{\bar\theta=0}$ is the topological susceptibility of QCD)  
and induces mixing of the QCD axion field with the $\eta'$ and the rest of the $0^-$ mesons. 
By virtue of this mixing, the QCD axion develops model-independently couplings to hadrons and, most importantly, a coupling to two photons,  
\be
\label{twophoton}
{\cal L}_{A\gamma} = -\frac{\alpha C_{A\gamma}}{8\pi f_A} F_{\mu\nu} \widetilde F^{\mu\nu} A = 
-\frac{g_{A\gamma}}{4} F_{\mu\nu} \widetilde F^{\mu\nu} A = g_{A\gamma} \vec E\cdot \vec B A.
\ee
Here $F_{\mu\nu}$ is the electromagnetic field-strength, $\widetilde F^{\mu\nu}$ its dual and $\vec E,\vec B$ are the electric and magnetic fields, respectively. The coupling constant $g_{A\gamma}$ has units of an inverse energy scale, which is  $1/f_A$ except for an electromagnetic loop factor required for photon emission, $\alpha/2\pi$, and the mixing coefficient $C_{A\gamma} =-1.92$~\cite{diCortona:2015ldu}.  
The parameter $C_{A\gamma}$ can receive additional model-dependent additive contributions. 
These are usually of the order of $1$ in simple models~\cite{Kaplan:1985dv,Cheng:1995fd,Kim:1998va,DiLuzio:2016sbl} 
but can be much larger in some engineered cases~\cite{Farina:2016tgd,Agrawal:2017cmd}. 
Note that both the mass \eqref{mass} and couplings, like \eqref{twophoton}, are inversely proportional to $f_A$. Ratios of coupling/mass are independent of $f_A$ and only depend on QCD physics and a few hopefully O(1) model-dependent parameters so the properties of the QCD axion are reasonably constrained.   

The PQ solution to the strong CP problem is a minimal one, and is appealing for two reasons. The first is that the predicted axion particle can be experimentally confirmed or rejected (as opposed to some alternatives of the strong CP problem, e.g.~ \cite{Pospelov:2005pr}). So far QCD axions with  $f_a\gtrsim 10^6$~GeV have already been robustly excluded by solar axion searches,  stellar evolution, cosmology and laboratory  experiments. The second is that the QCD axion with a large decay constant $f_a\gg 10^6$~GeV is an excellent cold dark matter (CDM) candidate. 
The combination of the strong CP problem and the resounding evidence of CDM in the Universe motivate very strongly the search for a low mass QCD axion. 


Indeed, the popularity of the QCD axion has made generations of theorists to employ the word ``axion" for other, equally-hypothetical, particles that are not  related to QCD, the strong CP problem or the PQ mechanism simply because they have certain similarities to the QCD axion. 
The similarities often exploited are: being a low-mass $0^-$ particle, a pseudo-Goldstone boson of an ``axial" symmetry,  featuring anomalous couplings to topological-charge densities like \eqref{axicoup} for general non-abelian gauge-bosons, featuring anomalous couplings to two-photons, being low-mass and couple to two photons, to name a few. 
We will give more details below, but here we just note that the abuse is so overwhelming that we must bow to it and introduce here a disclaimer about nomenclature. Henceforth we will denote the true QCD ``axion" as ``QCD axion" or ``the axion", particles with similar properties as axion-like particles (ALPs) and the whole family as simply ``axions". The QCD axion will be $A$ and a generic ALP will be denoted as $a$. Couplings like \eqref{twophoton} will be generic for ALPs but the mass relation is exclusive of the QCD axion. 
It is extraordinary that axions appear so profusely in extensions of the SM. They are often encountered as pseudo-Nambu-Goldstone bosons of new global symmetries (spontaneously broken at a high energy scales) and in holomorphic theories of dynamical couplings, string theory being a prime example~\cite{Witten:1984dg,Conlon:2006tq,Svrcek:2006yi,Arvanitaki:2009fg}. 
Because of the similar properties of the QCD axion and ALPs, axion experiments have the potential to discover more than the solution of the strong CP problem and the dark matter of the Universe. 
An example is the search for solar axions and the proposed International Axion Observatory (IAXO), the focus of this monograph.

The two-photon coupling allows the production of axions in the collision of photons and charged particles via the Primakoff-effect $\gamma+q\to a+q$. 
The corresponding solar axion flux can be calculated to be 
$1.6\times 10^{21}C_{a\gamma}^2 (10^8 {\rm GeV}/f_a)^2/({\rm m}^2{\rm year})$ with a mean axion energy $\langle E_a\rangle=4.2$ keV.  
The axion coupling to electrons can be responsible for a copious flux $\sim 2\times 10^{24} C_{a e}^2 (10^8 {\rm GeV}/f_a)^{2}/({\rm m}^2{\rm year})$ with  $\langle E_a\rangle=2.3$ keV through the ABC processes~\cite{Redondo:2013wwa}. 
Note that although $C_{ae}$ is of order unity in some important models like DFSZ~\cite{Dine:1981rt,Zhitnitsky:1980tq} or general Grand Unification Theory axions (GUTs)~\cite{Ernst:2018bib}, it is absent at tree-level in other models, and the loop-induced contribution is very small. Fortunately, one can always resort to the former Primakoff flux. 
The solar axion flux is indeed copious even for very small values of $f_a$ because the Sun is huge, but the chances of detecting solar axions on Earth-based experiments are severely hampered by a small detection probability. 
Naive estimates of the natural value of detection cross sections of order $\sigma\sim g_{a\gamma}^2\sim 5\times 10^{-54} C_{a\gamma}^2 (10^8 {\rm GeV}/f_a)^{2}$ were initially discouraging. However, P. Sikivie presented in 1982 an experimental concept that uses the low mass of axions to boost the detection probability by making it coherent over macroscopic magnetic fields, the axion ``helioscope"~\cite{Sikivie:1983ip,VanBibber:1987rq}.


Axions travelling through a transversely polarised $B$-field can convert into photons at any point along the magnetic region. 
The photon polarisation is aligned to the external $B$-field due to the $\vec E \cdot \vec B$ coupling. 
The conversion probability can be understood as axion-photon oscillations~\cite{Raffelt:1987im} with oscillation length $\lambda_a=8\pi E/ \sqrt{m_a^4+(2g_{a\gamma }B E)^2}$ ($E$ is the axion energy), 
but also as the square of the coherent sum of the conversion amplitudes at each point along the line of sight~\cite{Redondo:2010dp}.
After a homogeneous magnetic length $L$, the probability of conversion is  

\be
P(a\to \gamma) = \frac{(2g_{a\gamma }B E)^2}{m_a^4+(2g_{a\gamma }B E)^2}\sin^2\frac{\sqrt{m_a^4+(2g_{a\gamma }B E)^2}L}{4 E},    
\ee
which is coherent ($\propto L^2$) if the magnetic length is smaller than the oscillation length $L \lesssim \lambda_a$. For small $m_a$ this can be very large and so can be the enhancement! For instance, for the parameters proposed for IAXO, this conversion probability is

\be
P(a\to \gamma) \simeq 10^{-19}C_{a\gamma}^2\(\frac{10^8\rm GeV}{f_a}\)^2\(\frac{B}{3\, \rm T}\)^2\(\frac{L}{20\, \rm m}\)^2, 
\ee 
and stays in the coherent regime for $m_a\lesssim 16$~meV, making the IAXO search for solar axions realistic.
Insisting on an incoherent detection scheme, one would 
need hundreds of kilometres of low-background instrumented detectors to obtain a similar conversion probability with $\sigma\sim g_{a\gamma}^2$. 
This is far too long compared to the most ambitious direct DM experiments looking for WIMP recoils, which are also parasitically used to search for solar axions.
Their virtue is that their detection does not rely on macroscopic coherence and therefore they are relatively insensitive to the axion mass. Unfortunately, 
they will not be in the foreseeable future as sensitive as indirect stellar evolution constraints, or as current helioscopes, see e.g.~\cite{Bernabei:2001ny,Li:2015tsa}.

For helioscopes, the smallness of the axion mass is a virtue, as the coherent length can be macroscopic and the use of long and intense magnets can greatly enhance the axion detection probability with respect to incoherent detection. This so-called coherent inverse Primakoff-process is at the core of the power of the Helioscope technique to find solar axions with IAXO but it can lead to other very interesting phenomena. 

Indeed, the most mature experimental efforts to detect axions in lab experiments~\cite{Graham:2015ouw,Irastorza:2018dyq} rely on their conversion to photons in macroscopic magnetic fields~\cite{Sikivie:1983ip,Raffelt:1987im}.
This approach includes the search for galactic axion dark matter~\cite{Sikivie:1983ip} and the photon regeneration experiments (``shining light through a wall'')~\cite{VanBibber:1987rq,Redondo:2010dp,Bahre:2013ywa}. 
There is a large complementarity between the different strategies. While relic axion searches enjoy high sensitivity in terms of $\gagamma$, they typically accomplish this only for a narrow $m_a$ range and rely on the assumption that the DM is mostly composed of axions. Laboratory experiments are free from cosmological or astrophysical assumptions on the production of ALPs, but they have a comparatively reduced sensitivity. Helioscopes enjoy an attractive compromise between axion source dependency (the solar flux is a robust prediction from any axion model) and competitive sensitivity, as will be shown. 
Indeed, a next generation helioscope is the only demonstrated technique that can discover QCD axions in the meV mass range~\cite{Raffelt:2011ft}. 
The CERN Axion Solar Telescope (CAST)~\cite{Zioutas:2004hi,Andriamonje:2007ew,Arik:2008mq,Arik:2013nya}
, in operation at CERN for more than a decade, represents the current state-of-the-art of the axion helioscope technique. 
The International Axion Observatory (IAXO) has been recently proposed as a follow-up of CAST, scaling the helioscope concept to the largest size realistically allowed. IAXO will be largely based on experience and concepts developed by CAST. 

In this paper we provide an updated description of the physics case of IAXO. We organise our paper by reviewing the contexts in which the discovery of the QCD axion or an ALP in the parameter space accessible by IAXO could have strong implications in other contexts of theoretical particle physics, cosmology and astrophysics. 
We set the stage in section~\ref{sec:StringAxions} with a detailed theoretical background for axions in extensions of the SM, reviewing recent advances in axion model building and axion couplings, and placing particular emphasis on stringy axions.

Section~\ref{sec:axiondm} is devoted to axions as dark matter. Indeed, a great deal of the appeal of axions is that they can be produced in the early Universe via non-thermal processes (realignment mechanism and decays of cosmic strings and domain walls~\cite{Bae:2008ue,Visinelli:2011wa}), which makes them well motivated cold dark matter candidates.
The relic density depends on $f_a,m_a$ but also on the initial conditions and the cosmological history until today. For the pure QCD axion, one can distinguish between two broad classes: 
\begin{itemize}
\item If the PQ symmetry spontaneously breaks before inflation and is not restored afterwards,  
the axion field becomes homogeneous during inflation at some a-priori-unknown VEV.  
A broad range of axion masses can produce the correct relic density just adjusting to the adequate initial conditions, including the $m_A\sim$~meV values that could be discovered with IAXO.  
\item If the symmetry is restored after inflation, initial conditions are reset and the QCD axion becomes coherent only at very small scales compared to today's horizon. An average over disconnected patches removes the uncertainty of initial conditions but the quantity of axions radiated from cosmic strings is still uncertain and those radiated from domain wall (DW) decay is model dependent. Models with short-lived DWs ($N_{\rm DW}=1$) can give the totality of the DM for $m_A\gtrsim 26 \,\, \mu$eV, the uncertainty in the calculation encompassing the $m_A\sim $~meV values accessible by IAXO.  
In models with long-lived DWs, the DM yield can be greatly increased, which points to $m_A\sim $~meV as the most favoured mass range to account for all the DM in the Universe.
\end{itemize}
We discuss the status of the meV QCD axion as a DM candidate in detail in section~\ref{sec:axiondm}. 
General ALP models have ample freedom to become DM candidates in a broad range of values of $f_a$ and $m_a$~\cite{Arias:2012az}. 

Axions can be also produced from thermal processes or decays of heavy particles, contributing to the radiation energy density as effective relativistic degrees of freedom in the early Universe and/or today. Current cosmological observations present some tension that could suggest the presence of this Dark Radiation, in addition to the known neutrino species. For some values of the relevant parameters, this axionic Dark Radiation could give observable signals in helioscopes like IAXO. This scenario is described in section~\ref{sec:AxionDR}.

The fact that ALPs have masses protected from large radiative corrections is very suggestive to use them as candidates for the inflaton.
A recent scenario showed that, for an adequate potential and coupling to photons (for reheating purposes), an axion accessible to IAXO might indeed be responsible for cosmic inflation.
This interplay between ALPs and inflation is reviewed in section~\ref{sec:inflation}. 

The existence of axions can have very important consequences in astrophysics. 
Indeed, the properties of well-known stellar systems have been providing the strongest constraints on axion properties for a long time~\cite{Raffelt:1999tx}.
More intriguing axion effects may account for unexplained astrophysical observations. Two of these cases deserve special consideration: the anomalous cooling rate observed in a number of stellar systems and the excessive transparency of the intergalactic medium to very high energy (VHE) photons.  
In both contexts, the existence of very light axions with properties at reach of IAXO has been repeatedly invoked as an explanation, although claims will be taken with caution, as more conservative explanations cannot be excluded. The observational situation as well as the potential interpretation in terms or axions are reviewed for both cases in sections~\ref{sec:cooling_hints} and~\ref{sec:transparency}, respectively.

To conclude the physics case update, we briefly review the conceptual design of IAXO~\cite{Armengaud:2014gea} in section~\ref{sec:iaxo}, present updated sensitivity projections of the experiment and conclude in section~\ref{sec:conclusions}.

%% file: sections/stringyaxions.tex

General Relativity (GR) is a perturbatively unitary theory only up to the Planck scale. 
Whether or not GR is non-perturbatively unitary is still an open question but the requirement of a Wilsonian (perturbative) unitarization of GR has a very promising venue in string theory where point particles are replaced by extended objects: strings.
For theoretical consistency, string theory requires additional spatial dimensions beyond the known three. 
Experiments and observations constrain these additional dimensions to be compact and very small. 
The compactification schemes determine many of the properties of the low-energy four-dimensional effective description.

The field content in the higher dimensional theory includes many anti-symmetric tensor fields ($p$-forms) which may be integrated over any non-trivial cycle in the compactification geometry. In four-dimensions, the values of such integrals appear as dynamical scalar fields, and the gauge symmetry of the $p$-forms translates into a global shift symmetry making them candidate axions~\cite{Witten:1984dg,Conlon:2006tq,Svrcek:2006yi,Arvanitaki:2009fg}. These are commonly known as `closed string' axions. 
In addition, D-branes (places where open strings can end) that fill four-dimensional spacetime can support additional `open string' axions, which can realise the PQ-mechanism~\cite{Choi:2011xt} or appear as independent ALPs.

In some string models, the mass spectrum and couplings of these scalars to ordinary matter can be explicitly computed \cite{Choi:2006qj, Cicoli:2012sz}. This serves as the starting point for several interesting applications to cosmology, astrophysics and particle physics. 
In particular, one linear combination of these scalars will play the r\^ole of the standard QCD axion if coupled to the QCD sector realised on D-branes, i.e. if  leading to the low-energy coupling~\eqref{axicoup}. 
All other scalars are instead ALPs, which might behave very similarly to the QCD axion but whose mass and decay constant are not related by QCD scales but by some other detail of the model. Effectively, mass and couplings become independent parameters.  
Hence the corresponding parameter space is much wider, featuring for instance regions where ALPs can successfully: ($i$) drive inflation with the production of observable primordial gravity waves \cite{Pajer:2013fsa}, ($ii$) account for the observed dark matter content of the Universe \cite{Sikivie:2006ni, Arias:2012az}, ($iii$) contribute to dark radiation by behaving as extra neutrino-like degrees of freedom \cite{Cicoli:2012aq, Higaki:2012ar}, or ($iv$) become the longitudinal component of extra $U(1)$ gauge bosons with mass well below the string scale which could belong to either the visible or a hidden sector \cite{Ghilencea:2002da, Goodsell:2009xc, Cicoli:2011yh}.

It is worth mentioning that some of these potential axions might acquire large masses~\cite{Cicoli:2012sz}. 
For instance, they could be simply excluded in the compactifications leading to the SM gauge group \cite{Grimm:2004uq}. 
Moreover, they could be eaten up by anomalous $U(1)$'s with masses of order the string scale in the Green-Schwarz mechanism of anomaly cancellation \cite{Allahverdi:2014ppa}). 
Finally, they could acquire a very large mass if they are stabilised in a supersymmetric way since they would become as massive as the corresponding supersymmetric partners, the so-called \textit{saxions}, 
which have to be heavier than $\sim$~50~TeV to decay before primordial nucleosynthesis and avoid altering the successful Big Bang predictions \cite{Coughlan:1983ci, Banks:1993en, deCarlos:1993wie}.

For each of the above discussed `closed string' axions, there is generically also a `modulus' scalar field that parametrises the size or shape of the compactification manifold. In contrast to the `closed string' axions, moduli fields do not enjoy a shift symmetry, are not in general protected from large quantum corrections to their mass and therefore might be easily too heavy to be considered axions.  
However, the imaginary part $a$ of a \textit{moduli} field $\Phi= \phi\,+ \,{\rm i}\, a$ is exactly shift symmetric and thus is a natural candidate for an axion. 

The real part, the saxion $\phi$, parametrises either the size or the shape of the extra dimensions and comes from the dimensional reduction of the ten-dimensional metric. Hence the saxion is a gauge singlet with only gravitational couplings to ordinary matter. Moreover, as already mentioned, $\phi$ does not enjoy a shift symmetry, and so can be lifted and become massive by any kind of perturbative effect. Therefore it is crucial to study moduli stabilisation in order to work out the axion mass spectrum for phenomenological applications. There are two benchmark scenarios \cite{Conlon:2006tq, Cicoli:2012sz}: 
\begin{itemize}
\item The moduli are fixed by non-perturbative effects. Since the same mechanism gives mass to both $\phi$ and $a$, they will naturally tend to acquire masses  of the same order of magnitude, which is constrained by the saxion phenomenology to  $m_a\sim m_\phi \gtrsim 50$ TeV, too massive for our purposes. 
\item The saxion masses are fixed by perturbative effects. 
Axions will remain massless because they are protected by their perturbative shift symmetry, except for the non-perturbative effects which are now unconstrained by saxion phenomenology.
In this last case, the actual value of the axion mass is model-dependent. 
However, since the non-perturbative effects depend on exponentials of volume factors, the axions $a$ can be exponentially lighter than saxions and thus naturally very light.  
\end{itemize}

Let us stress that fixing all the moduli via non-perturbative effects is highly non-generic because of various technical issues that can forbid non-perturbative effects \cite{Blumenhagen:2009qh}. Therefore we conclude that the low-energy four-dimensional limit of string compactifications generically gives rise to the presence of multiple light axions which can include the QCD axion plus one or more axions whose mass spectrum and couplings depend on the microscopic details of the physics and the geometry of the extra dimensions \cite{Cicoli:2012sz}. The axions which acquire small masses are of course of phenomenological interest. 

The decay constants of stringy axions depend on their microscopic origin as closed or open string modes. 
In particular, the size of $f_a$ is determined by the geometry of the compactified dimensions. 
The closed string axion coupling $f_a$ can be either of order the compactification scale (the Kaluza-Klein scale $M_{\rm KK}$) for strings associated with internal cycles in the extra dimensions bulk, or it can be of order the string scale $M_s$, for strings associated with resolutions of local singularities~\cite{Cicoli:2012sz,Conlon:2006tq}.

Given that $M_s \sim g_s^{1/4}\,M_{\rm Pl} / \sqrt{\mathcal{V}}$ and $M_{\rm KK} \sim M_s / \mathcal{V}^{1/6}$, where $M_{\rm Pl} = 2.435 \cdot 10^{18} \,\text{GeV}$ denotes the (reduced) Planck mass, $g_s$ the string coupling, and $\mathcal{V}$ the internal volume in string units, compactifications at large volume and weak string coupling ---both preferable properties of string models under perturbative control---  predict $f_a \ll M_{\rm Pl}$.

In models where all the moduli are safely heavier than $50$ TeV, $M_s$ and $M_{\rm KK}$ tend to be around the GUT scale which is therefore the natural value to expect for the decay constant of closed string axions: $f_a^{\rm closed}\sim 10^{16}$\,GeV.\footnote{A significantly lower $f_a^{\rm closed}$ could be achieved if the cycle supporting the axion is located in a highly warped region of the compactification, where the effective action is however under less control.} Notice also that in the closed string axions the effective PQ symmetry is always broken in the 4D effective field theory below the Kaluza-Klein scale.
However, for sufficiently large $\mathcal{V}$, the decay constant of closed string axions can fall within the intermediate scale window \cite{Conlon:2006tq} and thus the reach of IAXO. However, in these models with relatively low string and Kaluza-Klein scale, some moduli tend to get lighter than $50$ TeV and cause cosmological problems. 

Let us turn now to open string axions. They arise as the phases of complex fields charged under anomalous $U(1)$s that could live on a stack of D-branes \cite{Choi:2011xt}. In the process of anomaly cancellation, the $U(1)$ acquires a mass of order $M_s$ by eating up a closed string axion \cite{Green:1984sg}. 
Therefore, the effective field theory below the string scale contains an effective global $U(1)$ PQ-like symmetry.
The axion decay constant is then set by the magnitude of the radial part of the charged open string mode. In turn, in supersymmetric theories this radial part is set by a model-dependent Fayet-Iliopolous term. For models with D-branes at singularities where the SM sector is sequestered from the sources of supersymmetry breaking in the bulk \cite{Blumenhagen:2009gk, Aparicio:2014wxa}, the decay constant of open string axions can be suppressed with respect to $M_s$ due to either sequestering effects or a mixing of sequestering and U(1) kinetic mixing, leading to an intermediate scale $f_a^{\rm open}$. In some of these models the resulting decay constant can be as small as $f_a^{\rm open}\sim 10^8$ GeV \cite{Cicoli:2012sz, Cicoli:2017zbx}, in an interesting region testable by IAXO.

In sum, axions are ubiquitous in string theory, and while most axions may be rendered heavy, some may remain light in the low-energy theory and contribute to a variety of physical processes that can be accessible to axion experiments like IAXO.

%% file: sections/axionmodels.tex

The low-energy description of the properties and interactions of an axion is strongly determined by the requirement of possessing a global shift symmetry,  
\be
a \to a +\epsilon f_a . 
\ee
If the symmetry is exact, the axion will be massless, as a mass term ${\cal L}_a\propto m_a^2 a^2$ does not respect the shift symmetry. 
Low mass axions can exist if the symmetry is very weakly broken. There are two conceptually different ways of achieving this breaking. 
On the one hand, the shift symmetry can simply be explicitly broken by some term in the Lagrangian that is extremely small. In these 
cases there is a set of couplings or parameters $\{p\}$ which, when taken to zero, leave the Lagrangian density invariant under the shift symmetry. 
In this case, the axion mass can receive radiative corrections from the high-energy sector of the theory, hinting to a possible hierarchy problem. However, they will always be proportional to these coefficients. If those are very small, the theory can have low mass axions and still be technically natural. 
On the other hand, the shift symmetry might be violated only by anomalous couplings like~\eqref{axicoup}. 
These soft-breaking terms can and do generate non-trivial potentials for axions but do not generate UV-sensitive radiative corrections to their mass. 
The resulting low mass axion is therefore naturally light. 
In practice it is often assumed that global symmetries are not respected by quantum gravity effects because sufficiently classical black-holes have no hair and thus can swallow arbitrary global charges. Therefore, unless the global symmetry is a remnant from a gauge symmetry or has some other means of protection, one assumes that quantum gravity will generate some kind of explicit symmetry breaking and thus a small potential for axions (and thus a mass). 
This kind of reasoning will play a role in Sec.~\ref{sec:axiondm}. 

In either case, the Lagrangian of a low-mass axion can be divided into a shift-symmetric and a shift-breaking part. The second must be subdominant to consider our axion naturally light. We assume that our axion is a periodic angular field defined in $a \in (-\pi,\pi)v_a$ where $v_a$ is a scale related to the spontaneous symmetry breaking of the shift symmetry. 
Below $v_a$ and the electroweak scales, but above the QCD confining scale the Lagrangian of an axion can be written as 
\bea
{\cal L}_{a} &=& \frac{1}{2}(\partial_\mu a)(\partial^\mu a) +
\sum_{ij} \frac{c_{aij}}{2v_a} (\bar \psi_i \gamma^\mu\gamma_5 \psi_j) \partial_\mu a
-\frac{{\calE} \alpha}{2\pi v_a} \frac{F_{\mu\nu} \widetilde F^{\mu\nu}}{4} a  \\ 
&& -\frac{{\calN} \alpha}{2\pi v_a} \frac{G_{\mu\nu} \widetilde G^{\mu\nu}}{4} a + ... 
\eea
where $c_{aij}'s$ are dimensionless coupling coefficients, $\psi_i$ are the SM quarks and leptons,  $\calE$ and $\calN$ are the electromagnetic and colour anomalies of the shift symmetry and the ellipsis stands for additional explicit symmetry breaking terms.  
If $\calN\neq 0$ and additional explicit breaking terms are sufficiently small, the axion becomes the QCD axion. 
If we compare with \eqref{axicoup}, we find that we should define $f_A = v_A/\calN$. 
In this case, it is important to note that $V_{\rm QCD}(\bar\theta)$ is 2$\pi$ periodic in $\bar\theta$ and so it will in $A/f_A$. Therefore, there are $\calN$ physically different values of the QCD axion VEV that minimise the QCD potential and cancel CP violation, $\bar\theta - \langle A\rangle/f_A= 0,1,2,...N-1$. 
Once again, this will be crucial for our section on axion CDM, Sec.~\ref{sec:axiondm}. 

Below the QCD confining scale, $\Lambda_{\rm QCD}$, quarks and gluons are no longer adequate degrees of freedom and any axion coupling to $G\widetilde G$ will mix with the neutral mesons and acquire a non-trivial potential. We define the low-energy Lagrangian for a generic axion as 
\bea
\label{alplowenergyL}
{\cal L}_{a} &=& \frac{1}{2}(\partial_\mu a)(\partial^\mu a) +
\sum_{ij} \frac{C_{aij}}{2f_a} (\bar \psi_i \gamma^\mu\gamma_5 \psi_j) \partial_\mu a
-\frac{C_{a\gamma} \alpha}{2\pi f_a} \frac{F_{\mu\nu} \widetilde F^{\mu\nu}}{4} a   \\ \nonumber
&& \quad -V_{\rm QCD} \(\bar\theta - \frac{\calN a}{v_a} \) + ... 
\eea
where the ellipsis stands now for shift-symmetry couplings to mesons and explicit symmetry breaking terms. 
We also define the phenomenological coupling constants 
\be
g_{aff'} = C_{aff'}\frac{m_f+m_{f'}}{2 f_a}  \quad , \quad g_{a\gamma} = \frac{\alpha}{2\pi f_a} C_{a\gamma},  
\ee
where $m_f$ are SM fermion masses. 

The distinction between $f_a$ and $v_a$ is only relevant for the QCD axion in our context so we will use $f_a=v_a$ for ALPs (we remove $V_{\rm QCD}$ but we allow for other sources of ALP mass in the ellipsis). 
The coupling constant $C_{a\gamma}$ has a model-dependent contribution $\propto \calE$ and a model-independent contribution from meson-mixing just for the QCD axion~\cite{diCortona:2015ldu}, 
\bea
\label{cagcoupling}
C_{A\gamma} =& -1.92(3) + \frac{{\calE}}{{\calN}} \quad &(\text{QCD axion}),   \\ \nonumber 
C_{a\gamma} =& \calE  \quad &(\text{ALP}).
\eea
The couplings to fermions are very similar. 
They are inherited from the high-energy coefficients $c_{aij}$ except for a model-independent part which arises for the QCD axion coupling to hadrons. 
The relevant couplings for the low-energy phenomenology discussed here are to protons, neutrons and electrons. 
The case of an ALP is quite unconstrained so we quote here those of the QCD axion~\cite{diCortona:2015ldu}, 
\bea
\label{fermioncouplings}
C_{Ap} &=& -0.47(3) + \Delta C_{App} \quad ,\quad  C_{An} = -0.02(3) + \Delta C_{Ann} \\ 
C_{Ae} &=& \frac{\alpha^2 C_{A\gamma}}{3\pi}\log\(\frac{\Lambda_{\rm QCD}}{m_e}\) + c_{aee} 
\eea
where $ \Delta C_{App}, \Delta C_{Ann}$ are the model-dependent parts that arise from the model-dependent QCD axion couplings to SM quarks. 
The model-independent axion-electron coupling is zero at tree-level so we have included the loop correction arising from the axion-photon coupling~\cite{Srednicki:1985xd}, although it is typically very small. Similar loop effects appear for the proton and neutron couplings but are not very relevant. 

If SM fermions do not transform under the PQ symmetry, QCD axions do not couple with them at tree level ($c's$ are all zero).  
These are called ``hadronic axions'', of which the Kim-Shifman-Vainshtein-Zakharov 
(KSVZ)~\cite{Kim:1979if,Shifman:1979if} model is an often quoted example. 
A recent work~\cite{DiLuzio:2016sbl,DiLuzio:2017pfr} has classified a wealth of these models constrained by their cosmological viability. 
Simple models involve one extra charged singlet under the SM gauge group and a new heavy coloured fermion and populate the range 
$0.25 < |C_{A\gamma}|<12.75$. The range broadens when multiple coloured new fermions are allowed~\cite{DiLuzio:2016sbl} but only very special charge assignments and field contents lead to significantly different values of $C_{A\gamma}$. 
Therefore, it turns out that this study provides a very nice bracketing of the photon coupling in axion models, encompassing  predictions beyond the KSVZ-type constructions. Therefore, henceforth we will use the above-mentioned range when we plot the predictions for axion models for the axion-photon coupling.

Let us now consider the cases where SM fermions transform under the PQ symmetry. The simplest flavour preserving models involve two-Higgs-doublet models:  
 the Peccei-Quinn model itself where $f_a\sim v_{\rm EW}$---ruled out long ago--- and the Dine-Fischler-Srednicki-Zhitnitsky (DFSZ)~\cite{Dine:1981rt,Zhitnitsky:1980tq}, where an additional SM singlet is responsible for the very large value of $f_a$.  
Two existing DSFZ variants have ${\calE}/{\calN}=8/3$ (type-I) and 2/3 (type-II). 
Their coupling to electrons is $C_{Ae}=\cos^2\beta /3,\sin^2\beta /3$ respectively, depending of the ratio of the two Higgs particle VEVs $\tan\beta=v_2/v_1$. 
These potentially large couplings to electrons make these models especially interesting for explaining the stellar cooling anomalies and to be detected by IAXO. 

Introducing new SM Higgs doublets/singlets, new PQ fermions and/or non-flavour diagonal couplings gives rise to 
axions featuring a broad range of couplings to SM particles, including flavour violating ones~\cite{Wilczek:1982rv}, which are strongly constrained by rare decays. Reference~\cite{DiLuzio:2017ogq} demonstrated that, by adjusting the couplings to quarks and leptons, it is possible to arrange the model-dependent couplings to cancel the model-independent contributions to the proton, neutron and electron couplings to the QCD axion, the so-called \emph{astrophobic} axion.  This model and those accidentally similar are not severely constrained by stellar evolution (indeed supernova, neutron star, white dwarf and red-giant constraints become essentially irrelevant) and might be most advantageously discovered by IAXO itself.  
It is worth mentioning that other recently proposed models/constructions like the minimal flavour violating axion~\cite{Arias-Aragon:2017eww}, the axi-flavon~\cite{Calibbi:2016hwq} and flaxion~\cite{Ema:2016ops} do not seem to be particularly astrophobic in general, although a recent study focused on minimality found two interesting exceptions~\cite{Bjorkeroth:2018ipq}.

Another interesting QCD axion model is SMASH \cite{Ballesteros:2016xej,Ballesteros:2016euj}, together with other axi-Majoron models~\cite{Dias:2014osa}. 
SMASH was born as a minimal self-consistent model solving the most pressing issues of particle physics and cosmology: neutrino masses, inflation, dark matter, baryogenesis, the strong CP problem and the stability of the Higgs potential. SMASH is a KSVZ-type QCD axion model where the PQ symmetry mixes with lepton number and its spontaneous breaking gives a majorana mass to right-handed neutrinos. Note that the QCD axion here plays the role of the majoron as well. The SMASH axion as well as other majoron models does not have tree-level coupling to electrons, but they feature a potentially large radiative correction with right-handed neutrinos in the loop. Another interesting connection between the QCD axion and neutrinos is the Ma-axion~\cite{Ma:2017vdv}. 

At the end of the day, concerning solar axion detection, the main important model dependency is whether axions have couplings to electrons with $C_{Ae}\sim {\cal O}(1)$ like DFSZ, or the couplings to electrons are loop-suppressed like in KSVZ.
In the event of a discovery, IAXO could measure the QCD axion mass as well as the photon and electron couplings independently~\cite{Dafni:2018tvj,Jaeckel:2018mbn}. 
Therefore it can potentially distinguish between generic possibilities and pinpoint very conspicuous models in optimistic conditions. 
We review this possibility in Sec.~\ref{sec:measaxipara}.

%% file: sections/darkmatter.tex

The QCD axion is regarded as one of the best motivated candidates to be the DM of the Universe.
Its couplings to SM particles are suppressed due to a large decay constant $f_A$, which ensures the stability and collisionless properties of DM.
Furthermore, the axions are produced non-thermally in the early Universe, and hence they are ``cold" in the sense that
their velocity dispersion is small enough to fit the observed large scale structure.

The cosmological production mechanism of axion DM, called the vacuum realignment mechanism,
was first discussed in~\cite{Preskill:1982cy,Abbott:1982af,Dine:1982ah}.
In the early Universe, the axion field may have an expectation value which is different from
the vacuum expectation value (VEV) at the minimum of the effective potential at the present Universe.
The effective potential of the axion field arises due to instantons: non-perturbative and topologically non-trivial configurations of the gluon fields in QCD, 
which become increasingly relevant as QCD becomes non-perturbative, i.e. as the temperature of the Universe approaches the confining phase-transition around $\mathcal{O}(0.16)\,\mathrm{GeV}$.
When the QCD potential becomes sizeable, the axion field starts to oscillate around the minimum of the effective potential,
and such a coherent oscillation of the classical axion field contributes to the matter energy density of the Universe.
Taking into account this production mechanism, it was concluded in~\cite{Preskill:1982cy,Abbott:1982af,Dine:1982ah} that the QCD axion
DM would overclose the Universe ($\Omega_{A}>1$) if the decay constant takes values above $f_A\approx \mathcal{O}(10^{12})\,\mathrm{GeV}$ corresponding to the axion mass of $m_A\approx \mathcal{O}(10^{-6})\,\mathrm{eV}$.

Although the early discussion described above strongly motivated us to consider the axion as a DM candidate,
the estimate of the relevant mass range was somewhat simplistic and should be refined in light of recent developments of
theoretical and observational cosmology. 
First of all, the precise estimation of cosmological parameters revealed that DM constitutes only a fraction of
the total energy density of the Universe. According to the recent results of the Planck collaboration, the matter density parameter is
determined as $\Omega_{\rm c} h^2= 0.1206\pm 0.0021$ (Planck TT+lowE)~\cite{Aghanim:2018eyx}, where $h$ represents the value of the Hubble parameter today,
$H_0=100 h\,\mathrm{km}\,\mathrm{sec}^{-1}\mathrm{Mpc}^{-1}$.
An order of magnitude improvement in the accuarcy of $\Omega_{\rm c}$ leads to a tighter
constraint on the decay constant $f_A\lesssim \mathcal{O}(10^{11})\,\mathrm{GeV}$ for axion DM
produced by the realignment mechanism~\cite{Wantz:2009it}.
Furthermore, the axion DM abundance estimate is not so straightforward if we follow the evolution of the axion field
in the context of inflationary cosmology.
If the Peccei-Quinn (PQ) symmetry is restored after inflation,
topological defects such as strings and domain walls are formed, and they could also produce significant amounts of cold axions~\cite{Davis:1986xc,Lyth:1991bb}.
On the other hand, if the PQ symmetry is never restored after inflation, such defect contributions can be neglected
since their density is significantly diluted due to the exponential expansion during the inflationary epoch.

\begin{figure}[htbp]
\includegraphics[scale=0.44]{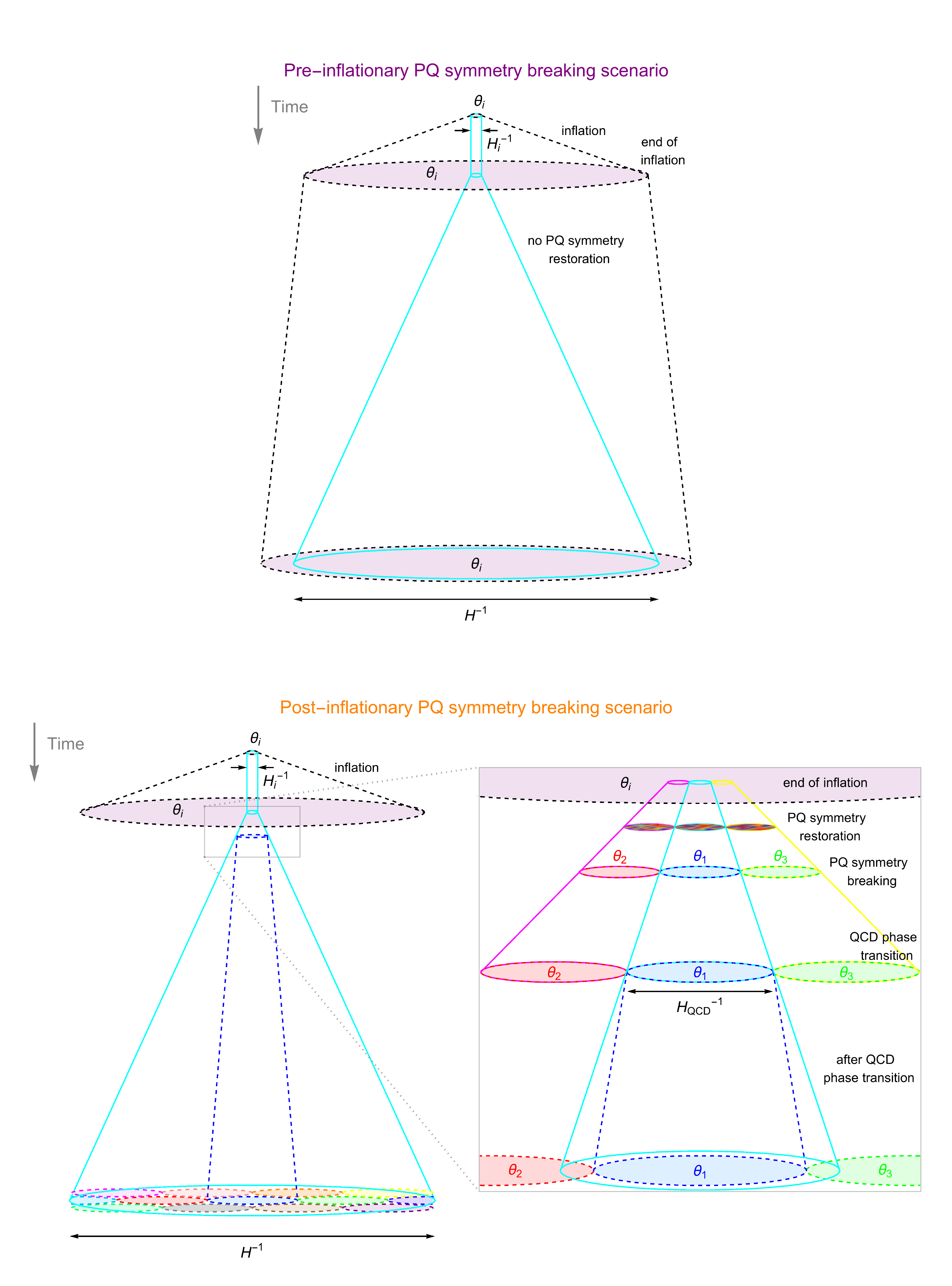}
\caption{A schematic view of the time evolution of the spatial distribution of $\theta$ field in the pre-inflationary PQ symmetry breaking scenario (top) and the post-inflationary PQ symmetry breaking scenario (bottom). In the pre-inflationary scenario, inflation makes the initial value of $\theta = \theta_i$ uniform in all the observable Universe. In the post-inflationary scenario, the Universe ends up containing many different patches that had different values of $\theta$ at the time of QCD phase transition. We also expect that topological defects such as strings and domain walls form around the borders of the patches. }
\label{theta_evolution}
\end{figure}

The difference between the two possible scenarios is sketched in Fig.~\ref{theta_evolution}.
Here we show how the spatial distribution of an angular field $\theta (x)$ evolves over time, where we introduce the dimensionless axion field 
\begin{equation}
\theta (x) = \frac{A (x)}{f_A}, 
\end{equation}
which ranges from $-\pi$ to $\pi$. 
Let us assume that the PQ symmetry has been broken at a very early stage, and that the $\theta$
field takes a certain initial value $\theta_i$ within the Hubble radius $\sim H_I^{-1}$ at that time. If  inflation occurs at a later time, the physical scale at which $\theta$ takes a uniform value $\theta_i$
exponentially grows, while the Hubble radius remains almost constant.
After inflation, the Hubble radius grows with time, $H^{-1}\sim t$.
As shown in the upper panel of Fig.~\ref{theta_evolution},
if the PQ symmetry is never restored after inflation, the size of the patch of the Universe
in which $\theta$ takes the value $\theta_i$ can be much larger than the Hubble radius even at the present time.
In this case, we can assume that the axion field has the same initial value $\theta_i$
throughout the observable Universe at the onset of its coherent oscillation.
We call this scenario the \emph{pre-inflationary} PQ symmetry breaking scenario.

The situation is drastically different if the PQ symmetry is restored after inflation, which is shown in the lower panel of Fig.~\ref{theta_evolution}.
If the reheating temperature after inflation is large enough , the $\theta$ field has large fluctuations after inflation, taking random values from $-\pi$ to $\pi$ on microscopic scales and restoring the PQ symmetry.
After  the Universe expands and cools enough, the PQ symmetry is spontaneously broken and $\theta$ takes
a certain value $\theta_1$ within the Hubble radius at that time.
Note that, however, $\theta$ takes different values ($\theta_2$, $\theta_3$,$\dots$) outside the Hubble radius, since such regions are causally disconnected.
The Hubble radius continues to expand as time goes on, and the field relaxes to the uniform value within the new Hubble radius in order to minimize the gradient energy.
Such behaviour continues until the time of the QCD phase transition, at which the mass energy of the axion field cannot be neglected and it starts to oscillate
around the minimum of the potential. The typical scale of spatial variation of the axion field at the onset of the oscillation
is given by the Hubble radius at the time of the QCD phase transition $\sim H_{\rm QCD}^{-1}$.
Such a scale (typically a few comoving mpc if the Universe is radiation dominated at the time) becomes shorter than the Hubble radius at later times, and we expect that our present observable Universe contains many different patches which had different values of $\theta$
at the time of the QCD phase transition.
Furthermore, we also expect that topological defects such as strings and domain walls are formed around the boundaries of different patches.
Therefore, the relic axion density should be estimated by summing over all possible field configurations, which include various initial values for $\theta$ at the onset of the coherent oscillation
and those produced by the collapse of topological defects.
We call this scenario the \emph{post-inflationary} PQ symmetry breaking scenario.

Since there is a conceptual difference in the spatial distribution of the axion field between the pre-inflationary and post-inflationary PQ symmetry breaking scenarios,
estimates of the axion DM abundance will be different accordingly.
In the following subsections, we will discuss these two scenarios separately.

\subsection{The pre-inflationary Peccei-Quinn symmetry breaking scenario}

First, let us consider the pre-inflationary PQ symmetry breaking scenario.
In this case, we can ignore the contribution from topological defects, and the relic axion DM abundance can be estimated
by simply following the evolution of the homogeneous axion field $\theta$, which can be described by the standard realignment mechanism.
As mentioned earlier, the axion field has a certain initial value $\theta_i$ throughout the observable Universe,
and the relic axion abundance depends on this initial value as well as on $f_A$.

The axion potential arises from QCD non-perturbative effects and it is thus extremely suppressed at
high temperatures due to the asymptotic freedom of the QCD running coupling. The axion is therefore effectively massless
at high temperatures. As the temperature of the Universe approaches the QCD confinement scale, the axion mass becomes
significant and starts influencing the axion field evolution. 
This phenomenon can be modelled by the following $T-$dependent potential,
\begin{equation}
V_{\rm QCD} (A,T) = \chi(T)\left[1-\cos\left(\frac{A}{f_A}\right)\right], \label{finite_T_potential_axion}
\end{equation}
where we emphasise that $\chi=\chi(T)$ depends on $T$. Note that we have redefined $A/f_A \to A/f_A-\bar\theta$ for simplicity.  
Recently, lattice calculations of the topological susceptibility in full QCD became available~\cite{Borsanyi:2016ksw},
whose behavior at high temperatures is close to the power law $\chi(T) \propto T^{-n}$ with $n = 8.16$
predicted by the dilute instanton gas approximation~\cite{Pisarski:1980md,Gross:1980br,Borsanyi:2015cka}.
Below the QCD phase transition, $\chi$ becomes a constant $\chi_0=(75.44(34)\rm  MeV)^4$~\K{\cite{Gorghetto:2018ocs}.}
In what follows, we estimate the axion DM abundance by using the latest lattice QCD results~\cite{Borsanyi:2016ksw}.\footnote{
The state of the art calculations of the topological susceptibility $\chi(T)$ based on lattice simulations of full QCD
result in some discrepancy among different numerical treatments~\cite{Borsanyi:2016ksw,Bonati:2015vqz,Petreczky:2016vrs}.
In particular, the result of~\cite{Bonati:2015vqz}, which shows a much milder temperature dependence,
leads to an order of magnitude smaller value for the axion mass explaining the total DM abundance
compared with that obtained in~\cite{Borsanyi:2016ksw}.
Such a result might be interpreted as huge lattice artifacts caused by strong cutoff effects~\cite{Borsanyi:2016ksw,Petreczky:2016vrs}.}

\begin{figure}[htbp]
\centering
\includegraphics[width=0.48\textwidth]{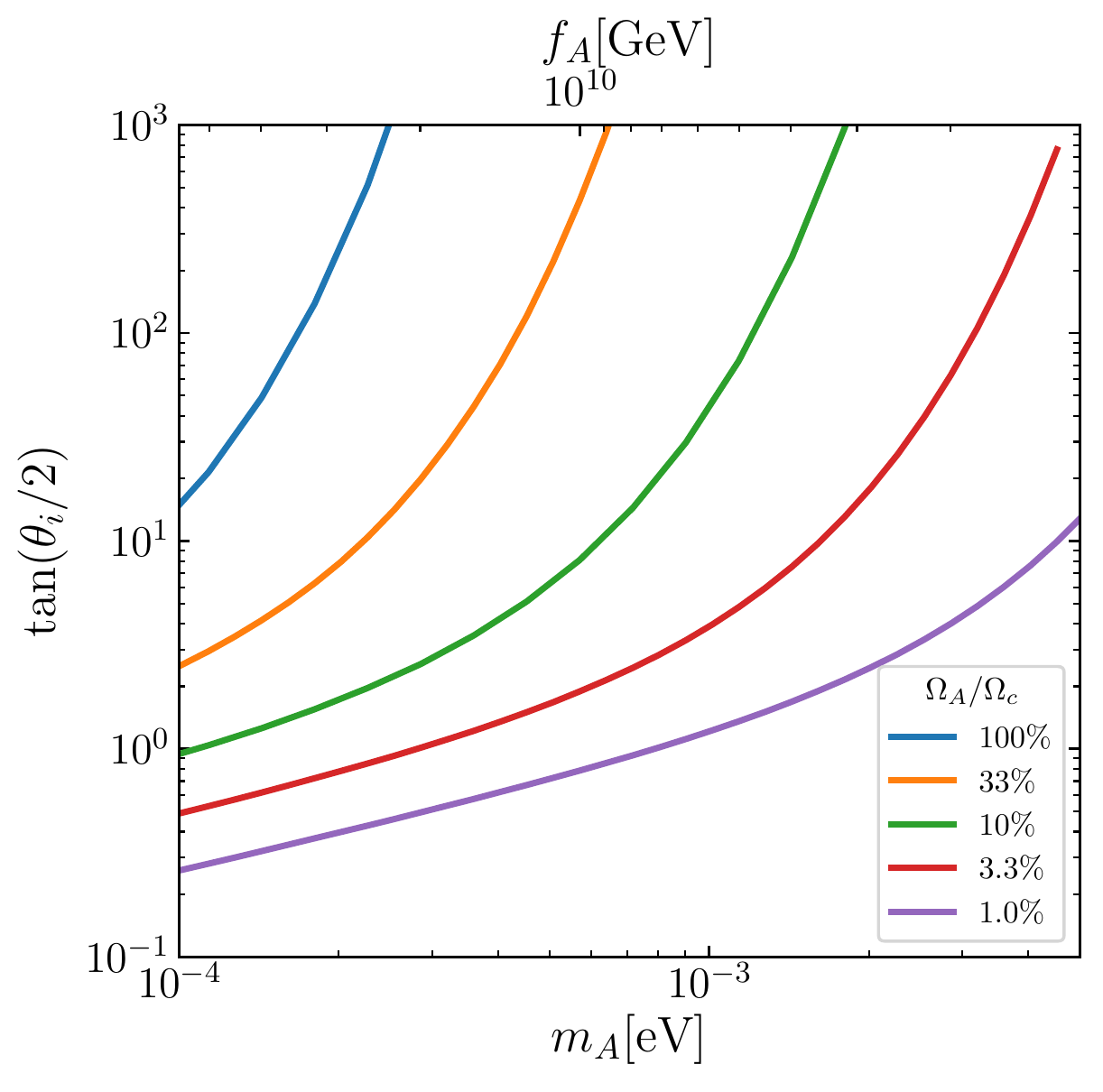}
\includegraphics[width=0.48\textwidth]{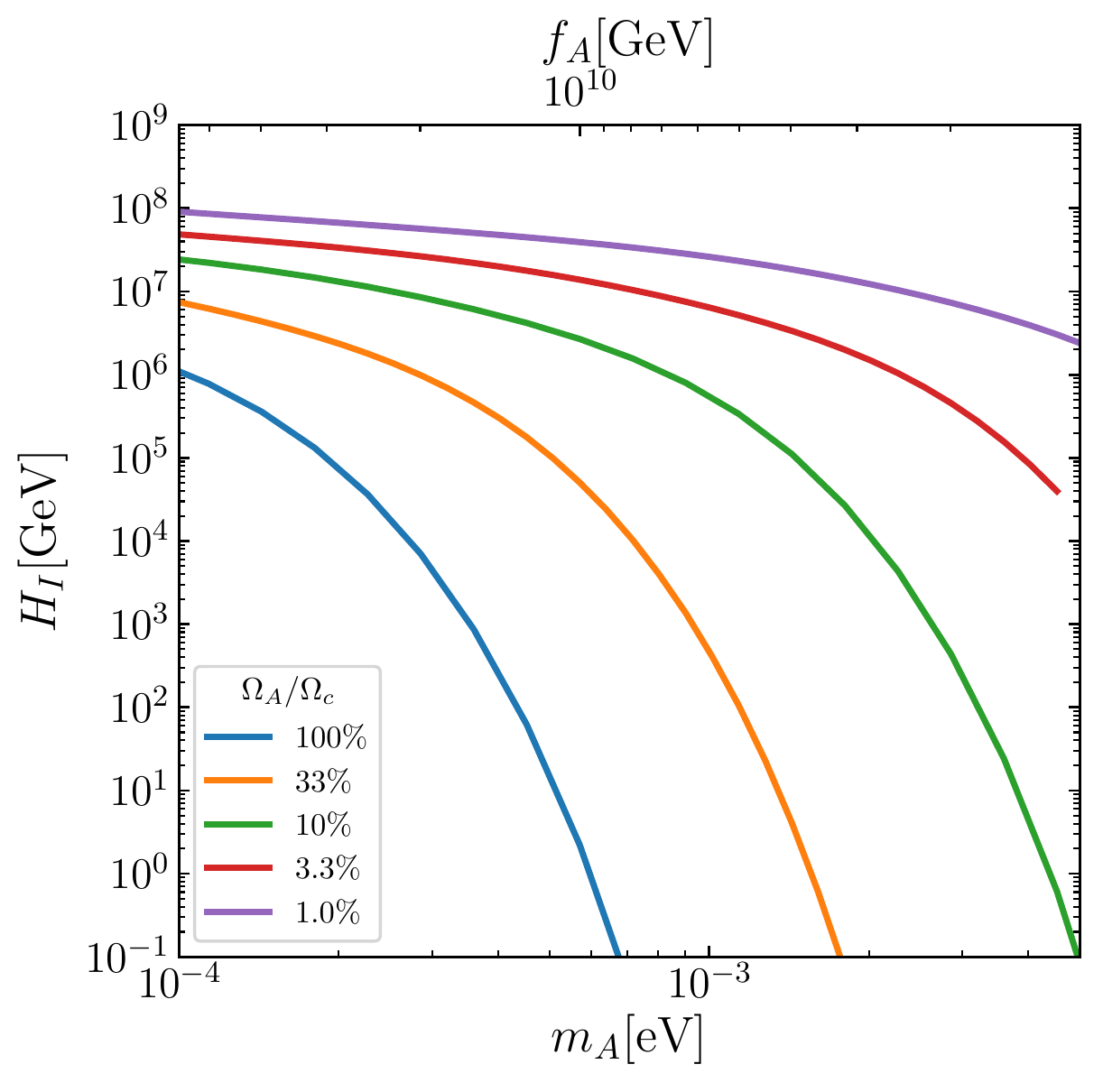}
\caption{Left: Value of the misalignment angle after inflation (initial conditions of the axion DM field) required to obtain today
different fractions of the observed DM abundance.
Right: Upper bound on the Hubble constant during inflation from Planck's absence of isocurvature fluctuations for axion DM having such DM fractions.
These bounds can be evaded in a number of scenarios~\cite{Linde:1991km,Kasuya:2009up,Folkerts:2013tua,Jeong:2013xta}.
Both calculations are performed with the topological susceptibility and equation of state of the SM from \cite{Borsanyi:2016ksw}.
}
\label{fig_theta1}
\end{figure}

The zero momentum mode of the axion field obeys the following field equation,
\begin{equation}
\ddot{\theta} + 3 H\dot{\theta} + m_A^2(T)\sin\theta = 0. \label{homogeneous_axion_field_eq}
\end{equation}
The realignment production of axions occurs when the Hubble friction becomes smaller than the potential force, {\it i.e.} $3H\lesssim m_A$.
At that time the axion field starts to oscillate around the minimum of the potential.
The solution of Eq.~\eqref{homogeneous_axion_field_eq} guarantees the conservation of the axion number in the comoving volume,
from which we can obtain the present density of axion DM. 
Assuming radiation domination during the onset of oscillations, and the harmonic approximation  $\sin\theta\sim \theta$ one finds~\cite{Ballesteros:2016xej}, 
\begin{equation}
\Omega_{A,\rm{real}}h^2 \approx 0.35 \left(\frac{\theta_i}{0.001}\right)^2\times
\left\{
\begin{array}{ll}
\left(\frac{f_A}{3\times 10^{17}\,\mathrm{GeV}}\right)^{1.17} & \quad \text{for}\quad f_A\lesssim 3\times 10^{17}\,\mathrm{GeV},\\
\left(\frac{f_A}{3\times 10^{17}\,\mathrm{GeV}}\right)^{1.54} & \quad \text{for}\quad f_A\gtrsim 3\times 10^{17}\,\mathrm{GeV}.
\end{array}
\right.
\label{omega_a_real_pre}
\end{equation}
The case $f_A \gtrsim 3\times 10^{17}\,\mathrm{GeV}$ corresponds the onset of axion oscillations happening
when the axion mass has already reached its $T=0$ value ($\chi = \chi_0$).
In the case of interest to Helioscope detection, $f_a \ll 3\times 10^{17}\,\mathrm{GeV}$,  the onset of axion oscillations happens before the confining phase transition, when $\chi$ depends very strongly on the temperature. 
Unfortunately, Eq.~\eqref{omega_a_real_pre} is based on
an approximate solution, which is valid as long as $|\theta_i|\ll \pi$ (corresponding to $f_a \gtrsim \mathcal{O}(10^{11})\,\mathrm{GeV}$).
If $|\theta_i|$ takes a larger value, the factor $\theta_i^2$ in Eq.~\eqref{omega_a_real_pre} is replaced by a $\theta_i$-dependent correction term, which can be quantitatively estimated by solving the non-linear field equation [Eq.~\eqref{homogeneous_axion_field_eq}] numerically.
The anharmonic correction factor becomes very large in the limit $|\theta_i|\to \pi$, and in such a case
$\Omega_{A,\rm{real}}h^2$ can account for the total cold DM abundance for the lower values of the decay constant $f_A$ accessible to IAXO. 
We have computed the required value of $\theta_i$ to obtain $100\%,33\%,10\%,3.3\%$ and $1\%$ of the observed DM abundance and plot them in 
Fig. \ref{fig_theta1}. One can see that very high fine-tunings of $\theta_i$ are required to have $100\%$ for $f_A\sim 10^9$ GeV, already noticed by~\cite{Wantz:2009it}, but the values are quite reasonable for a few $\%$. It is precisely in these meV-mass scenarios, in which the direct axion DM detection is more difficult and one predicts typically a small DM abundance, where IAXO can be the only way of discovering the QCD axion.

In the pre-inflationary scenario, quantum fluctuations of an axion field during inflation lead to isocurvature fluctuations in its realignment dark matter density that are imprinted in the temperature fluctuations of the cosmic microwave background, whose amplitude is strongly constrained by observations~\cite{Linde:1985yf,Seckel:1985tj}. This constraint applies to any ALP DM candidate, not only to QCD axions. 
The upper limits on isocurvature fluctuations result in a constraint that relates the Hubble scale during inflation $H_I$ (which gives the typical size of the fluctuations in the $a$ field), the fraction of axion cold DM to the total, $r_{\rm aDM}=\Omega_{a,\rm tot} h^2/0.12$, and the decay constant $f_a$ (during inflation)  that might depend on anharmonic effects if $\theta_i$ is large. In the simplest cosmological scenario, the constraint is~\cite{Kobayashi:2013nva}
\be
r_{\rm aDM} \frac{d \ln \Omega_a}{d \theta_i} \frac{H_I}{2\pi f_a} \lesssim 10^{-5}. 
\ee
The case of the QCD axion is shown in Fig. \ref{fig_theta1} (right) as an upper bound on $H_I$ for different values of $r_{\rm ADM}$. 
Note that the above graph uses our numerical calculations of $\theta_i$ based on~\cite{Borsanyi:2016ksw}, which is specific to the QCD axion.  
One can see that values of $f_A\sim 10^9$ GeV and a significant contribution of axion DM $r_{\rm ADM}\sim{\cal O}(1)$, require quite low $H_I$. 
Note that a bound on $H_I$ constrains the maximum possible reheating temperature to be $T_{\rm RH}\sim \sqrt{H_I M_{\rm Pl}}$ and this is constrained to be $T_{\rm RH}\gtrsim 5$ MeV by Big Bang nucleosynthesis~\cite{Kawasaki:1999na,Kawasaki:2000en,Hannestad:2004px,Ichikawa:2005vw} and Cosmic Microwave Background (CMB) arguments~\cite{Ichikawa:2006vm,deSalas:2015glj} but otherwise there is no phenomenological prejudice against small scale inflation. 
Indeed, a recent paper turned this argument around and showed that the 
quantum fluctuations during a low-$H_I$ period of inflation may also be responsible for generating a large enough misalignment angle to give rise to QCD axion or ALP DM~\cite{Graham:2018jyp,Guth:2018hsa}.

Future CMB polarisation probes like CMB-S4 and LiteBIRD can measure the CMB polarisation effect of gravitational waves from which we can obtain the value of $H_I$. However, due to the moderate sensitivity improvement with respect to the current limits the measured values will not be far from $H_I\sim 10^{14}$ GeV. 
Thus, if a future CMB polarisation experiment measures $H_I$, we will be forced to consider high scale inflation, which limits even tiny fractions of axion DM in this pre-inflation scenario. It is worth noting here that the isocurvature spectrum can be reduced in a number of non-minimal models~\cite{Linde:1991km,Kasuya:2009up,Folkerts:2013tua,Jeong:2013xta},  so pre-inflationary phenomenology is not completely eliminated in this scenario.
However, we might say that the simplest axion models of high-scale inflation favour the scenario where the PQ symmetry is broken after inflation, which we discuss in the following subsection.

\subsection{The post-inflationary Peccei-Quinn symmetry breaking scenario}

The axion appears as a pseudo Nambu-Goldstone boson in the low energy effective theory,
and such a description is only valid at energies below the scale of symmetry breaking $v_A$.
Once the symmetry is restored, we cannot use the effective field theory description in terms of the axion field $A(x)$.
Instead, we can consider the evolution of a gauge singlet complex scalar field $\Phi(x)$ (the PQ field for QCD axions), which transforms as $\Phi\to\Phi e^{i\alpha}$ with $\alpha$ being a real constant parameter under the global U(1)$_{\rm PQ}$ symmetry.
The PQ symmetry is spontaneously broken when the PQ field acquires a VEV $|\langle\Phi\rangle| = v_{\rm PQ}/\sqrt{2}$,
and after that the axion field $A(x)$ is identified as an angular direction of the PQ field, {\it i.e.} $\Phi = (v_{\rm PQ}/\sqrt{2}) \exp(i A(x)/v_{\rm PQ})$. 
The VEV of $\Phi$, $v_{\rm PQ}$ here plays the role of $v_A$. 
The Lagrangian for the PQ field is given by
\begin{equation}
(\sqrt{-g})^{-1}\mathcal{L} = \partial_{\mu}\Phi \partial^{\mu}\Phi^* -V_{\rm eff}(\Phi,T),
\end{equation}
where $V_{\rm eff}(\Phi,T)$ represents the finite-temperature effective potential for the PQ field,
\begin{equation}
V_{\rm eff}(\Phi,T) = \lambda\left(|\Phi|^2-\frac{v_{\rm PQ}^2}{2}\right)^2 + \frac{\lambda}{3}T^2|\Phi|^2.
\end{equation}
If we assume that the reheating temperature $T_R$ after inflation is sufficiently high, $T_R\gg v_{\rm PQ}$,
the PQ symmetry is restored ($\langle\Phi\rangle =0$) at early times.
Subsequently, it is broken when the temperature drops below $T\lesssim v_{\rm PQ}$.\footnote{The PQ symmetry may also be
restored non-thermally due to non-perturbative field dynamics after inflation~\K{\cite{Tkachev:1995md,Kasuya:1998td,Tkachev:1998dc,Kasuya:1999hy,Kawasaki:2013iha,Ballesteros:2016xej}}.
The evolution of topological defects in this case is similar to that in the thermally restored case once they enter the scaling regime,
and there is no significant difference in the prediction of DM abundance between the two cases.
However, in the non-thermally restored case, relativistic axions can be abundantly produced, which tends to violate dark radiation constraints
if such axions are not thermalized~\cite{Ballesteros:2016xej}.}

When the U(1)$_{\rm PQ}$ symmetry is spontaneously broken, vortex-like objects called strings are formed due to the Kibble mechanism~\cite{Kibble:1976sj}.
After their formation, the string network evolves according to an approximate scaling solution~\cite{Kibble:1984hp,Bennett:1985qt,Gorghetto:2018myk}, where the typical length scale of long strings is given by the cosmic time.
In order to maintain the scaling property, the energy stored in strings is dissipated as radiation of massless axions,
which leads to a further contribution to the present axion DM abundance~\cite{Davis:1986xc}.

\begin{figure}[htbp]
\centering
\includegraphics[width=0.7\textwidth]{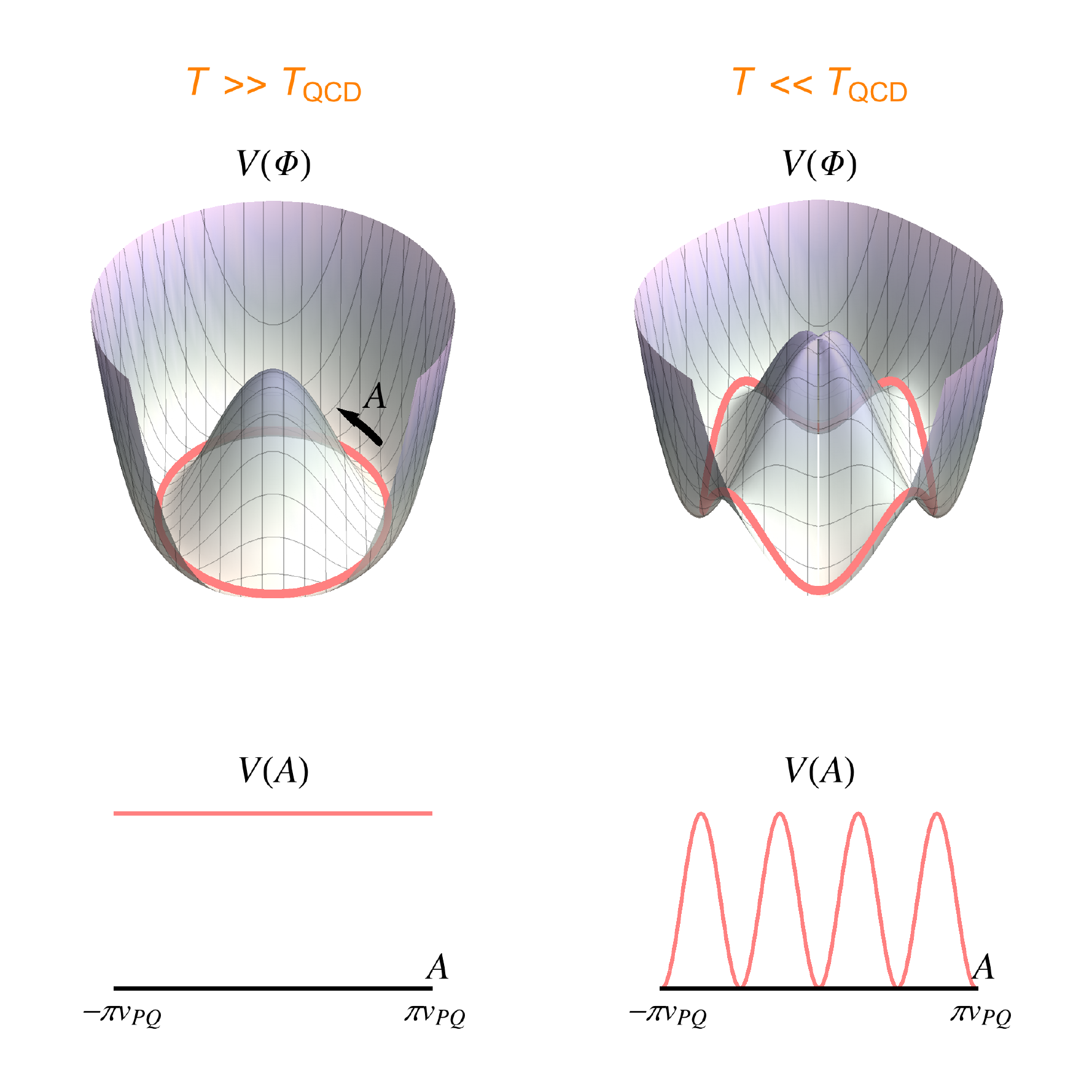}
\caption{Effective potential of the PQ field (top) and that of the axion field (bottom).
The left (right) panel shows the potential at the temperature much higher (lower) than the critical temperature of the QCD phase transition.
The axion field corresponds to the direction along the bottom of the potential $V(\Phi)$ (pink lines), and such a direction is exactly flat at high temperatures.
At low temperatures, the periodic axion potential appears due to the presence of the topological susceptibility $\chi(T)$, which gives rise to
$N_{\rm DW}$ different minima. In these figures, we choose $N_{\rm DW}=4$.}
\label{fig_potential}
\end{figure}

In addition to the dynamics of strings, we must take into account that of domain walls, which appear at the epoch of the QCD phase transition.
The appearance of domain walls is understood through the effective potential of the axion field [Eq.~\eqref{finite_T_potential_axion}] written in terms of $v_{\rm PQ}$, 
\begin{equation}
V_{\rm QCD} (A,T) = \chi(T)\left[1-\cos\left(\frac{N_{\rm DW}A}{v_{\rm PQ}}\right)\right], \label{finite_T_potential_axion_NDW}
\end{equation}
where $N_{\rm DW}=\calN$ is the colour-anomaly of the PQ symmetry, which turns out to be a positive integer and it is called the ``domain wall number" in this context.
The above potential explicitly breaks the U(1)$_{\rm PQ}$ symmetry into its discrete subgroup $Z_{N_{\rm DW}}$, in which
the QCD axion field transforms as $A\to A+2\pi v_{\rm PQ} k/N_{\rm DW}$ ($k=0,1,\dots,N_{\rm DW}-1$).
Such a symmetry breaking effect is irrelevant at early times since it vanishes at high temperature, $\chi(T)\to0$,
but it gradually arises as the Universe cools. This effect is schematically shown in Fig.~\ref{fig_potential}.
Once the Hubble friction becomes smaller than the potential force,
the VEV field settles down into one of the $N_{\rm DW}$ degenerate vacua $\langle A\rangle_k = 2\pi v_{\rm PQ} k/N_{\rm DW}$.
Since the field value $\langle A(x)\rangle$
must be uncorrelated over distances larger than the Hubble radius at the QCD phase transition $\sim H_{\rm QCD}^{-1}$,
the QCD axion field relaxes into different vacua from one Hubble volume to another. 
Continuity of the VEV demands that there must be a sheet-like boundary between these regions  in which the energy density is as high as $V_{\rm QCD}\approx 2\chi(T)$.
Such a non-trivial field configuration is called a domain wall, and its formation is an inevitable consequence of the QCD axion model~\cite{Sikivie:1982qv}.

\begin{figure}[htbp]
\begin{center}
\includegraphics[scale=.4]{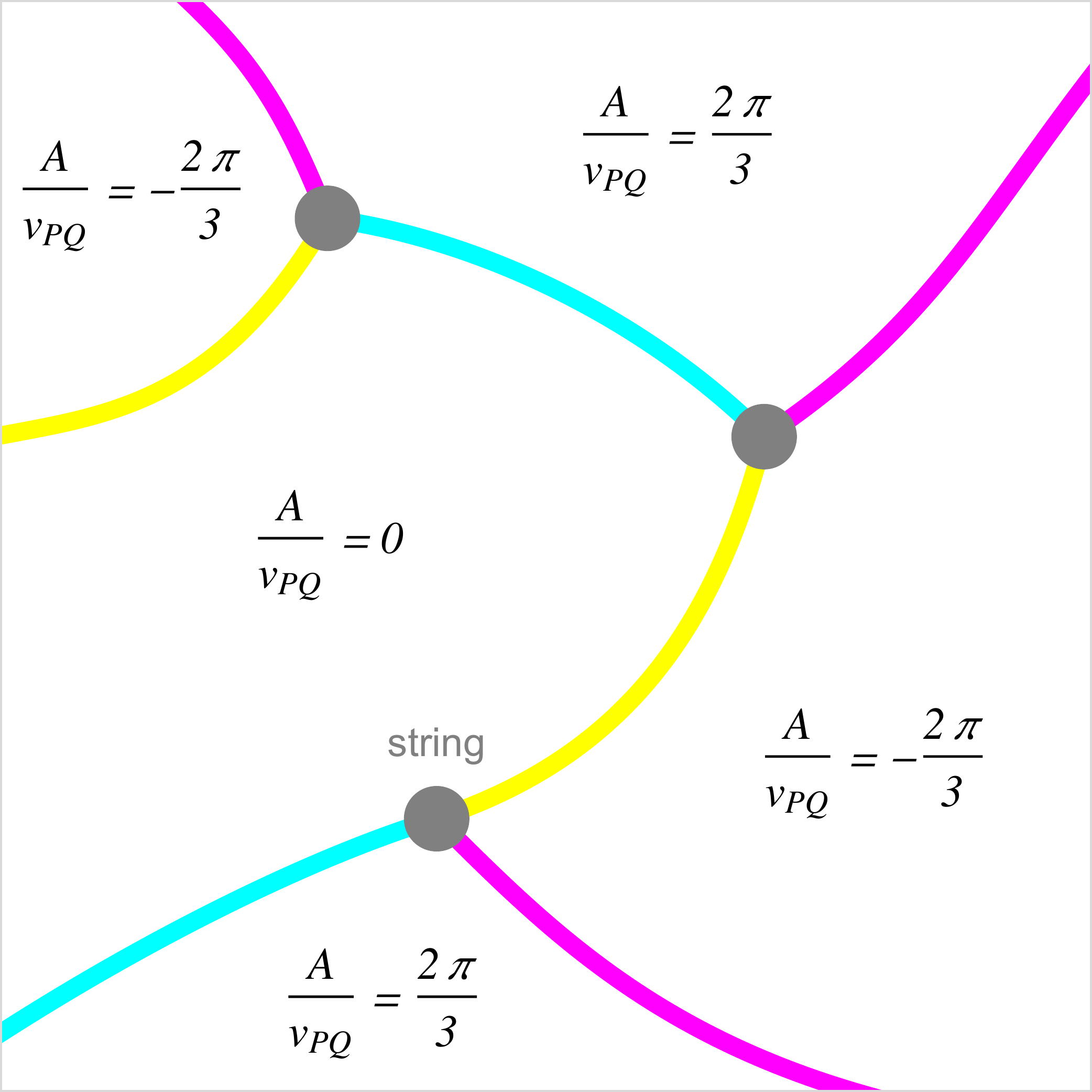}
\caption{A top view of string-wall systems with $N_{\rm DW}=3$.
Gray circles represent the cores of strings, which are attached by three domain walls.
Coloured lines correspond to the position of the centre of domain walls.
Domain walls exist around boundary of three disconnected vacua, in which the axion field has a value
$A/v_{\rm PQ}=0$, $2\pi/3$, and $-2\pi/3$. 
Around the centre of domain walls, it has a value $A/v_{\rm PQ}=\pi/3$ (cyan), $\pi$ (magenta), and $-\pi/3$  (yellow).}
\label{string_wall_section}
\end{center}
\end{figure}

Figure~\ref{string_wall_section} illustrates a two-dimensional section of topological defects in the axion model with $N_{\rm DW}=3$.
We note that in every case strings are attached by $N_{\rm DW}$ domain walls, since the value of the phase of the PQ field $\langle A(x)\rangle/v_{\rm PQ}$ must continuously change from $-\pi$ to $\pi$ around the string core.
Therefore, we expect that hybrid networks of strings and domain walls, called string-wall systems, are formed at around the epoch of the QCD phase transition.

The domain wall number $N_{\rm DW}$ determines the number of degenerate vacua in the effective potential of the $A$ field.
We have already mentioned that in models where only one global U(1)$_{\rm PQ}$ symmetry exists, the value of $N_{\rm DW}$ is determined by
the colour anomaly coefficient, ${\calN}$. In simple models this coincides with the number of new quark flavours that transform under the global U(1)$_{\rm PQ}$ symmetry~\cite{Sikivie:1982qv}. For instance, in the KSVZ model there is one hypothetical heavy quark which transforms under U(1)$_{\rm PQ}$ and $N_{\rm DW}={\cal N}=1$ while in DFSZ-I all of the standard model quarks transform under U(1)$_{\rm PQ}$ and we have $N_{\rm DW}=6$. 

The subsequent evolution of the string-wall systems differs according to whether $N_{\rm DW}=1$ or $N_{\rm DW}>1$.
If $N_{\rm DW}=1$, strings are pulled by one domain wall, which causes the disintegration into smaller pieces of a wall bounded by a string~\cite{Vilenkin:1982ks}.
Therefore, these string-wall systems are unstable, and they collapse soon after their formation.
On the other hand, if $N_{\rm DW}>1$, the tension force of domain walls acts on strings from $N_{\rm DW}$
different directions, which makes the systems stable. Once such stable string-wall systems are formed,
their evolution also obeys the scaling solution, in which the typical distance between two neighbouring walls
is given by the Hubble radius. Since the energy density of such scaling domain wall networks decays slower than
that of radiation or matter, it eventually overcloses the Universe, which conflicts with many observational results~\cite{Zeldovich:1974uw,Sikivie:1982qv}.
Therefore, QCD axion models with $N_{\rm DW}>1$ suffer from the cosmological domain wall problem if the PQ symmetry is broken after inflation.

One possible solution to the domain wall problem is to consider models where the PQ symmetry is violated by something more than the colour anomaly. Indeed, a small explicit symmetry breaking term in the Lagrangian \eqref{alplowenergyL} leads to a small energy bias between degenerate vacua~\cite{Vilenkin:1981zs,Sikivie:1982qv,Gelmini:1988sf}. The small bias acts as a pressure force on the walls, squashing the false vacuum region and allowing the string-wall network to collapse.
The magnitude of the bias must be large enough such that the collapse of domain walls occurs before they overclose the Universe,
but must be sufficiently small because it also generically shifts the minimum of the overall effective potential field away from the CP 
conserving minimum, which spoils the solution to the strong CP problem. 

If we consider the PQ symmetry to be an accidental symmetry at energies $\sim v_{\rm PQ}$, the explicit symmetry breaking terms appearing from dynamics at much higher energy scales $\Lambda$ could be represented by higher dimensional PQ violating operators like
\begin{equation}
\Delta V = c_N\Lambda^4\left(\frac{\Phi}{\Lambda}\right)^{N}+h.c., \label{PQ_violating_operators}
\end{equation}
where $c_N$ is a dimensionless constant.
With this parametrisation, even if $\Lambda$ is as high as the Planck scale, one needs to suppress the modulus of the coefficients $c_N$ with $N<8$ 
or so to avoid fine-tuning the phases. 
This is naturally realised in scenarios where the PQ symmetry breaking operators are constrained to fulfill fundamental exact discrete symmetries~\cite{Choi:2009jt,Ringwald:2015dsf}. 

In summary, there are two distinct possibilities in the post-inflationary PQ symmetry breaking scenario according to the domain wall number $N_{\rm DW}$.
The string-wall systems are short-lived in the case with $N_{\rm DW}=1$, while they are long-lived in the case with $N_{\rm DW}>1$.
In both cases the estimation of the relic QCD axion DM abundance is quantitatively different from the conventional case, since we must take
account of the contribution from the collapse of string-wall systems.
In the next subsection, we briefly review the recent results on the axion DM abundance and relevant mass ranges in these two scenarios.

\subsection{Production of dark matter axions from topological defects}

In the post-inflationary PQ symmetry breaking scenario, the present total QCD axion DM abundance can be somewhat artificially split into a sum of three contributions, 1) the contribution from the realignment mechanism, 2) that from global strings, and 3) that from the decay of string-wall systems,
\begin{equation}
\label{Omega_a_tot}
\Omega_{A,\rm{tot}}h^2 = \Omega_{A,\rm{real}}h^2 + \Omega_{A,\rm{string}}h^2 + \Omega_{A,\rm{dec}}h^2. 
\end{equation}
We note now that ALPs can also develop string-wall networks and be produced in similar ways than QCD axions if they are endowed with periodic potentials with one or several minima. The QCD axion is peculiar in that we know its potential and its temperature dependence. 

\begin{figure}[htbp]
\includegraphics[width=1.0\textwidth]{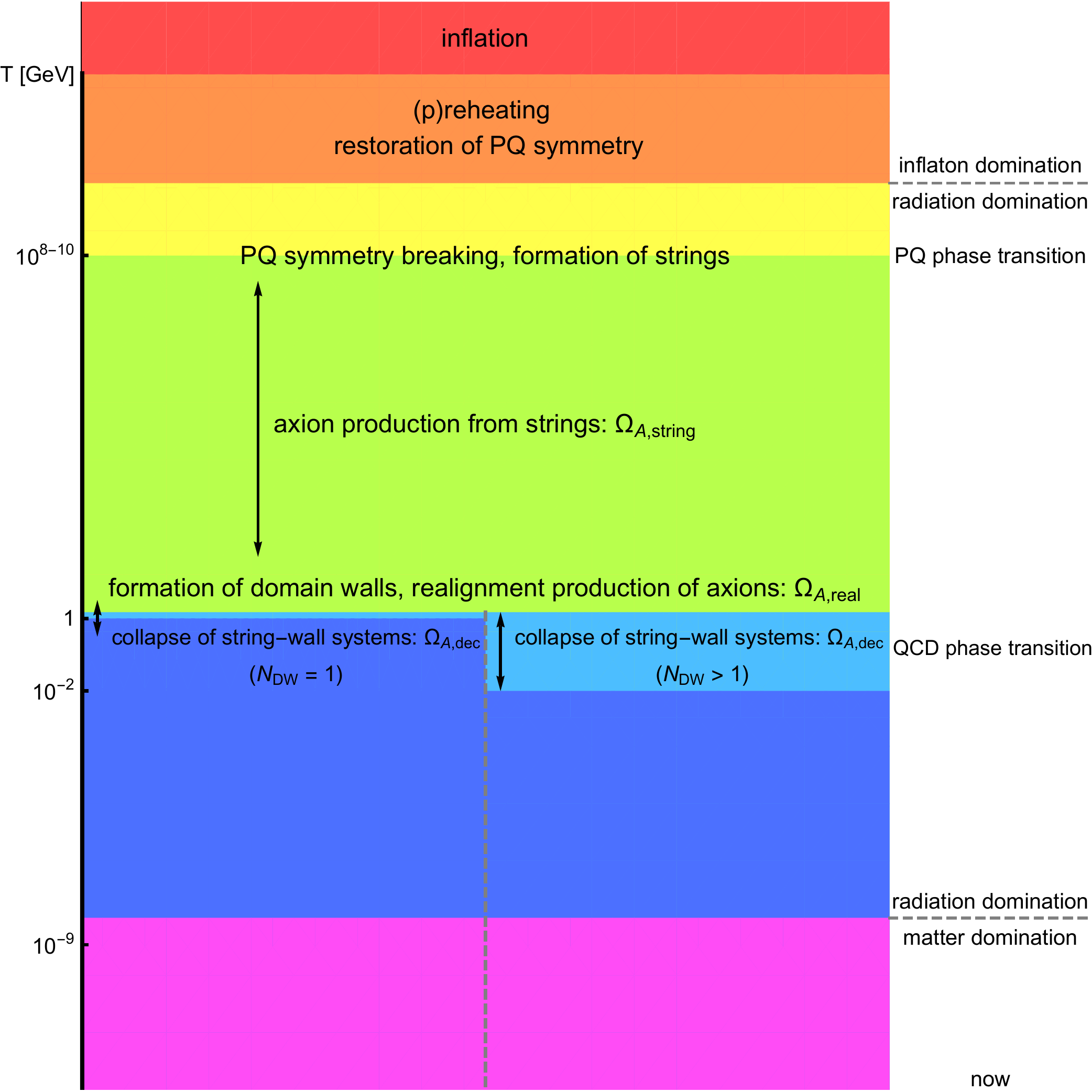}
\caption{A schematic diagram of the thermal history of the Universe in the post-inflationary PQ symmetry breaking scenario.
Two possibilities for the evolution of string-wall systems, the case with $N_{\rm DW}=1$ and that with $N_{\rm DW}>1$,
are parallelly shown in the region below $T\approx 1\mathrm{GeV}$.}
\label{fig_timeline}
\end{figure}

We summarize the timeline of the history of the early Universe in Fig.~\ref{fig_timeline}.
Here, it is assumed that inflation has happened at sufficiently high energy scale,
and that the subsequent reheating is so efficient that the PQ symmetry is restored after inflation.
The PQ symmetry is spontaneously broken when the temperature of the Universe becomes $T\sim v_{\rm PQ}$.
At that time strings are formed, and they continue to produce massless axions until the epoch of the QCD phase transition.

The simple picture of the strings decaying into massless axions does not hold once the axion mass becomes non-negligible,
which corresponds to the temperature of the Universe of $T\approx 1\,\mathrm{GeV}$, and at that time
axions are also produced from the realignment mechanism.
The realignment contribution can be estimated in a similar way to Eq.~\eqref{omega_a_real_pre}, but in this case
we must take an average over all possible values of the initial angle $\theta$,
since the value of $\theta$ is different for each Hubble volume.
After performing such averaging procedure, the contribution from the realignment mechanism becomes~\cite{Ballesteros:2016xej}
\begin{equation}
\Omega_{A,\rm{real}}h^2 \approx (3.8\pm 0.6)\times 10^{-3}\times \left(\frac{f_A}{10^{10}\,\mathrm{GeV}}\right)^{1.165},
\label{omega_a_real_post}
\end{equation}
where the uncertainty originates from the estimation of the topological susceptibility.
This result assumes infinite distance from the strings and is thus probably an overestimate~\cite{Ballesteros:2016xej}. 

For temperatures below $T\lesssim 1\,\mathrm{GeV}$, domain walls are formed due to the appearance of the QCD potential~\eqref{finite_T_potential_axion_NDW}.
The collapse of the string-wall systems gives an additional contribution to the cold DM abundance, which we denote by $\Omega_{A,\rm{dec}}$.
Since the contribution from the string-wall systems is different according to whether $N_{\rm DW}=1$ or $N_{\rm DW}>1$,
in the following subsections we will discuss these two cases separately.

\vspace{0.5cm}

There has been a lot of controversy on the estimation of $\Omega_{A,\rm{string}}h^2$~\cite{Harari:1987ht,Davis:1989nj,Dabholkar:1989ju,Hagmann:1990mj,Battye:1993jv,Battye:1994au,Yamaguchi:1998gx,Hagmann:2000ja} and $\Omega_{A,\rm{dec}}h^2$~\cite{Nagasawa:1994qu,Chang:1998tb},
which arises from the difficulty in understanding the energy loss process of topological defects and analyzing the spectrum of the axion
produced from them in a quantitative way. 
A straightforward approach to this problem is to perform first principle field theory simulations of topological defects in the expanding Universe~\cite{Yamaguchi:1998gx}.
Along this line, the computational methods to estimate the energy spectrum of radiated axions have been developed in~\cite{Hiramatsu:2010yu,Hiramatsu:2012gg,Hiramatsu:2012sc,Kawasaki:2014sqa,Fleury:2015aca,Fleury:2016xrz,Klaer:2017qhr,Klaer:2017ond,Gorghetto:2018myk,Vaquero:2018tib}.
In the following, we review these results and discuss their uncertainties.

\subsubsection{Models with $N_{\rm DW}=1$}
\label{subsubsec:axiondm_NDW_eq_1}

Let us first consider the case with $N_{\rm DW}=1$.
In this case, the collapse of the string-wall systems occurs immediately after the formation, and the parameter dependence of the total DM abundance
is similar to that of the realignment contribution~\eqref{omega_a_real_post}.

The number density of axions
produced from strings at time $t_s$ before the formation of domain walls can be estimated as~\cite{Kawasaki:2014sqa}
\begin{equation}
n_A(t_s) \simeq \frac{\xi v_{\rm PQ}^2}{\epsilon t_s}\left[\ln\left(\frac{\sqrt{\lambda}v_{\rm PQ}t_s}{\sqrt{\xi}}\right)-3\right], \label{n_a_string}
\end{equation}
where the parameters $\xi$ and $\epsilon$ are defined in terms of the energy density of strings $\rho_{\rm string}$ and the mean energy of axions
produced from strings $\langle E_a \rangle$, respectively,
\begin{align}
\xi = \frac{\rho_{\rm string}(t_s)}{(\mu_{\rm string}/t_s^2)}, \quad
\epsilon = \frac{\langle E_a \rangle}{(2\pi/t_s)},
\end{align}
with $\mu_{\rm string} = \pi v_{\rm PQ}^2\ln(\sqrt{\lambda}v_{\rm PQ}t_s/\sqrt{\xi})$ being the tension of strings.
The results of field theory simulations indicate that the network of strings evolves toward the scaling solution,
in which $\xi$ takes an almost constant value of $\mathcal{O}(1)$~\cite{Yamaguchi:1998gx,Hiramatsu:2010yu}.
The results of the simulations also show that the mean energy of axions produced from strings is comparable to the Hubble scale at the production time,
and $\epsilon$ takes a value of $\mathcal{O}(1)$~\cite{Hiramatsu:2010yu,Kawasaki:2014sqa}.
This fact implies that most of the axions produced from strings
become non-relativistic during the radiation-dominated era, and they contribute to the cold DM abundance.
We note that the estimate~\eqref{n_a_string} implies that the axion abundance scales linearly with $\xi$.
This dependence originates from the assumptions used to derive Eq.~\eqref{n_a_string},
that the energy density of strings obeys that of the scaling solution
and that the total comoving energy of the system consisting of strings and axion radiations is conserved. 
Adopting the values $\xi = 1.0\pm 0.5$ and $\epsilon = 4.02 \pm 0.70$ suggested in~\cite{Kawasaki:2014sqa} and multiplying Eq.~\eqref{n_a_string}
by an appropriate dilution factor, we estimate the contribution from strings as\footnote{Errors shown in Eqs.~\eqref{omega_a_string},~\eqref{omega_a_dec_short}, and~\eqref{omega_a_tot_short}
are not the standard uncertainty estimate ({\it i.e.} those from the propagation of uncertainty law), but maximum and minimum values
obtained by using the results of numerical simulations.}
\begin{equation}
\Omega_{A,\rm{string}}h^2 \approx 7.8^{+6.3}_{-4.5}\times 10^{-3}\times \left(\frac{f_A}{10^{10}\,\mathrm{GeV}}\right)^{1.165}.
\label{omega_a_string}
\end{equation}

In addition to the above contribution of axions produced from strings, there is a contribution from those produced from the collapse of string-wall systems. 
The results of numerical simulations in~\cite{Kawasaki:2014sqa} indicate that the corresponding abundance is slightly smaller than~\eqref{omega_a_string},
\begin{equation}
\Omega_{A,\rm{dec}}h^2 \approx 3.9^{+2.3}_{-2.1}\times 10^{-3}\times \left(\frac{f_A}{10^{10}\,\mathrm{GeV}}\right)^{1.165} \quad(N_{\rm DW}=1).
\label{omega_a_dec_short}
\end{equation}
If we adopt the estimates shown in Eqs.~\eqref{omega_a_real_post},~\eqref{omega_a_string} and~\eqref{omega_a_dec_short},
the total axion abundance~\eqref{Omega_a_tot} reads
\begin{equation}
\Omega_{A,\rm{tot}}h^2 \approx 1.6^{+1.0}_{-0.7}\times 10^{-2}\times \left(\frac{f_A}{10^{10}\,\mathrm{GeV}}\right)^{1.165}.
\label{omega_a_tot_short}
\end{equation}
Requiring that it explains the cold DM abundance observed today $\Omega_{A,\rm{tot}}h^2 = \Omega_{\rm c}h^2\simeq 0.12$, we obtain
predictions for the decay constant, $f_A \approx (3.8\textendash 9.9)\times 10^{10}\,\mathrm{GeV}$, 
or the axion mass, $m_A \approx (0.6\textendash 1.5)\times 10^{-4}\,\mathrm{eV}$.


Our estimate so far is based on the result of numerical simulations performed in~\cite{Kawasaki:2014sqa}, 
but there are some debates on the interpretation of the simulation results.
In particular, the conventional simulation method~\cite{Kawasaki:2014sqa} cannot realise a large logarithmic enhancement factor
in the string tension $\mu_{\rm string}$, and it was pointed out that the effects of such high string tension 
may further modify the estimate of the DM abundance~\cite{Fleury:2015aca,Fleury:2016xrz}.

Recent studies have reported small (seemingly logarithmic) corrections to the scaling solution~\cite{Fleury:2015aca,Gorghetto:2018myk,Vaquero:2018tib}, which indicate a slow increase of the string length parameter $\xi$. The authors of~\cite{Gorghetto:2018myk} have studied the uncertainty due to the extrapolation of the simulation results to realistic values of the string tension and emphasize a much broader uncertainty. The extrapolation includes the uncertainty on the mean energy of radiated axions as well as the increase of the string length parameter. The main source of the uncertainty on the axion abundance is the ambiguity in the analysis of the spectrum of axions produced from strings. It was pointed out in~\cite{Gorghetto:2018myk} that a different interpretation on the shape of the spectrum would alter the prediction for the axion DM abundance by a few orders of magnitude when extrapolated to realistic values of the string tension. In particular, if the mean energy is indeed comparable to the Hubble scale as claimed in~\cite{Davis:1986xc,Davis:1989nj,Battye:1994au,Yamaguchi:1998gx,Hiramatsu:2010yu,Kawasaki:2014sqa}, the axion DM abundance can be further enhanced compared to the estimate shown above due to the increase of the string length parameter. Although such infrared (IR) dominated spectrum is incompatible with recent simulation results~\cite{Fleury:2015aca,Gorghetto:2018myk,Vaquero:2018tib} obtained based on a string tension whose value is smaller than realistic ones, we cannot exclude the possibility that the energy of radiated axion is dominated by IR modes in realistic cases. Intriguingly, simulations performed in~\cite{Gorghetto:2018myk} shows some indication that the spectrum slightly changes towards an IR dominated shape with increasing the string tension, and more careful studies on the axion spectrum in simulations with larger dynamical ranges are warranted to confirm such a trend. We emphasize that even a small change of the shape of the spectrum could drastically enhance the axion DM abundance when extrapolated to realistic parameter values, and typically the axion DM mass becomes as large as $\mathcal{O}(\mathrm{meV})$ in such scenarios, which is accessible to IAXO.

In order to quantify the potentially large uncertainty on the string contribution, let us recast it to the following form,
\begin{equation}
\Omega_{A,\rm{string}}h^2 \approx 2.6\times 10^{-4}\times K\,\left(\frac{f_A}{10^{10}\,\mathrm{GeV}}\right)^{1.165},
\label{omega_a_string_K}
\end{equation}
where $K$ represents the axion production efficiency from strings,
which gives the number of axions produced until the radiation from strings is terminated at a time $t_e$
around the epoch of the QCD phase transition, {\it i.e.} $n_A(t_e) = K H(t_e)v_{\rm PQ}^2$,
and we have used the condition $m_A = 3H$ to estimate $t_e$.
The value of $K$ depends on the logarithmic correction to the string tension $\ln(\sqrt{\lambda}v_{\rm PQ}t_e/\sqrt{\xi})$,
which requires extrapolation over an enormous separation range between $\sqrt{\lambda}v_{\rm PQ}$ and $t_e^{-1}$.
The estimate shown in Eq.~\eqref{omega_a_string} corresponds to $K\approx 13\textendash 54$,
while it was claimed in~\cite{Gorghetto:2018myk} that its value can differ by many orders of magnitude
according to different interpretations of simulation results.
The extrapolation with an extremely IR dominated spectrum leads to a value of $K$ as large as $\lesssim 5 \times 10^3$~\cite{Gorghetto:2018myk}
which corresponds to the axion DM mass of $m_A \lesssim 4.4\,\mathrm{meV}$.

On the other hand, the authors of~\cite{Klaer:2017qhr,Klaer:2017ond} introduced an alternative technique to perform
direct simulations of string-wall networks with high effective string tension.
This study
concludes that axions emitted from strings and walls are largely irrelevant,
obtaining a DM QCD axion mass on the lower side, $m_A \approx (0.262 \pm 0.034)\times 10^{-4}\,\mathrm{eV}$~\cite{Klaer:2017ond}, 
that corresponds to the production efficiency $K \approx 13$ in Eq.~\eqref{omega_a_string_K}.
The results of these simulations also show that the string networks become denser for larger values of the string tension,
which leads to a larger value of $\xi \sim 4$~\cite{Klaer:2017qhr,Klaer:2017ond}.
Reinterpreting these results in light of energy conservation, we expect that the large contribution from the energy density of strings should be compensated by producing more energetic axions in order to realise smaller axion production efficiency, which leads to a smaller DM axion mass.
Indeed, if we adopt the value $\xi \sim 4$ in Eq.~\eqref{n_a_string} together with the axion production efficiency obtained in~\cite{Klaer:2017ond},
we obtain $\epsilon \sim 30\textendash 40$ for the mean energy of radiated axions,
which is an order of magnitude larger than the value $\epsilon = 4.02 \pm 0.70$ obtained in the conventional
field theory simulations~\cite{Kawasaki:2014sqa}.
This implies that physics at smaller scales can be relevant to the determination of the axion DM abundance.
We note that the new simulation method introduced in~\cite{Klaer:2017qhr} is based on an effective theory that breaks down at some smaller distance scales, and hence it is still not straightforward to figure out how the physics of small scale strings affects the axion production efficiency.
Further studies on the dynamics of string-wall networks are required to include precise modeling of physics at smaller distance scales. 

Regarding the fact that there remains the large uncertainty on the estimation of the axion abundance produced from strings,
here we treat $K \approx 13$ corresponding to the result of~\cite{Klaer:2017ond} as a lower limit and 
$K \lesssim 5\times 10^3$ obtained in~\cite{Gorghetto:2018myk} as an upper limit on the axion production efficiency.
This corresponds to the following range of the axion DM mass in the models with $N_{\rm DW} = 1$,
\begin{equation}
2.6 \times 10^{-5}\,\mathrm{eV} \lesssim m_A \lesssim  4.4 \times 10^{-3}\,\mathrm{eV}\quad (N_{\rm DW}=1),
\end{equation}
or the axion decay constant,
\begin{equation}
1.3\times 10^9\,\mathrm{GeV} \lesssim f_A \lesssim 2.2\times 10^{11}\,\mathrm{GeV} \quad(N_{\rm DW}=1).
\end{equation}
We again emphasize that the latest simulation results~\cite{Gorghetto:2018myk} show a trend that the IR modes are getting more important
for larger values of the string tension, and that the extrapolation with such a feature shows a preference for a higher axion DM mass close to 
the upper limit on $m_A$ shown above.
Note that this mass range is derived based on the assumption that axions produced from strings account for the total cold DM abundance.
In other words, IAXO will be able to probe the parameter space where axions can constitute 100\% of the observed cold DM abundance within the range of uncertainty.


\subsubsection{Models with $N_{\rm DW}>1$}

If $N_{\rm DW}>1$, the string-wall systems live longer than those in the case with $N_{\rm DW}=1$.
They eventually collapse due to the effect of the explicit symmetry breaking term.
The present abundance of axions produced from these string-wall systems can be estimated as
\begin{equation}
\Omega_{A,\rm{dec}}h^2 \approx 1.75\times\frac{C_d^{1/2}\mathcal{A}^{3/2}}{\tilde{\epsilon}_a N_{\rm DW}^2}\left(\frac{\Xi}{10^{-52}}\right)^{-1/2}\left(\frac{f_A}{10^{9}\,\mathrm{GeV}}\right)^{-1/2} \quad(N_{\rm DW}>1), \label{omega_a_dec_long}
\end{equation}
where
\begin{equation}
\Xi \equiv \frac{\Delta V}{2 v_{\rm PQ}^4} \label{Xi_def}
\end{equation}
parameterises the magnitude of the explicit symmetry breaking term $\Delta V$ that induces the energy bias
between different domains.\footnote{The result of numerical simulations for axionic domain walls with $N_{\rm DW}>1$
shows some deviation from the exact scaling solution, which leads to additional uncertainties of $\Omega_{a,\rm{dec}}h^2$~\cite{Kawasaki:2014sqa}.}
The dimensionless parameters $C_d$, $\mathcal{A}$, and $\tilde{\epsilon}_a$ appearing in Eq.~\eqref{omega_a_dec_long}
represent the decay time of the string-wall systems, the area of domain walls, and the mean energy of axions radiated from them,
whose values can be estimated from numerical simulations~\cite{Kawasaki:2014sqa}.
They generically take values of $\mathcal{O}(1)$ but slightly depend on $N_{\rm DW}$.

If the lifetime of the string-wall systems is not sufficiently long, one must take account of the realignment and string contributions
in addition to $\Omega_{A,\rm{dec}}h^2$.
For the string contribution, we extend Eq.~\eqref{omega_a_string_K} to the case with $N_{\rm DW}>1$,
\begin{equation}
\Omega_{A,\rm{string}}h^2 \approx 2.6\times 10^{-4}\times KN_{\rm DW}^2\,\left(\frac{f_A}{10^{10}\,\mathrm{GeV}}\right)^{1.165},
\label{omega_a_string_K_NDW}
\end{equation}
and assume that $K$ has a similar uncertainty as in the models with $N_{\rm DW}=1$.
The factor of $N_{\rm DW}^2$ just comes from the fact that the string tension is proportional to $v_{\rm PQ}^2$,
which becomes $N_{\rm DW}^2 f_A^2$ when we write everything in terms of the axion decay constant.

If we assume that the PQ symmetry breaking is protected by a discrete $Z_N$ symmetry,
and Planck-suppressed operators with $\Lambda = M_{\rm P}$,  the parameter $\Xi$ in Eq.~\eqref{Xi_def} is given 
by~\eqref{PQ_violating_operators}. 
In this case, we have the following parameterisation~\cite{Ringwald:2015dsf},
\begin{equation}
\Xi = \frac{|c_N|N_{\rm DW}^{N-4}}{(\sqrt{2})^N}\left(\frac{f_A}{M_{\rm P}}\right)^{N-4}.
\label{Xi_Planck_suppressed}
\end{equation}
It turns out that $N=9$ or $10$ is favoured if we require that the axions produced by the string-wall systems should not overclose the Universe and that the operator $\Delta V$ does not shift the minimum of the QCD potential~\eqref{finite_T_potential_axion} away from $\theta=0$ even for ${\cal O}(1)$ values of the phase and/or modulus of $c_N$.

Since the magnitude of the energy bias~\eqref{Xi_Planck_suppressed} becomes more suppressed for smaller values of $f_A$,
the lifetime of the string-wall systems becomes longer in the small $f_A$ range.
For such long-lived defects, the dominant contribution to the DM abundance is given by Eq.~\eqref{omega_a_dec_long},
which places a \emph{lower} bound on $f_A$ from the requirement that the axion abundance should not exceed the observed cold DM abundance.
On the other hand, if $f_A$ takes a larger value, the lifetime of the defects becomes shorter, and the DM abundance is determined by
the realignment and string contributions, which leads to an \emph{upper} bound on $f_A$.\footnote{As mentioned before, 
the realignment contribution shown in Eq.~\eqref{omega_a_real_post} can be an overestimate.
Instead of using it, here we adopt Eq.~\eqref{omega_a_string_K_NDW} with a lower value of $K \approx 13$
suggested in Sec.~\ref{subsubsec:axiondm_NDW_eq_1} to obtain upper limits on $f_A$ 
and lower limits on $m_A$ shown in Eqs.~\eqref{fA_range_NDW6_N9}-\eqref{mA_range_NDW6_N10}.}
Therefore, the value of $f_A$ is constrained to a finite range in the post-inflationary PQ symmetry breaking scenario with $N_{\rm DW}>1$.
For instance, the allowed value of $f_A$ in the models with $N_{\rm DW}=6$ and $N=9$ reads
\begin{equation}
4.4\times 10^7\,\mathrm{GeV} \lesssim f_A \lesssim 9.9\times 10^{9}\,\mathrm{GeV} \quad(N_{\rm DW}=6\ \text{and}\ N=9),
\label{fA_range_NDW6_N9}
\end{equation}
which corresponds to the mass range
\begin{equation}
5.8\times 10^{-4}\,\mathrm{eV} \lesssim m_A \lesssim 1.3\times 10^{-1}\,\mathrm{eV} \quad(N_{\rm DW}=6\ \text{and}\ N=9).
\label{mA_range_NDW6_N9}
\end{equation}
Similarly, for the models with $N_{\rm DW}=6$ and $N=10$, we have
\begin{equation}
1.3\times 10^9\,\mathrm{GeV} \lesssim f_A \lesssim 9.9\times 10^{9}\,\mathrm{GeV} \quad(N_{\rm DW}=6\ \text{and}\ N=10),
\label{fA_range_NDW6_N10}
\end{equation}
which corresponds to the mass range
\begin{equation}
5.8\times 10^{-4}\,\mathrm{eV} \lesssim m_A \lesssim 4.5\times 10^{-3}\,\mathrm{eV} \quad(N_{\rm DW}=6\ \text{and}\ N=10).
\label{mA_range_NDW6_N10}
\end{equation}

Since the relic QCD axion abundance depends on the coefficient $|c_N|$ in~\eqref{Xi_Planck_suppressed}
in addition to $f_A$, axions can explain the total cold DM abundance in the whole mass ranges described above, 
up to the tuning of the coupling parameter $c_N$ and the uncertainty on the production efficiency  $K$ from strings.
In particular, axions can account for the whole DM in the parameter range $f_A\approx \mathcal{O}(10^8\textendash10^9)\,\mathrm{GeV}$
and $m_A \approx \mathcal{O}(10^{-3}\textendash 10^{-2})\,\mathrm{eV}$ with a mild tuning of the phase of the parameter $c_N$~\cite{Ringwald:2015dsf,Giannotti:2017hny}. 
Interestingly, such a parameter range agrees with that indicated by the anomalous cooling of stars in various evolutionary stages, 
which will be discussed in the next section.

\begin{figure}[htbp]
\centering
\tikzsetnextfilename{pics/IAXO_meV_DM_plot}
\resizebox{\textwidth}{!}{\input{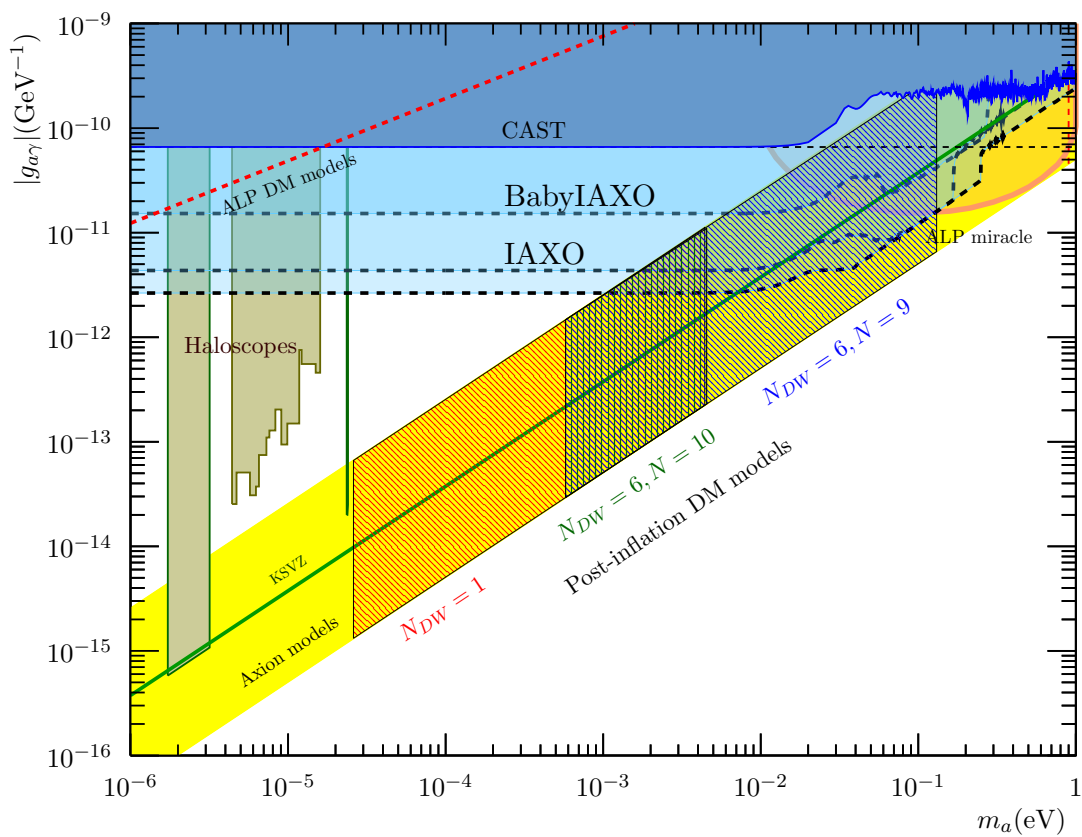}}
\caption{The predicted mass ranges in which QCD axions can account for the observed cold DM abundance
in the post-inflationary PQ symmetry breaking scenario.
The generic prediction of QCD axion models is plotted in the yellow region.
The sensitivity prospects of IAXO are also plotted.
The prediction of the post-inflationary PQ symmetry breaking scenario differs according to the value of $N_{\rm DW}$ and the structure of the
explicit symmetry breaking terms in the models with $N_{\rm DW}>1$.
Dashed regions correspond to the predictions (up to uncertainties in the estimation of the relic axion abundance)
of the models with $N_{\rm DW}=1$ (red), $N_{\rm DW}=6$ and $N=9$ (gray), and $N_{\rm DW}=6$ and $N=10$ (blue).}
\label{fig:DM_predictions}
\end{figure}

We plot the parameter range where QCD axions can account for the observed DM abundance for the models with $N_{\rm DW}=1$
and for the models with $N_{\rm DW}=6$ in Fig.~\ref{fig:DM_predictions}.
We see that the predicted value for $m_A$ can be much larger than the conventional estimation
$m_A\approx \mathcal{O}(10^{-6})\,\mathrm{eV}$ due to the contribution from strings and string-wall systems.
In particular, the axion can be DM in the meV mass range
both for the models with $N_{\rm DW}=1$ and those with $N_{\rm DW}>1$, 
and such a parameter range can be decisively probed by IAXO.

%% file: sections/darkradiation.tex

According to the standard cosmological model, $\Lambda$CDM, the energy budget of the Universe today consists of  (in order of decreasing magnitude): dark energy, dark matter, ordinary baryonic matter, and relic neutrino and photon radiation. However, the dark sector may be significantly richer than this phenomenological model suggests. Extensions of the Standard Model commonly include light particles in the dark sector, e.g.~axions and hidden photons.
Beyond their potential role as cold DM, these light particles may have been produced in the hot Big Bang from thermal processes or decays of heavy particles and would then contribute to the energy density of the Universe as a `dark radiation' component. The observational success of the $\Lambda$CDM model indicates that dark radiation, if it exists, must contribute to the total energy density today by an even smaller amount than the CMB photons and the C$\nu$B neutrinos. By convention, the dark radiation is parametrised phenomenologically as the `excess number of neutrino species', $\Delta N_{\rm eff}$:
\begin{equation}
\rho_{\rm d.r} = \rho_{rad} - \rho_\text{CMB} - \rho_{C\nu B} = \frac{7}{8} \left( \frac{4}{11}\right)^{4/3} \Delta N_{\rm eff}~ \rho_\text{CMB} \, ,
\end{equation}
where $\rho_{rad}$ denotes the total relativistic energy density. The CMB data from Planck (when combined with other experiments) is consistent with $\Delta N_{\rm eff} = 0$, but a significant dark radiation energy density is still allowed within the 68\% error bars of $\sigma(\Delta N_{\rm eff}) = 0.23$ \cite{Ade:2015xua}. 
Dark radiation is also capable of reducing the persistent tensions between local measurements of $H_0$ \cite{Riess:2016jrr} and the value inferred using $\Lambda$CDM and CMB data.
Dark radiation present at the time of Big Bang nucleosynthesis (BBN) would increase both the expansion rate and the freeze-out abundance of neutrons, and consequently affect the light element abundances. Observational determinations of the primordial deuterium abundance give no evidence for dark radiation with error bars comparable to Planck, $\sigma(\Delta N_{\rm eff}) = 0.28$ \cite{Cooke:2013cba}. Future CMB experiments will be significantly more sensitive to dark radiation: the ground based `Stage-3' and `Stage-4' CMB polarisation programmes state a projected sensitivity of $\sigma(\Delta N_{\rm eff}) = 0.06$ and $\sigma(\Delta N_{\rm eff}) = 0.02$, respectively.

\subsection{Thermally produced ALP dark radiation}

\begin{figure}[htbp]
\centering
\includegraphics[width=0.7\textwidth]{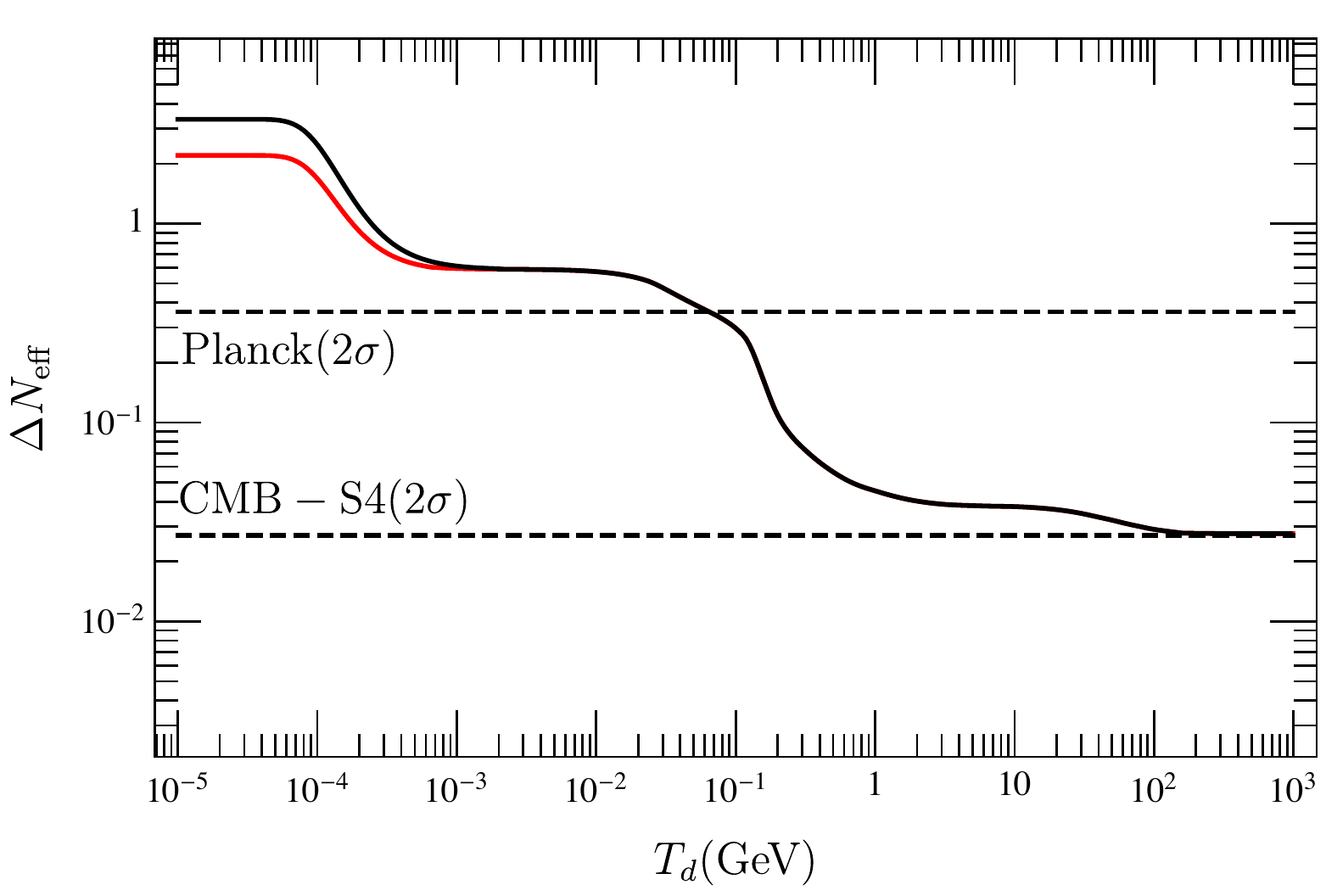}
\caption{Extra radiation in an axion as a function of its decoupling temperature assuming only the SM degrees of freedom. Dashed lines show the upper limits (2$\sigma$) from Planck and the future 4th generation CMB-S4.}
\label{DRTd}
\end{figure}

The definition of $\Delta N_{\rm eff}$ is chosen so that an additional  neutrino species that initially was in thermal equilibrium with the photon-baryon plasma and decoupled simultaneously with the Standard Model neutrinos gives $\Delta N_{\rm eff} = 1$. More generally, any thermally produced, light, feebly interacting particles that decouple at time $T_d$ contribute to the dark radiation by:
\begin{equation}
\Delta N_{\rm eff} = \left( \frac{g_{\star}(T_{\nu})}{g_{\star}(T_d)}\right)^{4/3} \times
\left\{
\begin{array}{l c l}
1 & ~~& {\rm Majorana~fermion},  \\
\frac{4}{7} && {\rm Scalar.}
\end{array}
\right.
\label{eq:NeffTh}
\end{equation}
Here $g_{\star}=g_{\star}(T)$ denotes the effective number of relativistic degrees of freedom in thermal equilibrium at temperature $T$. 
Assuming that the dark radiation component decoupled earlier than the regular neutrinos, the Standard Model contribution to $g_{\star}(t)$ ranges from  $g_{\star}(T_{\nu}) =  10.75$ to $g_{\star}(T \gtrsim 3m_{\rm top}) =106.75$ (see \cite{Borsanyi:2016ksw}) the latter figure being applicable at very early times when even the top quark was in thermal equilibrium. Beyond the Standard Model, $g_{\star}(T)$ may have received contributions from additional particles, and the Standard Model contribution in general only gives a lower bound on $g_{\star}(T)$. Hence, a thermally produced ALP that decouples before $T\sim 3 m_{\rm top}$ contributes to the dark radiation by:
\begin{equation}
\Delta N_{\rm eff} \leq 0.027 \, , 
\end{equation}
which saturates if $g_{\star}$ never received any contributions other than those of the Standard Model and the ALP.  
Figure~\ref{DRTd} shows the value of $\Delta N_{\rm eff}$ as a function of the decoupling temperature under the SM-only assumption. 

A thermal population of QCD axions can be always established through interactions with gluons at sufficiently high temperatures, which happens provided the reheating temperature satisfies $T_{\rm RH}> T_{d}$ where \cite{Graf:2010tv} (see also \cite{Masso:2002np, Salvio:2013iaa, Baumann:2016wac}):
 \begin{equation}
  T_d \approx 9.6\cdot 10^6\, {\rm GeV} \left( \frac{f_A}{10^{10}\, {\rm GeV}} \right)^{2.246} \, .
  \end{equation}
Moreover, the efficiency of the thermal production is enhanced by about 3 orders of magnitude if there is a direct coupling between axion and top quarks~\cite{Salvio:2013iaa}, and so in this case the required $T_{\rm RH}$ goes down by the same amount.

From analysis of Planck data and BBN constraints, the reheating temperature of the hot Big Bang plasma is only bounded from below, $T_{\rm RH} \geq 4.7 \cdot 10^{-3}$ GeV \cite{deSalas:2015glj}, but it is not known whether $T_{\rm RH}$ far exceeded this bound.\footnote{Note that a determination of the energy scale of inflation would not model-independently determine the initial temperature of the primordial hot Big Bang plasma, as the period between the end of inflation and BBN may be complicated and include multiple periods of matter and radiation domination, with $T_r$ corresponding to the maximal temperature of the final phase.}

Even if the reheating temperature is too low to fully thermalise the ALPs, small contributions to the energy density may arise from (non-equilibrium) ALP interactions with the thermal plasma, generically giving $\Delta N_{\rm eff} \ll 0.027$ \cite{Graf:2010tv}. 

\begin{figure}[htbp]
\centering
\includegraphics[width=0.63\textwidth]{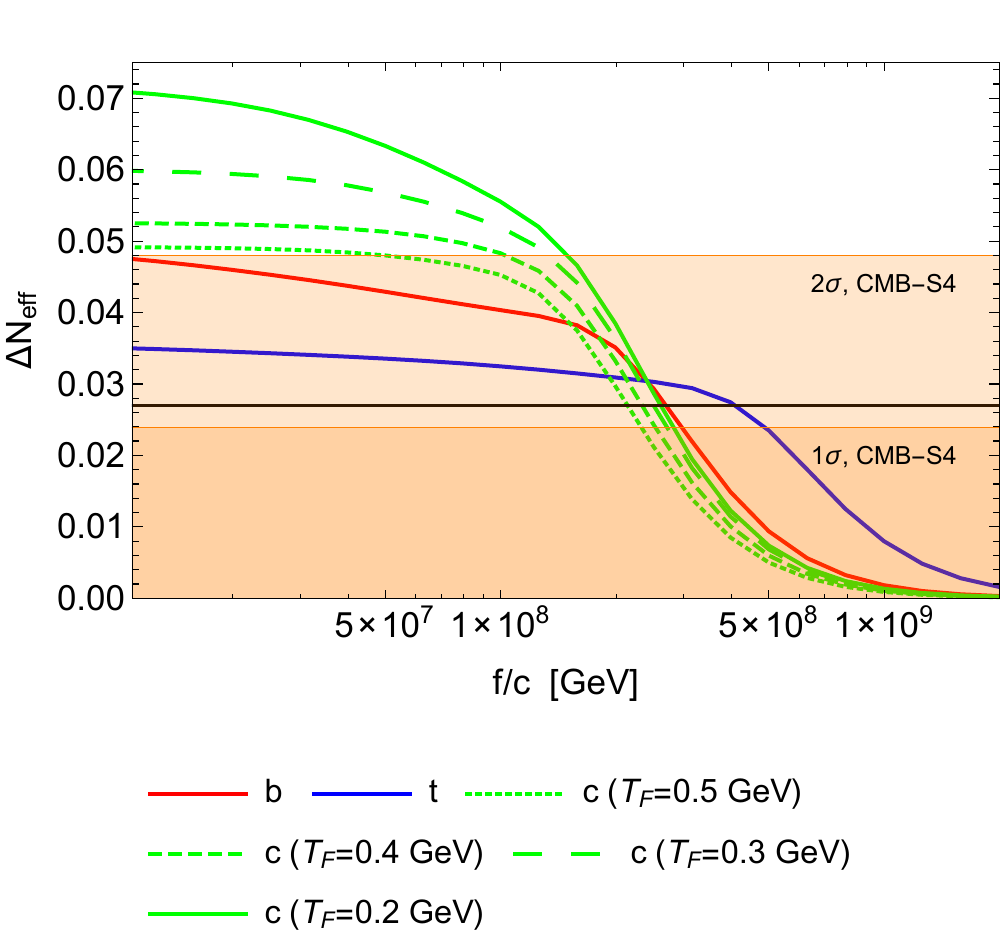}

\vspace{1.5cm}

\includegraphics[width=0.6\textwidth]{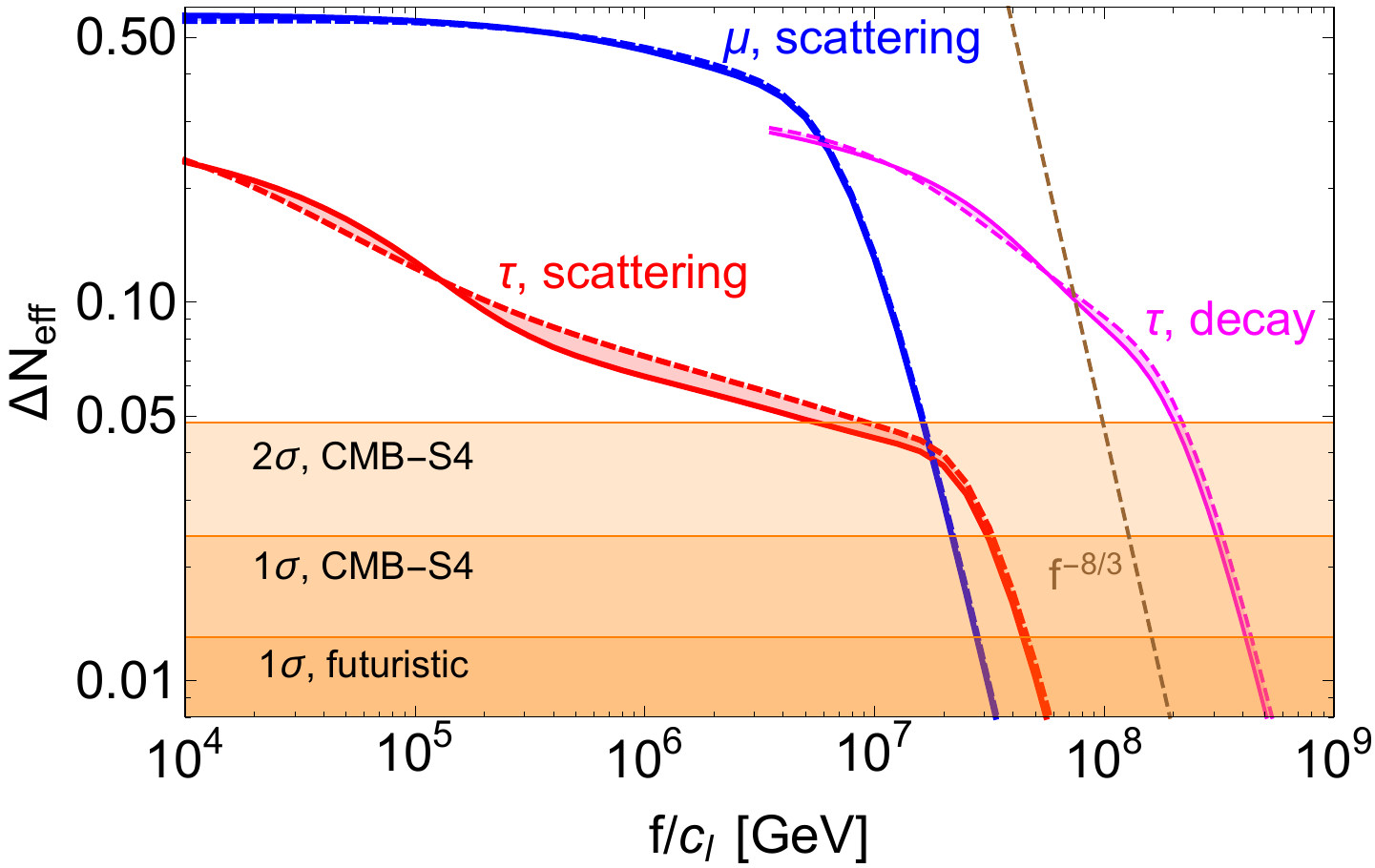}
\caption{Top: $\Delta N_\text{eff}$ as a function of $f/c_i$. The green, red and blue lines correspond to the predictions for the charm, bottom and top particle, respectively. The orange bands represent the $1\sigma$ and $2 \sigma$ forecasted contours of a next generation CMB experiment (CMB-S4). See also~\cite{Baumann:2017gkg} for similar forecasts combining CMB-S4 and large scale structure. Bottom: Contribution to $\Delta N_{\rm eff}$ from muon (blue) and tau (red) scattering as a function of $c_\ell/f$. Decays are possible for off-diagonal couplings; we show only results for tau decays (magenta), which are the only allowed ones in the above range for $c_{\ell \ell^\prime}/f$. Each process is shown as a band, parametrizing the uncertainty in the number of relativistic degrees of freedom: the straight line $g_*$ is taken from~\cite{Laine:2006cp} and the dashed line from~\cite{Borsanyi:2016ksw}. We also show the analytical expectation, $\Delta N_{\rm eff} \propto f^{-8/3}$, for non-thermalized axions. The orange bands represent the $1\sigma$ and $2 \sigma$ forecasted contours of CMB-S4, plus a more futuristic $1\sigma$ band, according to~\cite{Abazajian:2016yjj}.
}
\label{DeltaNeff}
\end{figure}

If axions have direct couplings to quarks and leptons, some of which are poorly constrained observationally, they can thermalize at temperatures between about 0.1 GeV and 100 GeV, for $f$ below $10^9$ GeV,  leading to larger values of $\Delta N_{\rm eff}$, up to about 0.05 for quarks and even up to 0.5 for leptons~\cite{Baumann:2016wac,Ferreira:2018vjj,DEramo:2018vss}, as shown in Fig.~\ref{DeltaNeff}. 
Such values are not upper bounds, but predictions, and should be visible by future CMB ‘Stage-4’ experiments, allowing for a very interesting interplay between CMB experiments and IAXO.  Large values of $\Delta N_{\rm eff}$  might even mitigate the present cosmological tension on Hubble constant measurements~\cite{DEramo:2018vss}, at about 3.5 $\sigma$ between CMB and Supernovae measurements~\cite{Aghanim:2018eyx, Bernal:2016gxb}, and claimed at 3.7 $\sigma$~\cite{Riess:2018uxu}, and even 4.4 $\sigma$~\cite{Riess:2019cxk} more recently.

\subsection{Non-thermal ALP dark radiation from modulus decay}

{\it Non-thermally} produced axions may potentially give much larger contributions to the  dark radiation energy density of the Universe. In fact, substantial levels of axionic dark radiation (corresponding to $\Delta N_{\rm eff} \sim {\cal O}(0.1$--$10)$) is a generic consequence of string theory models of inflation and its aftermath. The reason is very simple: when the reheating temperature is insufficient to thermalise them, \emph{direct decay} of the field driving reheating into ALPs will produce a non-thermal cosmic background of relativistic axions, and hence dark radiation.  While not specific to string theory, this scenario is a generic consequence of reheating scenarios involving weakly coupled moduli fields, as we will now discuss.

Compactifications of string theory include geometric deformation moduli and axions in general in the low-energy theory (cf.~section \ref{sec:StringAxions}). During inflation, the compactification geometry becomes slightly distorted due to the  energy density  of the inflaton, and moduli become displaced from their post-inflationary minima. After the end of inflation, the geometry will relax back to its final vacuum configuration, and will undergo oscillations as it settles down. Specifically, when the Hubble parameter $H$ has decreased to the same order of the mass of one of  the moduli fields, $m_\phi$, the modulus starts oscillating around its minimum with an initial amplitude set by the displacement acquired during inflation. The energy density of such a coherently oscillating field red-shifts only like matter, $\rho_{\phi} \sim R^{-3}$ ($R$ the scale factor of the Friedman-Robertson-Walker metric), and if the modulus is sufficiently long-lived, its energy density comes to dominate over any previous radiation component which red-shifts like $\rho_r \sim R^{-4}$.
This way, the most long-lived modulus field comes to dominate the energy density of the Universe, and the reheating temperature of the hot Big Bang plasma will be set at its decay. Moduli arise from higher-dimensional components of the metric tensor and generically couple with gravitational strength interactions, and their decay rate is typically given by,
\begin{equation}
\Gamma \sim \frac{1}{4 \pi} \frac{m_{\phi}^{3}}{M_{\rm Pl}^{2}} \, ,
\end{equation}
up to an ${\cal O}(1)$ constant.
Clearly, parametrically lighter moduli are more long-lived than heavier moduli. Assuming instantaneous reheating, $T_{\rm RH}$ is given by (for the more precise equation, see e.g. \cite{Conlon:2013isa}):
\begin{equation}
T_{\rm RH} \sim \sqrt{H M_{\rm Pl}} \approx \sqrt{\Gamma M_{\rm Pl}} \sim \frac{m_{\phi}^{3/2}}{M_{\rm Pl}^{1/2}}
= 20~{\rm MeV}~ \left( \frac{m_{\phi}}{100~{\rm TeV}}\right)^{3/2}
\, . \label{Trh}
\end{equation}
Hence,  the observational requirement of thermalisation before BBN and neutrino decoupling $T_{\rm RH} \geq 4.7 \cdot 10^{-3}$ GeV \cite{deSalas:2015glj}, translates into a generic bound on the mass of the lightest modulus: $m_{\phi}\gtrsim 50$ TeV.
The typical values of $m_{\phi}$ from string compactifications is not known in general, but some scenarios that also predict supersymmetry breaking `soft' terms at the TeV scale give  $m_\phi \sim 10^6$--$10^7$ GeV \cite{Choi:2005ge, Acharya:2008bk, Blumenhagen:2009gk, Aparicio:2014wxa} which gives from \eqref{Trh} $T_{\rm RH}\sim 1$--$10$ GeV. Larger minimum moduli masses (say, $m_{\phi}\gtrsim 10^{11}$ GeV for the QCD axion) may produce a thermal axion dark radiation, leading, in effect, to a suppression of $\Delta N_{\rm eff}$.

The moduli field $\phi$ generically has open decay channels into any sufficiently light degree of freedom. In particular, $\phi$ can decay into axions: $\phi \to a+a$.   Importantly, weakly coupled axions do not thermalise, and so are more energetic than the average particle in the hot Big Bang plasma,
\begin{equation}
E_a^{\rm initial} = \frac{m_{\phi}}{2} \gg  m_{\phi} \sqrt{\frac{m_{\phi}}{M_{\rm Pl}}} \, .
\end{equation}
As the Universe expands, the axion energy redshifts with the inverse scale factor. However, the ratio of the axion energy to the plasma/CMB temperature remains large until the present day \cite{Conlon:2013isa}:
\begin{equation}
\left(\frac{E_a}{T}\right)_{\rm now} = \left(\frac{4}{11}\right)^{1/3} \left(\frac{g_{\star}(t_{\nu})}{g_{\star}(t_{\rm RH})} \right)^{1/3} \left( \frac{E_a}{T}\right)_{\rm initial} \, .
\label{eq:drE}
\end{equation}
The levels of axions dark radiation produced directly from modulus decay is then determined by the branching ratio $B_a$ of $\phi$ decaying into axions, $B_a = \Gamma_{\phi \to aa}/\Gamma_{\phi \to {\rm anything}}$.  The relevant value of $\Delta N_{\rm eff}$ is given by \cite{Cicoli:2012aq, Higaki:2012ar}:
\begin{equation}
\Delta N_{\rm eff} = \frac{43}{7} \frac{B_a}{1-B_a} \left( \frac{g_{\star}(T_{\nu})}{g_{\star}(T_{\rm RH})}\right)^{1/3} \, . 
\end{equation}
The exceptionally sensitive CMB Stage-4 experiments will probe \emph{sub-percent} branching ratios into axions:
in the Standard Model (i.e.~for $g_{\star}(T_{\rm RH})$ in the range 10.75--106.75), a constraint on $\Delta N_{\rm eff} < 0.02$ corresponds to bounds on the branching ratio into axions $B_a < {\cal O}(0.3$--$0.7\%)$. Hence, precision constraints on dark radiation will provide an incredibly sensitive probe of the high-energy physics driving reheating. 

The spectrum of the axions produced from modulus decay differs from the thermal Boltzmann distribution. At decay, momentum conservation ensures that all axions have $E_a = m_{\phi}/2$, however, the decay of the modulus field  is not instantaneous, and, when observed today,  axions created early will be slightly more red-shifted than axions created at a later time. This results in a  `quasi-thermal' shape \cite{Conlon:2013isa}. For moduli masses of ${\cal O}(10^6)$ GeV, the characteristic energy of this relativistic `Cosmic axion Background' is around ${\cal O}(0.2)$ keV.

Particular realisations of this scenario have been considered for various string compactifications \cite{Cicoli:2012aq, Higaki:2012ar, Angus:2014bia, Hebecker:2014gka, Cicoli:2015bpq, Acharya:2015zfk}, where also the branching ratio into axions can be estimated. Commonly,  ${\cal O}(\Delta N_{\rm eff}) \sim 0.1$--$10$, and scenarios with many light axions (cf. the `axiverse' \cite{Arvanitaki:2009fg}) tend to produce levels of dark radiation significantly above current observational limits (see \cite{Acharya:2015zfk, Gorbunov:2017ayg} for suggestions on how to ameliorate this problem). Non-thermal axionic dark radiation can also be produced by other mechanisms than freeze-in and moduli decay, see e.g.~\cite{Mazumdar:2016nzr}. In sum, dark radiation is a very sensitive probe of the spectrum and dynamics relevant in the early Universe. The next generation of CMB experiments will detect the imprints of a Cosmic axion Background or rule out large classes of string compactifications.

\subsection{Additional observational hints of axionic dark radiation}

Relativistic ALPs can produce a number of additional signals beyond their effect on the energy density of the Universe. These arise through either direct scattering of the axions off matter, or through axion-photon conversion in background magnetic fields, as explained in section \ref{sec:transparency}.

\begin{itemize}

\item {\bf Scattering of ALPs off the thermal plasma}

Direct scattering of axions off the thermal plasma is a telltale feature of non-thermally produced axions which are naturally much more energetic than the plasma temperature, cf.~\eqref{eq:drE}. 
These high-energy axions  can access dynamical processes with a centre-of-mass energy $E_{\rm CoM} \gg T$, which in the thermal plasma would be suppressed by ${\rm exp}(-E_{\rm CoM}/T) \ll 1$. As these axions are non-thermal, the scattering rate is always $\Gamma \ll H$, yet rare scattering events can cause significant deviations from standard cosmology. For example,  scattering off photons during BBN is constrained by the observationally inferred primordial helium abundance  \cite{Conlon:2013isa}. However,  scattering off the pre-recombination thermal plasma in the red-shift range $z \sim 1100$ -- $2\cdot 10^6$ will produce very small spectral distortions of the CMB that will be hard to detect \cite{Marsh:2014gca}. Highly energetic ALP scattering may furthermore produce dark matter \cite{Conlon:2013isa}.

\item {\bf ALP-photon conversion in primordial magnetic fields}

ALP dark radiation may also be detected through ALP-photon conversion in astrophysical magnetic fields \cite{Conlon:2013txa}. Particularly interesting are hypothetical primordial cosmic magnetic fields spanning Mpc distances, and galaxy cluster magnetic fields, which tend to be coherent over kpc scales.

The magnitude of primordial intergalactic fields is unknown, but lies between $10^{-9}$ and $10^{-16}$ G \cite{Tavecchio:2010mk}.
In the presence of a substantial primordial, cosmic magnetic field, a highly relativistic Cosmic ALP Background can convert into energetic photons that will contribute to the reionization of the Universe. Such dark radiation  can be constrained by its contribution to the optical depth to recombination, $\tau$. Conservative estimates using Planck constraints on $\tau$ and $\Delta N_{\rm eff}$
indicate  that this could lead to a combined constraint of $g_{a \gamma} B \leq 10^{-18}~{\rm GeV}^{-1}$ nG \cite{Evoli:2016zhj}, where $B$ denotes the cosmic magnetic field, which is observationally constrained to $B \leq$ nG \cite{Ade:2015cva}. Hence, if a Cosmic ALP Background is detected by other means, it will provide strong constraints on the strength of cosmological magnetic fields, and vice-versa.

\item {\bf Soft X-ray excess from galaxy clusters}

The magnetic fields of galaxy clusters are even more interesting, as they are well constrained observationally.
The fields of galaxy clusters are typically $\mathcal{O}(1-10) \mu G$ in strength (e.g. see \cite{Boehringer:2016kqe}). These magnetic fields are measured via observations of Faraday rotation of radio sources located in or behind clusters. On passing through a magnetic field, the polarisation angle $\phi$ rotates with wavelength, $\lambda$, as,  
\be
\phi = \phi_0 + RM \lambda^2\,,
\ee
where the rotation measure is given by:
\be
RM =  812 \int \left( \frac{n_e}{1 \,\text{cm}^{-3}} \right) \left( \frac{B_{\parallel}}{1 \,\mu\text{G}} \right) \left(\frac{dl}{\text{kpc}}\right)
\ee
along the line of sight.
As the electron density in a cluster $n_e(r)$ is well-determined from X-ray measurements, radio observations of the Rotation Measure lead to a statistical characterisation of the magnetic field strength (although as the magnetic field reverses direction many times along the line of sight, it is not possible to know the exact 3-dimensional field configuration). The rate of variation in the Rotation Measure also allows the typical coherence lengths of the magnetic field to be estimated as 1 - 10 kpc, with the field extending throughout the $\sim 1 {\rm Mpc}$ extent of a galaxy cluster.

The typical electron density $n_e$ inside a galaxy cluster is between $10^{-3}$ and $10^{-2} {\rm cm}^{-3}$ and the typical extent of a galaxy cluster
is around 1 Mpc. For ALP masses less than around $10^{-12} \, {\rm eV}$ (the plasma mass within galaxy clusters), this implies that at X-ray energies
galaxy clusters provide ideal environments for ALP-photon conversion. Indeed, at X-ray energies
the ALP-photon interconversion probability on passing through a cluster is $\mathcal{O}(1)$ when $g_{a\gamma} \sim 10^{-11}\, {\rm GeV}^{-1}$.
The conversion probability is energy-dependent, and maximises at $E \gtrsim$ keV energies in the galaxy cluster environment.\footnote{Significant conversion probabilities  can also be obtained by `resonant' conversion of lower-energy ALPs into photons if the ALP mass is very close to the local plasma frequency.}
Detailed simulations of ALP-photon conversion passing through the Coma cluster at various energies can be found in \cite{Angus:2013sua}. These both
confirm these general results and also show the presence of a threshold energy for conversion to occur efficiently (for the Coma cluster, this is around 50 eV).

Non-thermally produced ALP dark radiation arising from the decay of a modulus with $m_{\phi} = 10^6$ GeV would currently have a mean energy of
around 200 eV \cite{Conlon:2013isa}. For $g_{a\gamma}$ in the range $10^{-13} {\rm GeV}^{-1}$ to $10^{-11} {\rm GeV}^{-1}$ this would lead
via ALP-photon conversion to an additional contribution to cluster X-ray spectra in the ${\cal O}(0.1$--1 keV) range \cite{Conlon:2013txa}.

In fact, an unidentified excess luminosity above the ICM spectrum at soft X-ray energies, ${\cal O}(\leq 0.4$ keV), has been observed by a
number of X-ray satellites -- and in particular EUVE and ROSAT -- from several clusters \cite{Bonamente:2002wh}.
This excess is especially strong and well established in the Coma cluster, where it extends from the centre of the cluster out to several megaparsecs away.
The soft excess is a somewhat unusual phenomenon, as older satellites such as ROSAT are far more sensitive to it than more recent ones such as XMM-Newton
and Chandra. This arises because ROSAT was designed for an all-sky survey, and so has a larger field of view and a much lower internal background than modern satellites which are designed to observe point sources. This gave ROSAT a far greater sensitivity to weak diffuse emission (such as the soft excess) than for more recent satellites.

It was shown in \cite{Conlon:2013txa, Angus:2013sua}, taking the nearby Coma cluster as a particular example for which well-developed models of the magnetic field exists, that the morphology and amplitude of the excess  could be explained by ALP-photon conversion of a Cosmic ALP Background. Consistent results were subsequently found for data from the outskirts of the Coma cluster \cite{Kraljic:2014yta}, and from several other galaxy clusters for which the magnetic field is relatively well-constrained from Faraday rotation measurements \cite{Powell:2014mda}. Hence, ALP dark radiation may explain this galaxy cluster soft X-ray excess in the parameter space region with $m_a< 10^{-12}$ eV and $g_{a\gamma} \sim 10^{-11} {\rm GeV}^{-1}$ which is within the reach of detectability of IAXO.

\item {\bf Direct detection with IAXO}

For obtaining the sensitivity of IAXO to dark radiation we parametrize the flux produced by the modulus $\phi$ partially decaying to ALPs as
\begin{equation}
\frac{d\Phi}{dE}=2\times 10^6 \times \Delta N_{\text{eff}} \times \frac{E}{E_*^3}\, e^{-(E/E_*)^2} \, \left[\frac{1}{\text{cm}^2\,\text{s}\,\text{keV}}\right]\,,
\label{eq:fluxDR}
\end{equation}  
where $E_*$ is the average ALP energy. 
Under the assumption that negligible background is achievable down to a threshold of 0.2\,keV, the sensitivity of IAXO to dark radiation axions/ALPS would be the one shown in Fig.~\ref{Fig:IAXO-DR}, as a function of $E_*$. More realistic prospects depend on the experimental parameters (background and threshold) actually achieved with the technologies of choice to extend IAXO energy window to lower values, and will be studied in the near future.

\begin{figure}[t]
\centerline{\includegraphics[width=0.6\textwidth]{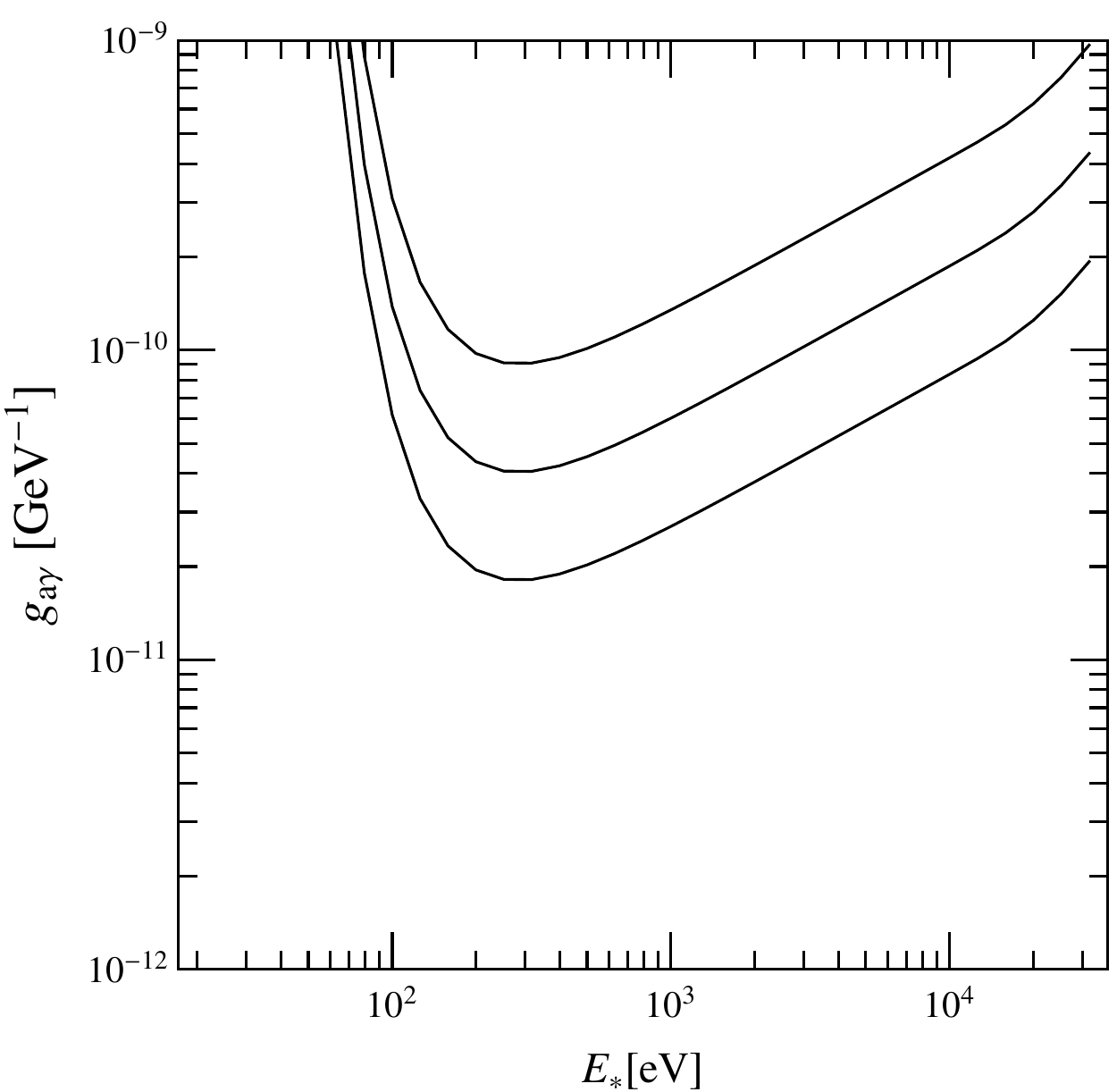}}
\caption{IAXO sensitivity to the axion/ALP two-photon coupling of cosmological dark radiation (DR). Lines correspond to 1 event/year for $\Delta N_{\text{eff}}= 0.1, 0.5,1.0$ from top to bottom. }
\label{Fig:IAXO-DR}
\end{figure}

\end{itemize}

%% file: sections/inflation.tex

The precise measurements of the CMB temperature and polarization anisotropies
provide strong support for the slow-roll inflation paradigm~\cite{Ade:2015lrj}. Successful inflation
requires an extremely flat inflaton potential for minimally coupled scalars, which is naturally realised
if the inflaton is an axion. 

The simplest axion inflation is the natural inflation with the following
potential~\cite{Freese:1990rb,Adams:1992bn},
\begin{align}
 V_{\rm inf}(a) =  \Lambda^4 \left(1 - \cos\left(\frac{a}{f_a} \right)\right),
\end{align}
where the inflation scale is set by the energy scale $\Lambda$.
The above potential leads to the large-field inflation where the inflaton field
excursion exceeds the Planck mass. In fact, the decay constant $f_a$
is constrained to be $f_a \gtrsim 5~M_{\rm Pl}$~\cite{Ade:2015lrj}.
Such a super-Planckian decay constant has been questioned from the perspective
 of a string theory~\cite{Rudelius:2014wla,delaFuente:2014aca,Rudelius:2015xta,Montero:2015ofa,Brown:2015iha} and quantum gravity in general\footnote{The cosmological constant cannot be larger than $M_P$.}.  Even if such a large decay
constant is justified, the predicted  spectral index $n_s$ and tensor-to-scalar ratio $r$
are not favoured by the current CMB observations.

From the effective field theory point of view, there are two possible simple ways to overcome both the trans-Planckian values of the decaying constant and the incompatibilities with current CMB data. The first one invokes the presence of multiple non-abelian gauge fields coupled to the axion in a shift invariant way.~\cite{Czerny:2014wza,Czerny:2014xja,Czerny:2014qqa,Croon:2014dma}. Here we assume that the axion potential consists of multiple cosine terms with different height and period.

In the minimal case, the  potential consists of two cosine terms~\cite{Czerny:2014wza}:
\begin{align}
\label{DIV}
V_{\rm inf}(a) = \Lambda^4\left( \cos\left(\frac{a}{ f_a}  + \vartheta \right)-\frac{\kappa}{ n^2}
\cos\left(\frac{n a}{f_a} \right)\right)+{\rm const.},
\end{align}
where $n (>1)$ is a rational number, $\kappa$ is a numerical coefficient, $\vartheta$ is a relative phase, and
the last term is a constant to realize the tiny cosmological constant at present.
 The quartic hilltop inflation is realized for  $\vartheta \approx 0$ and $\kappa \approx 1$.
The inflation takes place in the vicinity of the origin  where the curvature, or the effective mass,
is much smaller than the Hubble parameter during inflation.  After inflation the inflaton oscillates
about one of the potential minima, $a_{\rm min}$. Interestingly, if $n$ is an odd integer,  the axion mass at the potential maximum  and minimum is
equal in magnitude but has an opposite sign, i.e., the axion has a flat-top and flat-bottomed potential.
As a result, the axion is very light and cosmologically stable in the present vacuum,
and it may also account for DM.
This opens up a possibility to unify inflation and DM in terms of an axion which can
be searched for by experiments. In the following subsections we briefly summarize
implications of the axionic unification of inflation and DM~\cite{Daido:2017wwb}.

The second possibility is to change the axion propagator by non-minimally coupling to gravity. The only gravity-axion interaction conserving the tree-level shift invariance is Einstein tensor coupling \cite{Germani:2010hd}
\be
-\frac{G^{\alpha\beta}}{M^2}\partial_\mu a\partial_\nu a\ ,
\ee
where $M$ is a new mass scale\footnote{Note that this is not the unitarity violation scale \cite{Germani:2010hd}.}. 

\subsection{Axion hilltop inflation}
\label{sec:hilltop}
In the limit of $\vartheta = 0$ and $\kappa = 1$, the inflation model with (\ref{DIV}) reduces
to the quartic hilltop inflation, where the inflaton is massless both at the potential maximum and minimum. 
This case is ruled out by current observations implying a spectral index of order $n_s = 0.968 \pm 0.006$~\cite{Ade:2015lrj}.

The predicted spectral index can be increased to give a better fit to the observation
by allowing a non-zero $\vartheta$ in the small field scenario~\cite{Takahashi:2013cxa,Czerny:2014wza}. 
Interestingly, this implies a small axion mass,
\begin{align}
m_a^2 &\equiv V''(a_{\rm min}) \simeq
\left( \left(\frac{9(n^2-1)}{2} \right)^\frac{1}{6} \vartheta^\frac{1}{3} \frac{\Lambda^2}{f}\right)^2,
\end{align}
where we have approximated $|\vartheta| \ll 1$.
Then the observed spectral index  fixes  the inflaton mass as
\begin{align}
\label{m2H2}
 \frac{m_a^2}{H_I^2}  \sim  \frac{|V''(a_*)|}{H_I^2}  = {\cal O}(0.01),
\end{align}
where $H_I$ is the Hubble parameter during inflation, and
$a_*$ is the axion field value at the horizon exit of cosmological scales.
We emphasize that the second equality of (\ref{m2H2}) holds generally
in small-field inflation. A similar equality holds when $n_s$ is increased also by
allowing $\kappa \ne 1$. Thus, the Planck normalization of curvature perturbation and the observed spectral index
determine the relation between the decay constant and the axion mass~\cite{Daido:2017wwb,Daido:2017tbr}
\begin{equation}
\label{decay-mass}
f_a \simeq 5 \times10^7\,{\rm GeV}\,
\left(\frac{n}{3}\right)^\frac{1}{2}
\left(\frac{m_a}{1\,{\rm eV}}\right)^{\frac{1}{2}}.
\end{equation}
It seems that the QCD axion cannot behave as described here without changing radically the behaviour of the QCD potential and 
compromising the solution to the strong CP problem, however an ALP can very well realise this model. 
Moreover, if we couple the inflaton to photons we can obtain successful reheating. 
The coupling to photons is thus given as a function of the ALP mass as,
\begin{equation}
\label{eq:gagamma}
g_{a\gamma} \simeq 3 \times 10^{-11} C_{a\gamma}
\left(\frac{n}{3}\right)^{-\frac{1}{2}}
\left(\frac{m_a}{1\,{\rm eV}}\right)^{-\frac{1}{2}}  {\rm\,GeV}^{-1} .
\end{equation}

\subsection{Reheating of the ALP inflaton into photons}
\label{sec:reheating}
After inflation, the ALP oscillates about the potential minimum, and
the Universe is dominated by coherent oscillations of the ALP.
Since the ALP mass is much smaller than the typical curvature $\sim \Lambda^2/f$, the potential is 
well approximated by a quartic one. As the Universe expands,
the averaged oscillation amplitude, $a_0(t)$, gradually decreases inversely proportional to the scale factor.

The ALP decays and dissipates into plasma through its coupling to photons. First the ALP
decays into two photons with the rate,
\begin{equation}
\Gamma(a \to \gamma \gamma) = \frac{g_{a\gamma}^2}{64 \pi} m_{\rm eff}(t)^3,
\end{equation}
where $m_{\rm eff}(t) \equiv |V''(a_0(t))|^{1/2}$ is the effective mass of the ALP when the oscillation
amplitude is large. The decay proceeds until it is kinematically blocked by the thermal mass
of photons. Afterwards, the ALP gradually dissipates into plasma through scatterings of photons and electrons
off the ALP condensate. The dissipation rate is suppressed by the
effective ALP mass~\cite{Moroi:2014mqa}, and given by,
\begin{equation}
\Gamma_{{\rm dis},\gamma} = C_{{\rm dis},\gamma}
 \frac{g_{a \gamma}^2T^3}{8} \left( \frac{m_{\rm eff}^2}{ e^4T^2}\right)
\end{equation}
where $C_{{\rm dis},\gamma}$ is a numerical factor
of ${\cal O}(1-10)$ that represents uncertainties of the order-of-magnitude estimate
as well as the effects of tachyonic preheating and the subsequent
self-resonance~\cite{Felder:2000hj,Felder:2001kt,Micha:2004bv,Lozanov:2017hjm}.
When the plasma temperature exceeds the weak scale, one needs to take into account
the dissipation through similar couplings to the weak gauge bosons.

The effective ALP mass decreases as the averaged oscillation amplitude becomes smaller.
At a certain point, the dissipation becomes ineffective, and the reheating ends. To quantify the efficiency of the reheating,
we define $\xi$ as the ratio of the energy density of the remnant ALP to the total energy density
just after the reheating. For instance, if $\xi = 0.1$, $90$ \% of the initial ALP energy
dissipates into plasma, while the rest becomes the remnant.

Once the decay and dissipation rates are given, one can easily solve the Boltzmann equation
to evaluate $\xi$.  In order for the ALP to dissipate most of its energy, say, $\xi \lesssim {\cal O}(0.01)$,
it turns out that $g_{a \gamma}$ must be larger than ${\cal O}(10^{-11}){\rm \,GeV}^{-1}$,
taking account of the above mentioned uncertainties. In this case, the reheating ends in several Hubble times
after inflation, and the reheating temperature is estimated to be
\begin{equation}
T_{\rm RH} \simeq 40  \left(\frac{g_\star(T_{\rm RH})}{106.75}\right)^{-\frac{1}{4}} \left(\frac{m_a}{1{\rm\,eV}}\right)^\frac{1}{2}{\rm \,TeV}.
\end{equation}
In addition, ALPs are also thermalized in plasma, and its abundance is given by
\begin{equation}
\Delta N_{\rm eff} \simeq 0.03,
\end{equation}
where it decouples at a temperature slightly below the top quark mass.
Thermalized ALPs behave as dark radiation during
nucleosynthesis, and they become hot DM at a later time,
suppressing the matter power spectrum at small scales. Both effects can be searched for
by the future CMB and large-scale structure observations.

\subsection{The ALP miracle: axionic unification of inflaton and DM}
\label{sec:ALPmiracle}

After the reheating becomes ineffective, a small amount of the ALP is left over.
The remnant ALP is stable on cosmological time scales, and so, it contributes
to DM.  The energy density of the ALP remnant, $\rho_a$,  first decreases like radiation
since the potential is well approximated to be a quartic potential.  Then,
$\rho_a$ starts to decrease like matter when its oscillation amplitude becomes so small
that the ALP mass becomes non-negligible.
Assuming that the ALP remnant accounts for the observed DM density, the transition
temperature is given by
\begin{equation}
T_c \simeq  0.6\, \xi^{-1} \left(\frac{g_{\star s}(T_{\rm RH})}{g_{\star s}(T_c)}\right)^\frac{1}{3}
\,{\rm eV},
\end{equation}
where $T_c$ is the temperature at the transition. 
Since the ALP condensate behaves like dark radiation at $T > T_c$,  it
suppresses the small-scale matter power spectrum, which is
constrained by the SDSS and/or Lyman-$\alpha$ data. In order to be
consistent with the observed data,  the transition
should take place no later than the redshift $z_c = {\cal O}(10^5)$~\cite{Sarkar:2014bca}.
Then the ratio $\xi$ is bounded above:
\begin{equation}
\label{xi}
\xi \lesssim 0.02
\left(\frac{5 \times 10^5}{1+z_c}\right),
\end{equation}
where we substituted $g_{\star }(t_{\rm RH}) = 106.75$ and  $g_{\star S}(T_c) \simeq 3.909$.
Therefore, the small-scale matter power spectrum sets the lower bound on
the ALP-photon coupling,
\begin{equation}
g_{a \gamma} \gtrsim {\cal O}(10^{-11}){\rm \,GeV}^{-1}.
\label{lowerboundong}
\end{equation}

After the transition, the ALP remnant behaves like CDM.
The observed DM abundance is explained if the ALP mass is given by
\begin{align}
\label{mbelow}
m_a &\sim 0.1\, x^{-1} \left(\frac{\xi}{0.01}\right)^{-\frac{3}{4}}
 {\rm eV},
 \end{align}
where $x$ is a numerical factor of order unity which parametrizes the typical oscillation
amplitude of the ALP remnant as $a_{0}^{\rm (rem)} = x\,\xi^{1/4} f$.

Combining (\ref{eq:gagamma}), (\ref{xi}), (\ref{lowerboundong}), and (\ref{mbelow}), we arrive
at the unique parameter region,
\begin{align}
&0.01{\rm \,eV}  \lesssim m_a \lesssim 1 {\rm \,eV},\\
&g_{a\gamma} ={\cal O}(10^{-11}){\rm\,GeV}^{-1},
\end{align}
where both inflation and DM are simultaneously explained by the ALP.
It is highly non-trivial that such a viable region exists
without running afoul of the current experimental and observational limits.
We refer to this coincidence as the {\it ALP miracle}~\cite{Daido:2017wwb}. The region of the ALP parameter space consistent with the CMB observation has been more carefully evaluated in a recent work~\cite{Daido:2017tbr}, and it is shown in Fig.~\ref{fig:DM_predictions}. 
Interestingly, the predicted sweet spot of the ALP miracle significantly overlaps with the sensitivity
reach of IAXO.

%% file: sections/cooling.tex




Independent observations of diverse stellar systems have shown deviations from the predicted behaviour, indicating in all cases an over-efficient cooling~\cite{Ringwald:2015lqa,Giannotti:2015kwo,Giannotti:2017hny,DiVecchia:2019ejf}.
These deviations, often referred to as \textit{cooling anomalies}, have been observed in: \textit{1)} several pulsating white dwarfs (WDs), in which the cooling efficiency was extracted from the rate of the period change;
\textit{2)} the WD luminosity function (WDLF), which describes the distribution of WD as a function of their brightness;
\textit{3)} red giants branch (RGB) stars, in particular the luminosity of the tip of the branch;
\textit{4)} horizontal branch stars (HB) or, more precisely, the R-parameter, that is the ratio of the number of HB over RGB stars;
\textit{5)} helium burning supergiants, more specifically the ratio B/R of blue and red supergiants;
and \textit{6)} neutron stars.
Table~\ref{tab:anomalies} summarizes the results.
\begin{table}[t]
	\begin{center}
		\begin{tabular}{ l  c  l  l l}
			\hline \hline
			\textbf{Observable} &	& \textbf{Stellar System}		&  \textbf{Proposed Solution(s)}  				& \textbf{References}	\\ \hline
			Rate of	period		&	& WD Variables  				&  axion or ALP coupled to electrons;			& \cite{Corsico:2012ki,Corsico:2012sh,Corsico:2014mpa,Corsico:2016okh,Battich:2016htm}	\\
			change				& 	& 								&  neutrino magnetic moment; 					& \\ \hline
			Shape of WDLF		&	& WDs					   		&  axion or ALP coupled to electrons; 			 &\cite{Bertolami:2014noa,Bertolami:2014wua}\\ \hline
			Luminosity of 		&	& Globular Clusters		   		&  axion or ALP coupled to electrons;			 &\cite{Viaux:2013hca,Viaux:2013lha,Arceo-Diaz:2015pva}	\\
			the	RGB tip			&	&  (M5, $ \omega $-Centauri)	&  neutrino magnetic moment; 					&\\ \hline
			R-parameter			&	& Globular Clusters  	 	 	&  axion or ALP coupled to photons;  			&\cite{Ayala:2014pea,Straniero:2015nvc}\\ \hline
			B/R					&	& Open Clusters      			&  axion or ALP coupled to photons; 			 &\cite{Skillman:2002aa,McQuinn:2011bb,Friedland:2012hj,Carosi:2013rla}\\ \hline
			Neutron stars		&	& CAS A				 	  		&  axion or ALP coupled to neutrons.				&\cite{Leinson:2014ioa}\\ \hline \hline
		\end{tabular}
		\caption{Summary of anomalous cooling observations~\cite{Giannotti:2016hnk}.
			All the anomalies have, individually, about 1-2~$ \sigma $ statistical significance, except for the last two for which the significance has not been quantified.}
		\label{tab:anomalies}
	\end{center}
\end{table}

Furthermore, axions may have observable effects on the late evolutionary stages of massive stars, for example influencing the nucleosynthesis~\cite{Aoyama:2015asa}. 
Axions with couplings at reach of IAXO would also modify the threshold initial mass for which carbon is ignited in the stellar core (Mup ), ultimately shifting the required mass to explode as core collapse SN by 1-2 $M_{\odot}$, depending on the couplings~\cite{Dominguez:2017yhy,Dominguez:2017mia}.
Finally, a recent study indicates that axions reduce the final (pre-SN) luminosity expected for a star of a given initial mass~\cite{StranieroEtAl:2019}. Presently, observations are sparse but they do seem to indicate a preference for the axion scenario, with couplings somewhat below the recent CAST bound and likely accessible to IAXO and possibly BabyIAXO.

It is quite remarkable that such different stellar systems show, systematically, an excessive amount of energy loss, indicating a lack of understanding in the current modeling of stellar cooling.
Notice that the amount of anomalous energy loss is different in different objects.
For example, main sequence stars, such as our Sun, do not show this anomalous behaviour and, in general, the amount of exotic cooling indicates some correlation with the interior temperature of the stellar object.
The accumulation in the recent years of data from such different stellar objects offers a unique possibility to revise our understanding of the cooling mechanisms in stars.

An appealing explanation to these anomalous observations is to assume that the source of the exotic cooling is a new light, weakly interacting particle, produced in the stellar core and able to stream freely outside, carrying energy away, in much the same way as neutrinos do.
Interestingly, the peculiar dependence of the observed additional cooling on temperature and density is extremely selective on the possible particles which could account for all the observations, and indicate a clear preference for a pseudoscalar (axion-like) particle coupled to photons and matter~\cite{Giannotti:2015kwo}.
Indeed, given the large variations in temperature and density of the systems considered, it is quite remarkable that one particle alone can simultaneously address all the observed problems, a fact that, in our opinion, strengthens considerably the physics case for axions and ALPs.



\subsection{Summary of cooling anomalies}
\label{sec:summary_of_cooling_anomalies}

\subsubsection{Pulsating White Dwarfs}
Original hints to  cooling anomalies were derived from the observations, starting from Kepler et. al.~\cite{KeplerEtAl} in 1991, that the rate of period change $ \dot P /P $ of G117 - B15A, a pulsating WD, was larger than predicted by the standard pulsation theory~\cite{Isern:1992gia}.

After over 20 years of additional observations, the hint from G117 - B15A remains~\cite{Corsico:2012ki}.
Additionally, the WD variables R548~\cite{Corsico:2012sh},  PG 1351+489~\cite{Battich:2016htm}, L 19-2	(113)~\cite{Corsico:2016okh} and L 19-2	 (192)~\cite{Corsico:2016okh} have all shown similar anomalies,  with the observed rate of period change being always larger than expected.
The current results are reported in Table~\ref{tab:Pdot}.\footnote{The discrepancy has been calculated using the data for $ \dot P /P $ in the quoted references and does not account for possible unknown systematics.
	In particular, the high significance of the discrepancy for G117 - B15A could be due to the hypothesis that the particular oscillating mode examined (with period about 215 s) is trapped in the envelope (see, e.g., \cite{Corsico:2012ki}), an assumption which could be incorrect~\cite{Bertolami:2014wua}.
	Relaxing this hypothesis would significantly reduce the discrepancy~\cite{Corsico:2012ki} (see also discussion in~\cite{Giannotti:2015kwo}).}
\begin{table}[h]
	\begin{center}
		\begin{tabular}{  l  l l l l c p{5cm}}
			\hline\hline
			WD  			& class	&  $ P $[s] &  $ \dot{P}_{\rm obs} $[s/s]  &  $ \dot{P}_{\rm th} $[s/s] & discrepancy \\ \hline
			G117 - B15A 	&DA 	&  215 &  $ (4.19 \pm 0.73)\times 10^{-15} $ 	&  $ (1.25 \pm 0.09)\times 10^{-15} $ & $4\, \sigma $ \\
			R548 		 	&DA 	&  213 &  $ (3.3 \pm 1.1)\times 10^{-15} $ 	&  $ (1.1 \pm 0.09)\times 10^{-15} $ & $2\, \sigma $\\
			PG 1351+489		&DB 	&  489 & $ (2.0 \pm 0.9)\times 10^{-13} $ 		& $ (0.81 \pm 0.5)\times 10^{-13} $ & $1.1\, \sigma $\\
			L 19-2	(113) 	&DA 	&  113 &  $ (3.0\pm 0.6)\times 10^{-15} $ 	&  $ (1.42 \pm 0.85)\times 10^{-15} $ & $1.5\, \sigma $\\
			L 19-2 	(192) 	&DA 	&  192 &  $ (3.0\pm 0.6)\times 10^{-15} $ 	&  $ (2.41\pm 1.45) \times 10^{-15} $ & $0.4\, \sigma $ \\ \hline
		\end{tabular}
		\caption{Results for $\dot P $ measured and expected in WD variables.
		}
		\label{tab:Pdot}
	\end{center}
\end{table}

Since $ \dot P /P $ is practically proportional to the cooling rate $ \dot T /T $, the results seem to indicate that these WDs are cooling substantially faster than expected.
The unaccounted energy loss could be due to a novel particle, efficiently produced in the dense core of a WD and freely escaping carrying energy away.
Examples considered in the literature are axions produced through electron bremsstrahlung~\cite{Corsico:2012ki,Corsico:2016okh} (see sec.~\ref{sec:axion_interpretation}) and neutrinos with an anomalous large magnetic moment produced through plasmon decay~\cite{Corsico:2014mpa}.

\subsubsection{The White Dwarf Luminosity Function}

An anomalous behaviour was also observed in the WD luminosity function (WDLF), which describes the number distribution of WDs in brightness intervals.
The particular shape of this distribution depends on the lifetime of WDs in a specific luminosity bin and, consequently, on the efficiency of the cooling mechanisms.
A number of studies (see, e.g., \cite{Bertolami:2014wua}) showed that additional cooling provided by axions/ALPs coupled to electrons could improve considerably the fit, while even a large neutrino magnetic moment would not have any substantial effect on the WDLF~\cite{Bertolami:2014noa}.
This result can be attributed to the too steep temperature dependence of the plasmon decay into neutrinos~\cite{Giannotti:2015kwo}.

A more recent study of the hot part of the WDLF~\cite{Hansen:2015lqa} did not confirm this anomalous behaviour.
However, the hotter section of the WDLF has much larger observational errors and the ALP production would be almost completely hidden by standard neutrino cooling
in the hottest WDs.
Whatever the case, a decisive improvement in our understanding of the WDLF is expected in the next decade or so. 
Observations from the Gaia satellite have already increased the catalog of WDs by an order of magnitude with respect to SDSS, and LSST is expected to ultimately increase the census of the WDs to tens of millions~\cite{Drlica-Wagner:2019xan}.

\subsubsection{Globular Clusters}
\label{sec:stellarHints_GC}
Further hints to anomalous energy loss emerged in the recent analyses of Red Giant Branch (RGB) stars in the globular cluster M5~\cite{Viaux:2013hca,Viaux:2013lha} and, with less significance, in the globular cluster $ \omega $-Centauri~\cite{Arceo-Diaz:2015pva}.

The red giant is the evolutionary stage of low mass stars that follows the main sequence.
Stars in this stage have a He core and burn H in a shell.
During their evolution in the RGB, stars become brighter and brighter until they reach a tip in the color magnitude diagram at the time of the He-flash, corresponding to the He ignition in their core.
After the He-flash, the luminosity decreases and the stars move into the Horizontal Branch (HB) stage.

Additional cooling of the core during the RGB stage would delay the He ignition allowing the star to become brighter before moving into the HB phase.
The tip of the RGB is, therefore, a measure of the cooling efficiency during the RGB evolutionary stage.

The studies in~\cite{Viaux:2013hca,Viaux:2013lha,Arceo-Diaz:2015pva} showed a brighter than expected tip of the RGB, indicating a somewhat over-efficient cooling during the evolutionary phase preceding the helium flash.
In all cases, the significance is fairly low, $ \sim 1.2\, \sigma$ in the M5 analysis and less than $ 1\,\sigma $ in  $ \omega $-Centauri.
%
In spite of the low significance of these results, however, it is remarkable that RGB observations seem to confirm the need for additional cooling in agreement with the completely unrelated results from WDs.
These results can be improved using multi-band photometry of multiple globular clusters~\cite{Straniero:2018fbv}. 
A considerable reduction of the observational uncertainties, particularly those related to the clusters distances, are expected from the data of the Gaia satellite mission~\cite{pacino17}.
%
%

Moreover, an independent analysis~\cite{Ayala:2014pea} showed a disagreement between the observed and expected $ R $-parameter,
$R= {N_{\rm HB}}/{N_{\rm RGB}}$,
which compares the numbers of stars in the HB  ($N_{\rm HB}$) and in the upper portion of the RGB
($N_{\rm RGB}$).
Assuming Gaussian errors, the discrepancy between the observed, $ R=1.39\pm 0.03 $, and the expected,
$ R=1.47\pm 0.03 $, values is about $ 2\,\sigma $.
The result indicates a surplus of RGB with respect to the numerical prediction and can be interpreted as an anomalous cooling with a different degree of efficiency in the two evolutionary stages.
This anomaly is known as the $ R- $parameter or HB hint.

Assuming the result is due to new physics,
the ideal candidate to explain the low number of HB would be an ALP coupled to photons and produced through the Primakoff mechanism in the stellar core.
This process is fairly inefficient in the high density environment of the RG core and could have the effect of accelerating only the following HB stage (hence, HB hint), reducing therefore the number of HB versus RGB stars and explaining the result for $ R $~\cite{Ayala:2014pea,Straniero:2015nvc}.
A cooling mechanism efficient during the RGB stage could, however, also explain the discrepancy in the predicted and measured $ R- $parameter.
In particular, an ALP coupled to electrons and produced through electron bremsstrahlung or Compton could produce a similar effect~\cite{Giannotti:2015kwo}.
In general, we could expect a combination of the two mechanisms, as discussed in sec.~\ref{sec:axion_interpretation}.

\subsubsection{He-burning Supergiants}

An additional deviation from the standard cooling theory was observed in core He-burning stars of intermediate mass ($ M\sim 10M_\odot $).
The problem, in this case, is that numerical simulations predict a larger number ratio of blue (hot) over red (cold) supergiants (B/R), with respect to what is actually observed~\cite{Skillman:2002aa,McQuinn:2011bb}.
The predicted number would be lowered (alleviating or, perhaps, solving the B/R problem) in the hypothesis of an additional cooling channel efficient in the stellar core but not in the H-burning shell~\cite{Lauterborn:1971nva}, which could be provided by axions coupled to photons and produced through the Primakoff process in a way analogous to what discussed for HB stars~\cite{Friedland:2012hj,Carosi:2013rla,Giannotti:2014cpa}.
An exact prediction of the required additional cooling is, however, presently unavailable.

\subsubsection{Neutron Star in CAS A}

Finally, X-ray observations of the surface temperature of a neutron star in Cassiopeia A~\cite{Ho:2009mm,Heinke:2010cr,Shternin:2010qi} showed a cooling rate considerably faster than expected.
The effect seems to indicate the need for an additional energy loss roughly equal to the standard one.
This was interpreted in terms of an axion-neutron coupling~\cite{Leinson:2014ioa}
\begin{align}
	 g_{an}\simeq 4 \times 10^{-10} \,.
\end{align}
Although this result is compatible with well established limits from NS cooling~\cite{Keller:2012yr,Sedrakian:2015krq}, new results based on different assumptions of the NS micro-physics have challenged this picture~\cite{Hamaguchi:2018oqw,Beznogov:2018fda}. 
Moreover, the observed anomaly may also have origin in the phase transition of the neutron condensate into a multicomponent state~\cite{Leinson:2014cja}.
Given this controversy, and the general difficulty in the modelling of NS, we decided not to include this result in our global analysis of the cooling anomalies presented here.

\subsection{Cooling anomalies, axions and IAXO}
\label{sec:axion_interpretation}
As discussed above, axions/ALPs are favourite candidates, among the various new physics options, to explain the cooling anomalies~\cite{Giannotti:2015kwo}.
More recently, these have also been interpreted in terms of concrete QCD axions models~\cite{Giannotti:2017hny} such as KSVZ, DFSZ, and Axi-Majoron (A/J). Here we present the regions of the ALP and axion parameter space hinted by the anomalous observations and the IAXO potential to probe these areas.

We consider first the case, often discussed in the literature, of an ALP interacting only with photons.
In this case, the most relevant axion production mechanism is the Primakoff process,
\begin{eqnarray}
\gamma +Ze\to Ze+a\,,
\end{eqnarray}
which consists in the conversion of a photon into an ALP in the electric field of nuclei and electrons in the stellar core.
This process depends strongly on the environment temperature and is suppressed at high density (e.g., those characterizing the core of WDs and RGB stars) by the plasma frequency and degeneracy effects (see, e.g., \cite{Raffelt:1987yu}).
Thus, if we ignore the interaction with electrons, axions cannot provide a solution for the excessive cooling observed in WDs and RGB.
Instead, their main effect on the evolution of low mass stars would be to accelerate the HB phase while leaving essentially unchanged the RGB stage, and therefore to reduce the expected R-parameter.

\begin{figure}[t]
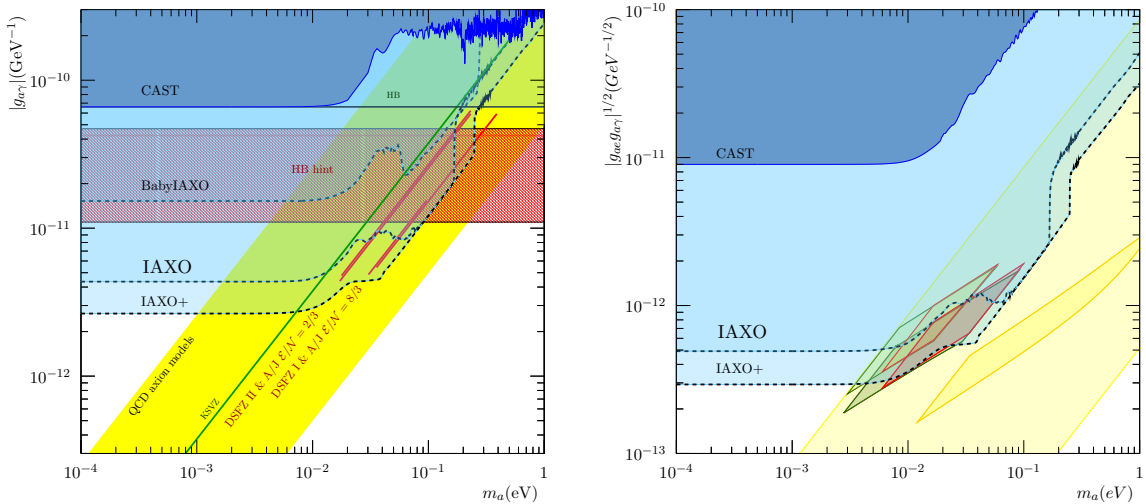

	\centering
    \tikzsetnextfilename{pics/IAXO_star_anomaly_zoom}
    \resizebox{0.49\linewidth}{!}{\input{pics/IAXO_star_anomaly_zoom.tex}}	
   \tikzsetnextfilename{pics/IAXO_star_anomaly_gae}
   \resizebox{0.49\linewidth}{!}{\input{pics/IAXO_star_anomaly_gae.tex}}
	\caption{
Summary of astrophysical hints in the ($\gagamma$,$m_a$) and ($\sqrt{\gagamma g_{ae}}$,$m_a$) planes compared with the expected sensitivity of IAXO. The 2\,$ \sigma $ regions corresponding to five explicit QCD axion models (DFSZ~I, DFSZ~II and A/J with $\cal{E}/\cal{N}=$ 2/3, 5/3 and 8/3) that can account for the WD/RGB/R anomalies following the work in~\cite{Giannotti:2017hny} are shown. Only the fraction of the regions for which solar axion production through the $\gagamma$ or $g_{ae}$ couplings is dominant, is shown on the left and right plots respectively. The unitarity constraints have also been imposed.
Left: 
The red-hashed region is the hint from HB stars~\cite{Ayala:2014pea,Straniero:2015nvc,Giannotti:2015dwa}, obtained in the assumption of interaction with photons only.
The yellow region shows the customary band for QCD axion models, see text for details.
The mentioned hinted regions are shown as red segments. Note that the DFSZ lines have been displaced 5\% upwards for visibility as they overlap with A/J models. The A/J model with  $\cal{E}/\cal{N}=$ 5/3 does not appear here because solar production via $g_{ae}$ is dominant for all this region.
Right: The IAXO sensitivity assumes only solar production through the electron coupling.
The 2\,$ \sigma $ hinted regions appear as diamond-shaped regions in yellow (A/J~$\cal{E}/\cal{N}=$~5/3), red (A/J~$\cal{E}/\cal{N}=$~8/3), orange (A/J~$\cal{E}/\cal{N}=$~2/3), light green (DFSZ~II) and dark green (DFSZ~I).
The wide yellow band encompasses all possible DFSZ models within unitarity constraints. As shown, the combined prospects of IAXO via the $\gagamma$ and $\gagamma g_{ae}$ channels will probe most of the hinted region, with the only exception of the A/J  $\cal{E}/\cal{N}=$~5/3 model, due to the extremely suppressed $\gagamma$ coupling.
	}
	\label{fig:hints}
\end{figure}

This property has been used to constrain the axion-photon coupling~\cite{Raffelt:1987yu,Ayala:2014pea,Straniero:2015nvc}.
The current $ 2\,\sigma $ bound, $ g_{a\gamma} < 0.65\times 10^{-10} $GeV$ ^{-1} $~\cite{Ayala:2014pea,Straniero:2015nvc}, is shown in Fig.~\ref{fig:hints} (left) by the solid black line labeled ``HB''.

The red-hashed region in the left panel of Fig.~\ref{fig:hints} is the hint from the $ R- $parameter (HB-hint)~\cite{Ayala:2014pea,Giannotti:2015dwa,Straniero:2015nvc} at $ 1\,\sigma $ confidence level: $ g_{a\gamma} =(0.29\pm 0.18) \times 10^{-10} $GeV$ ^{-1} $.
Notice that the bound and the hinted regions are obtained assuming a vanishing ALP-electron interaction.
The more general case will be discussed below.
As evident from the figure, IAXO is expected to have sufficient sensitivity to detect ALPs in this region, for masses below $m_a\sim 0.1$~eV.

Axions (or ALPs) interacting also with electrons have the potential to explain the additional hints from the WD pulsation, the WDLF, and RGB stars. Moreover, the axion-electron coupling opens up new axion production channels in the Sun, improving the discovery potential of IAXO.

The 1~$ \sigma $ intervals on the axion/ALPs coupling with electrons and photons derived from the observations of individual stellar systems are shown in Tab.~\ref{tab:WD_RGB_anomalies}.\footnote{The confidence intervals for the WD variables shown in Tab.~\ref{tab:WD_RGB_anomalies} have been derived from a likelihood analysis of the data in~\cite{Corsico:2012sh,Corsico:2014mpa,Corsico:2016okh} and are not given in the original references.	The hint from the WDLF is derived from the data in~\cite{Bertolami:2014wua}. The RGB hint refers to the M5 data~\cite{Viaux:2013lha} and is calculated from a likelihood analysis combining the observational and computational 1$ \sigma $ errors in quadrature. It has a slightly smaller error than the one reported in~\cite{Giannotti:2015kwo}, which was based on a conservative interpretation of Fig.~2 in~\cite{Viaux:2013lha}.
	}
%
\begin{table}[h]
	\begin{center}
		\begin{tabular}{| l  l |  l l |}
			\hline
			\textbf{observable} 					& \textbf{hint (1$ \sigma $)}		& \textbf{observable}			& \textbf{hint (1$ \sigma $)}					 \\ \hline
			WD G117 - B15A				&	 $ \alpha_{26}=1.89\pm 0.47 ~ $  	& 	WD R548 					&$ \alpha_{26}=1.84\pm 0.93 ~ $   	\\
			WD PG 1351+489 				&	 $ \alpha_{26}=0.36\pm 0.38~ $  	& 	WD L19-2   (113 s mode)		&	 $ \alpha_{26}=2.08\pm 1.35 ~ $   	\\
			WD L19-2  (192 s mode)		&	 $ \alpha_{26}=0.5\pm 1.2 ~ $   	& 	WDLF 						&$ \alpha_{26}=0.16^{+0.19}_{-0.14} ~ $   		 \\
			luminosity of RGB tip  		& $ \alpha_{26}=0.28^{+0.47}_{-0.24}~ $ &   R-parameter 				& $ g_{10}= 0.29\pm 0.18$	\\ \hline
		\end{tabular}
		\caption{
			Hints at $ 1~ \sigma $ from stellar cooling anomalies (from ref.~\cite{Giannotti:2016hnk}).
			Here, $ \alpha_{26} =(g_{ae}\times 10^{13})^2/4\pi$ and $ g_{10}=g_{a\gamma}\times 10^{10} $GeV.
			The hint on the R-parameter shown here assumes no ALP-electron interaction. }
		\label{tab:WD_RGB_anomalies}
	\end{center}
\end{table}
The most relevant axion production mechanism at high density is the bremsstrahlung process,
\begin{equation}
\label{Eq:ALP_bremm}
e+Ze\to Ze+e+a\,,
\end{equation}
which induces an additional energy loss rate proportional to $ T^4 $.
As shown in~\cite{Giannotti:2015kwo}, this temperature dependence is optimal to fit the WDLF and provides a reasonably good explanation for the observed excess cooling in DA and DB WD variables, whose internal temperatures differ by a factor of a few.
The combined analysis of all the observed WD variables gives a fairly good fit, $ \chi^2_{\rm min}/ $d.o.f$ =1.1 $, for $ g_{ae}=2.9\times 10^{-13} $ and favours the axion solution at $ 2\,\sigma $.

Moreover, the peculiar temperature dependence of the axion bremsstrahlung rate allows to account  for the excessive cooling observed in RGB stars~\cite{Viaux:2013lha} (which have a considerably larger internal temperature than WDs) with a comparable axion-electron coupling.
%
%
\begin{figure}[t]
	\begin{center}
		\includegraphics[width=7cm]{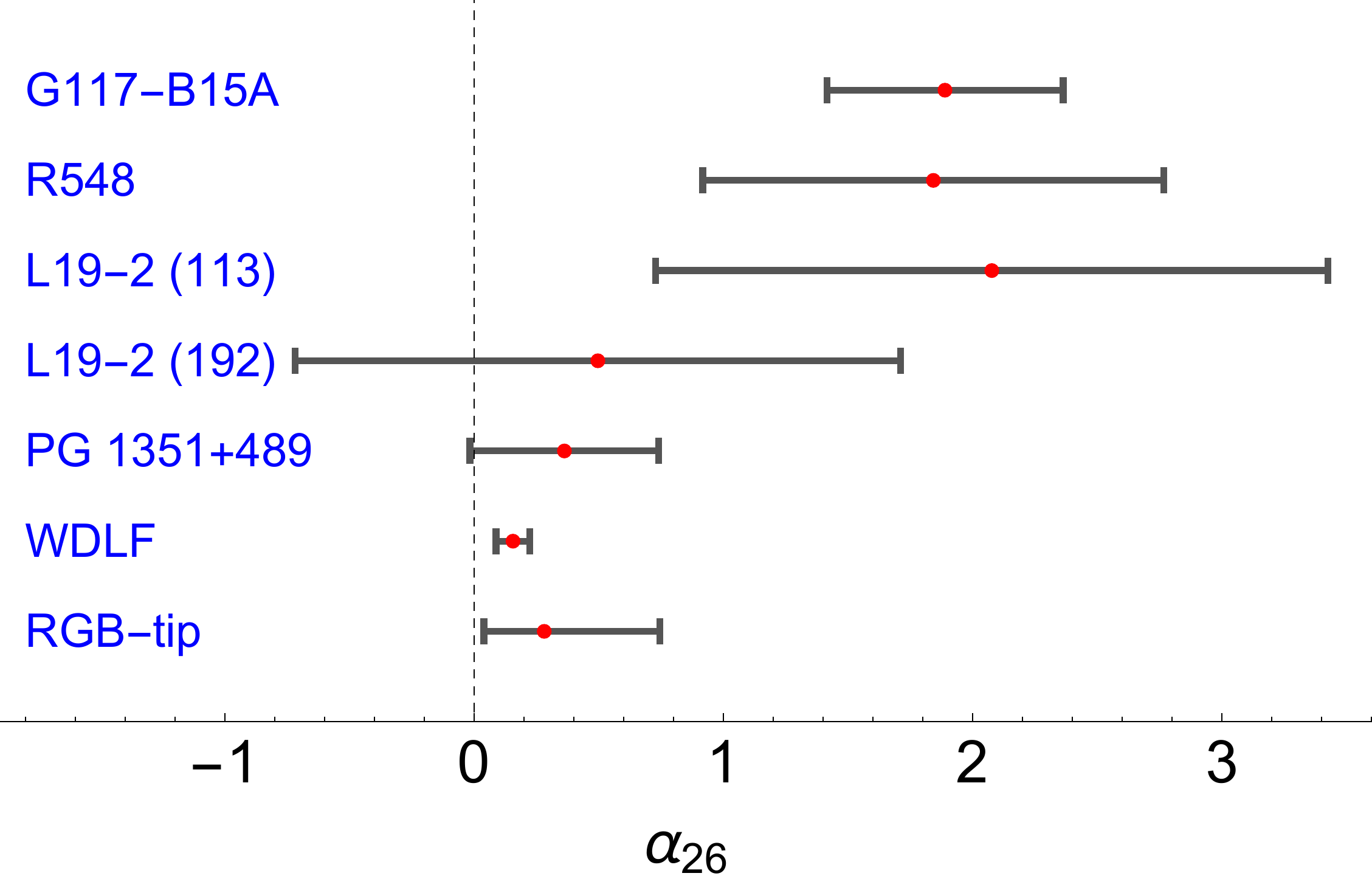}
		\hspace{0.3cm}
		\includegraphics[width=6cm]{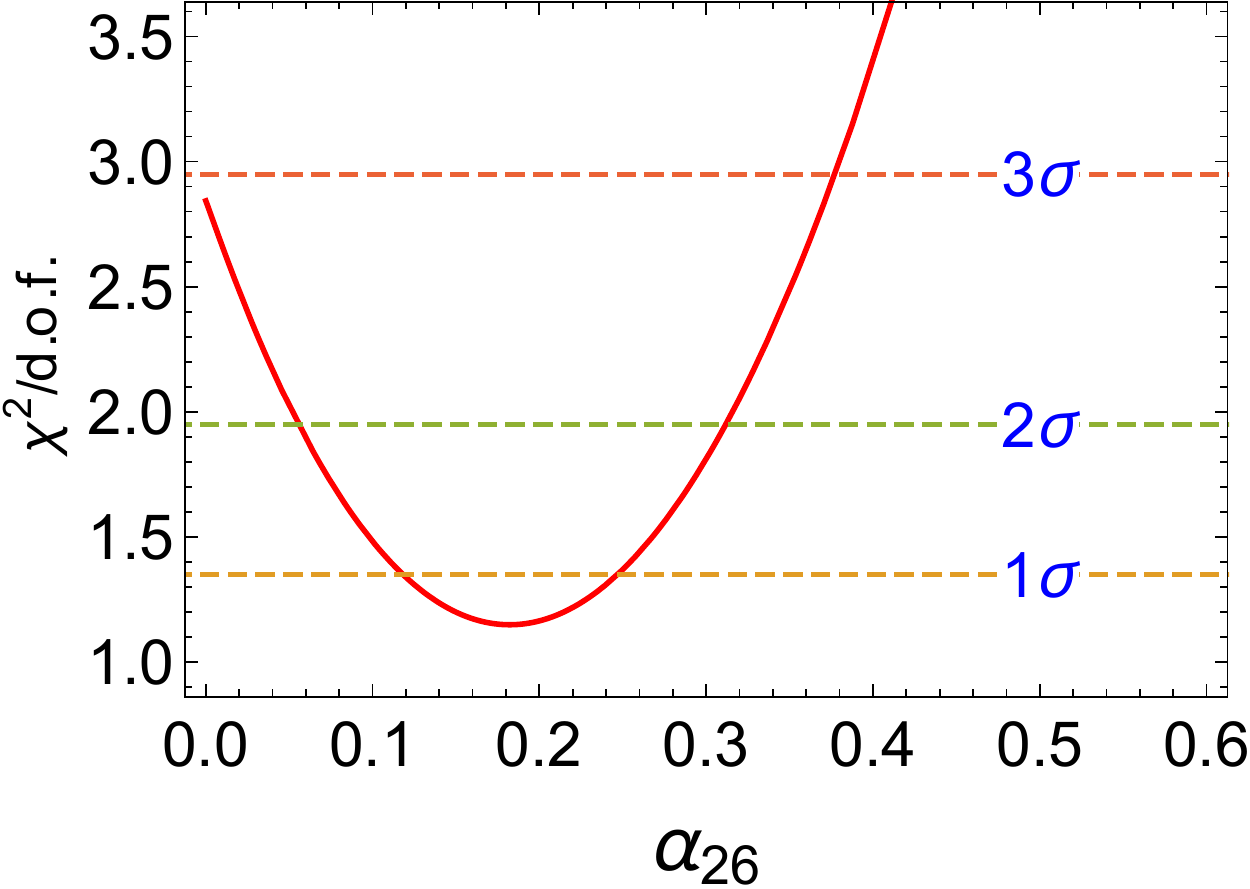}
		\caption{Hinted regions for the axion-electron coupling, with  $ \alpha_{26} =(g_{ae}\times 10^{13})^2/4\pi$.
			The segments in the left side plot show the $ 1\, \sigma $ intervals.
			On the right we show the fit of all the hints combined, excluding the G117-B15A hint, as discussed in~\cite{Giannotti:2015kwo}.	}
		\label{fig:alpha_26_hints}
	\end{center}
\end{figure}
The combination of the hints from the WD pulsation, the WDLF and RGB stars gives $ g_{ae}=1.51^{+0.25}_{-0.29} \times 10^{-13}$
with $ \chi^2_{\rm min}/ $d.o.f$ =1.1 $ and favors the axion solution at about $ 3\,\sigma $, see Fig.~\ref{fig:alpha_26_hints}.

A more recent independent analysis of the RGB luminosity tip in $ \omega $-Centauri~\cite{Arceo-Diaz:2015pva} confirms the hint to additional cooling, though this study has so far been carried out considering only the neutrino magnetic moment as source of exotic cooling.

\begin{figure}[t]
\centering
\tikzsetnextfilename{pics/IAXO_star_anomaly_gae_vs_gag}
\resizebox{0.5\linewidth}{!}{\input{pics/IAXO_star_anomaly_gae_vs_gag.tex}}

\caption{Combined analysis of the hints on $ g_{ae} $ and $ g_{a\gamma}$.
The light brown shaded region is the 1$ \sigma $ hint, with the best fit value $ g_{ae}=1.54\times 10^{-13} $, $ g_{a\gamma}=0.13\times 10^{-10} $ GeV$ ^{-1} $.
Also shown is the sensitivity of IAXO, assuming $m_a<0.01$ eV] for which the axion photon oscillation probability is independent of the mass. }
\label{fig:hinted_region}
\end{figure}
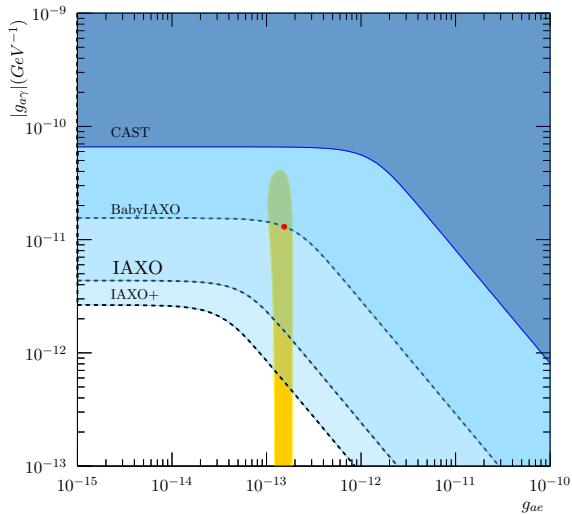

Finally, if we add also the analysis from the $ R- $parameter anomaly, we find the results presented in Fig.~\ref{fig:hinted_region}, where the light brown region shows the $ 1\,\sigma $ hinted area.
Notice that the $ R $-parameter depends, in general, on both the axion-electron and axion-photon couplings~\cite{Giannotti:2015kwo}.
This explains the slight left bending shape of the $ R-$parameter hinted region.
In particular, the quantitative analysis with the data at hand shows that it is possible to explain all the observed cooling hints, including the $ R $-parameter anomaly, even neglecting the axion-photon coupling, though there is a preference for non-vanishing couplings with both electrons and photons.
The best fit values, $g_{ae}=1.5\times 10^{-13}$ and $ g_{a\gamma}=0.13\times 10^{-10} $ GeV$ ^{-1} $, indicated in the figure with a red dot, are well within reach of IAXO though the $ 1\,\sigma $ hinted region extends, in this case, to lower axion-photon couplings.

The sensitivity of IAXO to axions coupled to both electrons and photons is shown in Fig.~\ref{fig:hints} (right). Here it is assumed that production of solar axions proceeds through the axion-electron coupling and detection through the photon coupling. If production via the Primakoff process dominates the solar flux the relevant graphical reference is Fig.~\ref{fig:hints} (left).
We can already note that the best fit value $g_{ae}=1.5\times 10^{-13}$ with $ g_{a\gamma}=0.13\times 10^{-10} $ GeV$ ^{-1}$ would be detectable by IAXO up to ALP masses of $0.1$ eV.
However, we recall again that within the $1\sigma$ hinted region, the coupling to photons could be zero and this would prevent any possibility of detection by IAXO.

\begin{figure}[t]
\begin{center}
\includegraphics[width=6cm]{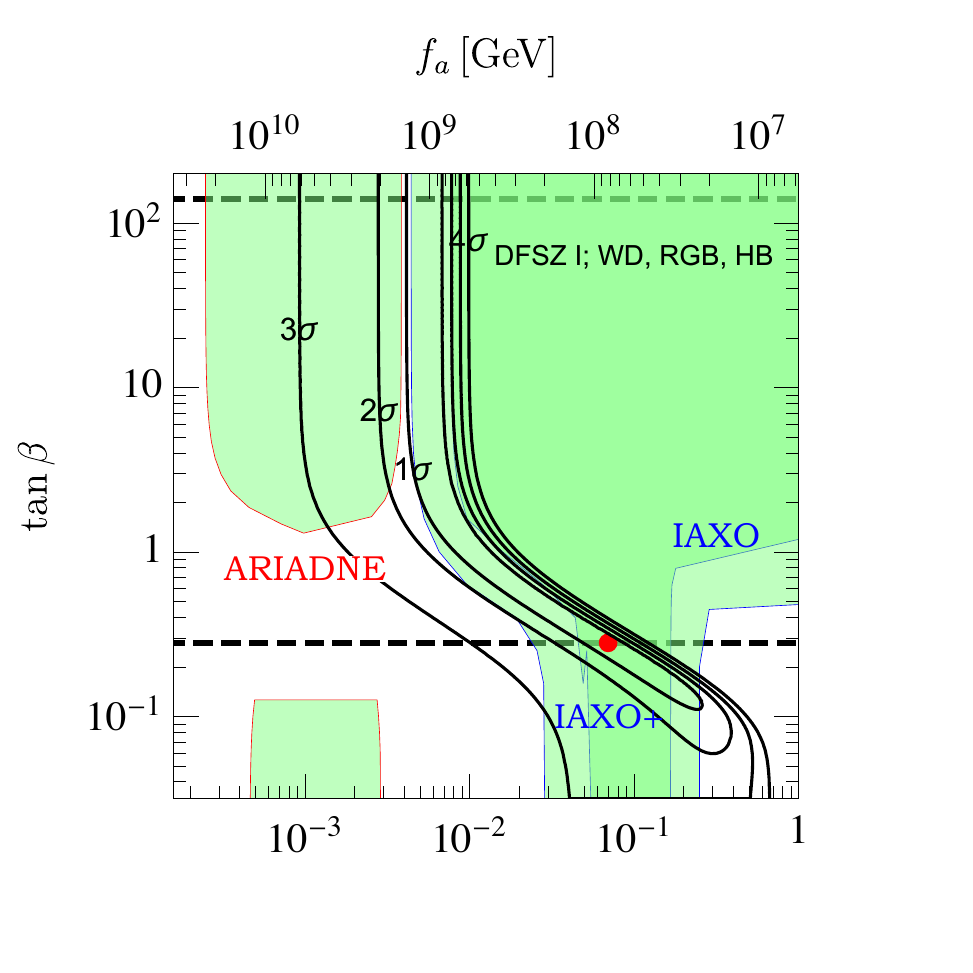}
\includegraphics[width=6cm]{pics/hints/DFSZ1WDRGBHB.pdf}
\vspace{-.7cm}
\includegraphics[width=5cm]{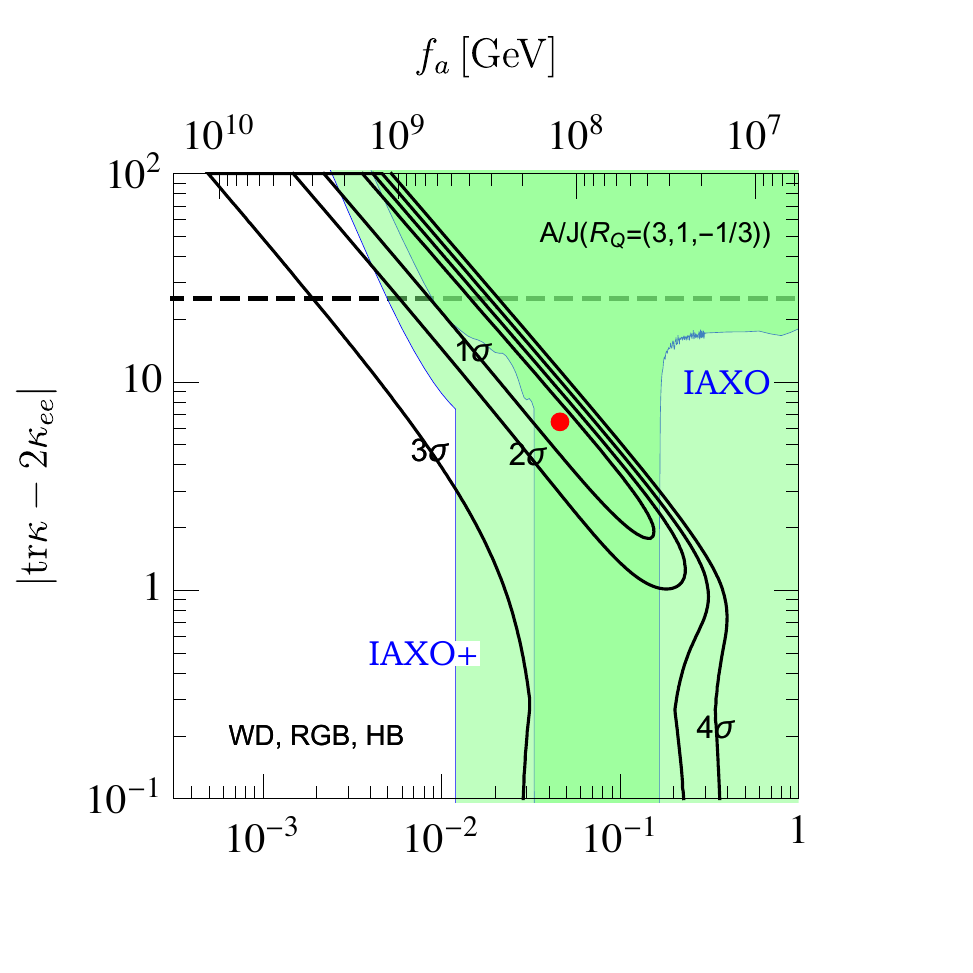}
\includegraphics[width=5cm]{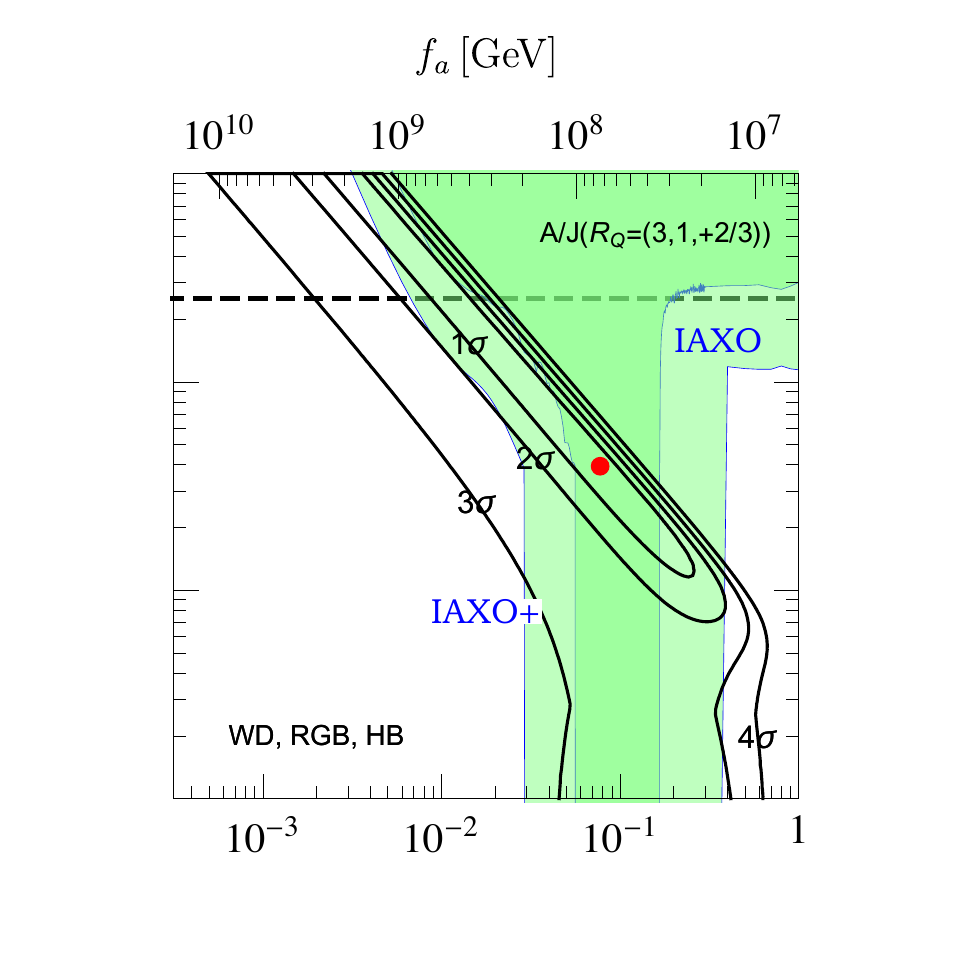}
\includegraphics[width=5cm]{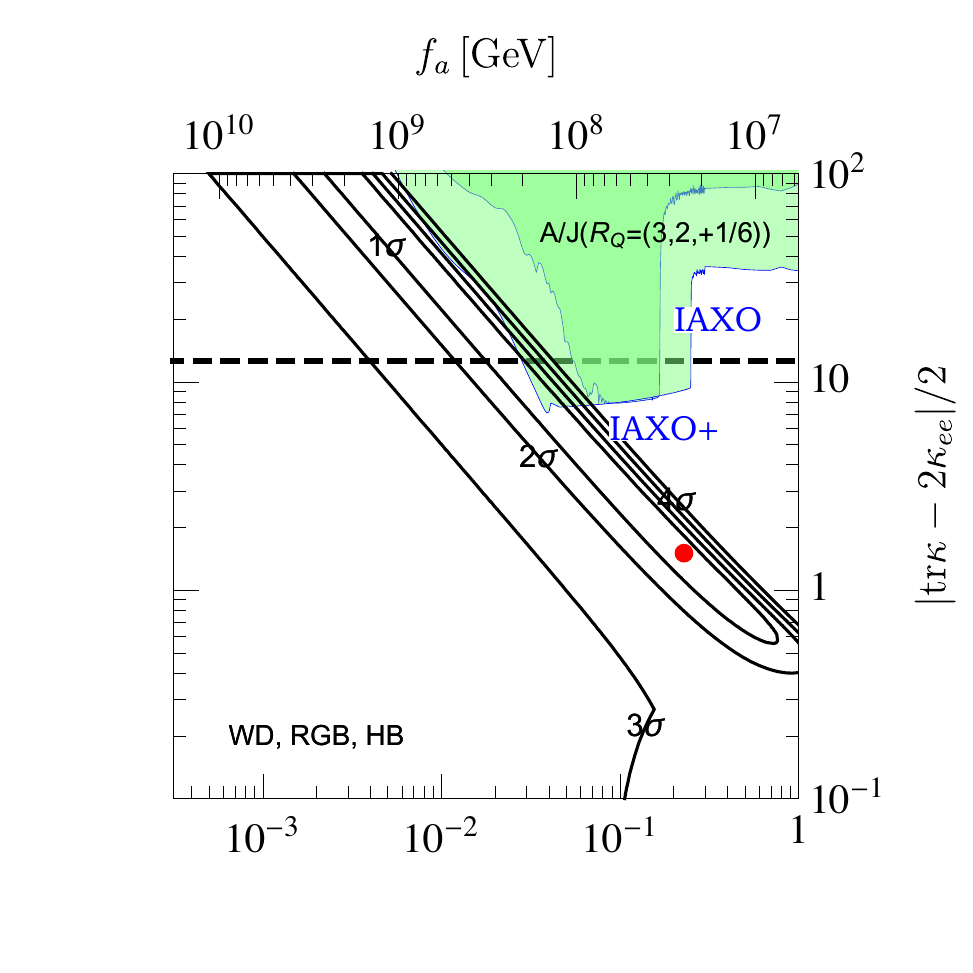}
\caption{Isocontours of the $\chi^2$ fit to the WD, RGB and HB anomalies in different axion models from~\cite{Giannotti:2017hny}. In DFSZ $\tan\beta$ is the ratio of the 2 Higgs doublet VEVs, which is constrained by unitarity in the Yukawa couplings to be within the dashed lines.
The parameter entering in the axion-electron coupling in the A/J models is the trace over RH Yukawa shown as  $|{\rm tr}\kappa-2\kappa_{ee}|$ and it also has an upper bound from unitarity.  The green regions are the sensitivities of IAXO and ARIADNE. See~\cite{Giannotti:2017hny} for details.
The lower axis gives the axion mass in units of eV.}
\label{fig:QCDaxionhintsSNless}
	\end{center}
\end{figure}

The situation is much better defined for QCD axion models, where we know that only fine-tuning can provide a small coupling to photons. Indeed, it is more generic to have a photon coupling than a sizeable electron coupling. Moreover, axions couple generically to protons and neutrons and this implies a strong constraint from the measured duration of the neutrino pulse of SN1987A (see, e.g., \cite{Raffelt:2006cw,Fischer:2016cyd,Chang:2018rso}).
Reference~\cite{Giannotti:2017hny} studied three classes of axion models, KSVZ, DFSZ and A/J, which differ on the origin of the axion coupling to electrons.
In KSVZ, this arises from a loop involving the photon coupling and turns out to be very small.
The interplay of the direct constraints on the photon coupling and the SN1987A bound does not allow a
KSVZ solution to the cooling anomalies~\cite{Giannotti:2017hny}.
The situation is very different in DFSZ I and DFSZ II models, which are two Higgs doublet models extended with an additional SM singlet scalar that allows a PQ symmetry spontaneously broken at high $f_a$ (see for instance~\cite{Dias:2014osa}).
DFSZ II has a larger $C_{a\gamma} = E/{\cal N}-1.92=2/3-1.92$ and will be thus easier to discover with IAXO than DFSZ I, for which ${\cal E}/{\cal N}=8/3$, implying a larger cancellation between the model dependent and model independent contributions to the photon coupling.
The regions where these models can account for the WB/RGB/HB hints
are depicted in Fig.~\ref{fig:hints} (right) and Fig.~\ref{fig:QCDaxionhintsSNless} (top panels).
In Fig.~\ref{fig:hints}, we are assuming production solely through the axion-electron coupling.
This condition is relaxed in Fig.~\ref{fig:QCDaxionhintsSNless}.

In Axi-Majoron models, the electron coupling arises through a loop involving right-handed neutrinos and can be large if the involved Yukawa couplings are large. The three different Axi-Majoron models considered consists in the addition of a new heavy quark and a SM scalar singlet and can be considered as variations of KSVZ where the PQ scalar field gives mass to RH neutrinos and breaks lepton number spontaneously. They differ in the SM charges of the heavy quark that give different couplings to photons.
The hinted regions for these models are shown in the right panel of Fig.~\ref{fig:hints} and in Fig.~\ref{fig:QCDaxionhintsSNless} (bottom panels).
Just like in the case of DFSZ axions, in Fig.~\ref{fig:hints} we are assuming production solely through the axion-electron coupling while we relaxed this conditions in Fig.~\ref{fig:QCDaxionhintsSNless}.

As shown in~\cite{Giannotti:2017hny}, DFSZ and Axi-Mjoron models can well explain the stellar hints even when the constraints from SN 1987A are accounted for.
However, some tension does exist in these models between the hinted values for the axion coupling with electrons and photons (from WD and globular cluster stars), and the bound on the axion-nucleus coupling extracted from the observed neutrino signal of SN 1987A.
Recent \textit{astrophobic} models~\cite{DiLuzio:2017ogq} relax further this tension and promise even better fits.
An quantitative analysis of these models in the contest of the cooling anomalies is in preparation.

In general, IAXO will be capable to find the hinted QCD axions in a sizable part of the parameter space, although it is with the upgraded IAXO configuration that most of the hinted parameter space will be covered.
The A/J model $R_Q=(3,2,+1/6)$ is an exception but it is already quite a tuned solution very close to the unitarity constraint~\cite{Giannotti:2017hny}.
In the region accessible by IAXO the solar axion emission happens mostly through the axion-electron coupling.
This is particularly true for the DFSZ I model.
In this case, the production rate induced by the axion-photon coupling would dominate for low values of tan$ \beta $, in a region almost entirely excluded by the unitarity constraints.

The regions hinted by the cooling anomalies, shown in Figs.~\ref{fig:hints} and \ref{fig:hinted_region}, are phenomenologically quite interesting.
They are largely accessible to IAXO and partially to ALPS II~\cite{Bahre:2013ywa}.
At high mass there is room for a QCD axion solution, in a range of parameters still partially accessible to IAXO, while at lower masses they point to ALPs which could provide the required CDM~\cite{Arias:2012az}. 
%
Finally, at slightly lower masses there is an overlap with the parameters required to explain the transparency hints~\cite{Meyer:2013pny}.
In all cases, IAXO shows a high potential to explore the relevant parameter space.

%% file: pics/IAXO_star_anomaly_gae_vs_gag.tex
\begin{tikzpicture}
\pgfdeclareplotmark{cross} {
\pgfpathmoveto{\pgfpoint{-0.3\pgfplotmarksize}{\pgfplotmarksize}}
\pgfpathlineto{\pgfpoint{+0.3\pgfplotmarksize}{\pgfplotmarksize}}
\pgfpathlineto{\pgfpoint{+0.3\pgfplotmarksize}{0.3\pgfplotmarksize}}
\pgfpathlineto{\pgfpoint{+1\pgfplotmarksize}{0.3\pgfplotmarksize}}
\pgfpathlineto{\pgfpoint{+1\pgfplotmarksize}{-0.3\pgfplotmarksize}}
\pgfpathlineto{\pgfpoint{+0.3\pgfplotmarksize}{-0.3\pgfplotmarksize}}
\pgfpathlineto{\pgfpoint{+0.3\pgfplotmarksize}{-1.\pgfplotmarksize}}
\pgfpathlineto{\pgfpoint{-0.3\pgfplotmarksize}{-1.\pgfplotmarksize}}
\pgfpathlineto{\pgfpoint{-0.3\pgfplotmarksize}{-0.3\pgfplotmarksize}}
\pgfpathlineto{\pgfpoint{-1.\pgfplotmarksize}{-0.3\pgfplotmarksize}}
\pgfpathlineto{\pgfpoint{-1.\pgfplotmarksize}{0.3\pgfplotmarksize}}
\pgfpathlineto{\pgfpoint{-0.3\pgfplotmarksize}{0.3\pgfplotmarksize}}
\pgfpathclose
\pgfusepathqstroke
}
\pgfdeclareplotmark{cross*} {
\pgfpathmoveto{\pgfpoint{-0.3\pgfplotmarksize}{\pgfplotmarksize}}
\pgfpathlineto{\pgfpoint{+0.3\pgfplotmarksize}{\pgfplotmarksize}}
\pgfpathlineto{\pgfpoint{+0.3\pgfplotmarksize}{0.3\pgfplotmarksize}}
\pgfpathlineto{\pgfpoint{+1\pgfplotmarksize}{0.3\pgfplotmarksize}}
\pgfpathlineto{\pgfpoint{+1\pgfplotmarksize}{-0.3\pgfplotmarksize}}
\pgfpathlineto{\pgfpoint{+0.3\pgfplotmarksize}{-0.3\pgfplotmarksize}}
\pgfpathlineto{\pgfpoint{+0.3\pgfplotmarksize}{-1.\pgfplotmarksize}}
\pgfpathlineto{\pgfpoint{-0.3\pgfplotmarksize}{-1.\pgfplotmarksize}}
\pgfpathlineto{\pgfpoint{-0.3\pgfplotmarksize}{-0.3\pgfplotmarksize}}
\pgfpathlineto{\pgfpoint{-1.\pgfplotmarksize}{-0.3\pgfplotmarksize}}
\pgfpathlineto{\pgfpoint{-1.\pgfplotmarksize}{0.3\pgfplotmarksize}}
\pgfpathlineto{\pgfpoint{-0.3\pgfplotmarksize}{0.3\pgfplotmarksize}}
\pgfpathclose
\pgfusepathqfillstroke
}
\pgfdeclareplotmark{newstar} {
\pgfpathmoveto{\pgfqpoint{0pt}{\pgfplotmarksize}}
\pgfpathlineto{\pgfqpointpolar{44}{0.5\pgfplotmarksize}}
\pgfpathlineto{\pgfqpointpolar{18}{\pgfplotmarksize}}
\pgfpathlineto{\pgfqpointpolar{-20}{0.5\pgfplotmarksize}}
\pgfpathlineto{\pgfqpointpolar{-54}{\pgfplotmarksize}}
\pgfpathlineto{\pgfqpointpolar{-90}{0.5\pgfplotmarksize}}
\pgfpathlineto{\pgfqpointpolar{234}{\pgfplotmarksize}}
\pgfpathlineto{\pgfqpointpolar{198}{0.5\pgfplotmarksize}}
\pgfpathlineto{\pgfqpointpolar{162}{\pgfplotmarksize}}
\pgfpathlineto{\pgfqpointpolar{134}{0.5\pgfplotmarksize}}
\pgfpathclose
\pgfusepathqstroke
}
\pgfdeclareplotmark{newstar*} {
\pgfpathmoveto{\pgfqpoint{0pt}{\pgfplotmarksize}}
\pgfpathlineto{\pgfqpointpolar{44}{0.5\pgfplotmarksize}}
\pgfpathlineto{\pgfqpointpolar{18}{\pgfplotmarksize}}
\pgfpathlineto{\pgfqpointpolar{-20}{0.5\pgfplotmarksize}}
\pgfpathlineto{\pgfqpointpolar{-54}{\pgfplotmarksize}}
\pgfpathlineto{\pgfqpointpolar{-90}{0.5\pgfplotmarksize}}
\pgfpathlineto{\pgfqpointpolar{234}{\pgfplotmarksize}}
\pgfpathlineto{\pgfqpointpolar{198}{0.5\pgfplotmarksize}}
\pgfpathlineto{\pgfqpointpolar{162}{\pgfplotmarksize}}
\pgfpathlineto{\pgfqpointpolar{134}{0.5\pgfplotmarksize}}
\pgfpathclose
\pgfusepathqfillstroke
}
\definecolor{c}{rgb}{1,1,1};
\draw [color=c, fill=c] (0,0) rectangle (20,19.3123);
\draw [color=c, fill=c] (2.8,2.70372) rectangle (19,18.3467);
\definecolor{c}{rgb}{0,0,0};
\draw [c,line width=0.9] (2.8,2.70372) -- (2.8,18.3467) -- (19,18.3467) -- (19,2.70372) -- (2.8,2.70372);
\definecolor{c}{rgb}{1,1,1};
\draw [color=c, fill=c] (2.8,2.70372) rectangle (19,18.3467);
\definecolor{c}{rgb}{0,0,0};
\draw [c,line width=0.9] (2.8,2.70372) -- (2.8,18.3467) -- (19,18.3467) -- (19,2.70372) -- (2.8,2.70372);
\draw [c,line width=0.9] (2.8,2.70372) -- (19,2.70372);
\draw [anchor= east] (19,1.29779) node[scale=1.52731, color=c, rotate=0]{$g_{ae}$};
\draw [c,line width=0.9] (2.8,3.17301) -- (2.8,2.70372);
\draw [anchor=base] (2.8,1.82984) node[scale=1.52731, color=c, rotate=0]{$10^{-15}$};
\draw [c,line width=0.9] (3.77534,2.93837) -- (3.77534,2.70372);
\draw [c,line width=0.9] (4.34588,2.93837) -- (4.34588,2.70372);
\draw [c,line width=0.9] (4.75068,2.93837) -- (4.75068,2.70372);
\draw [c,line width=0.9] (5.06467,2.93837) -- (5.06467,2.70372);
\draw [c,line width=0.9] (5.32121,2.93837) -- (5.32121,2.70372);
\draw [c,line width=0.9] (5.53812,2.93837) -- (5.53812,2.70372);
\draw [c,line width=0.9] (5.72601,2.93837) -- (5.72601,2.70372);
\draw [c,line width=0.9] (5.89175,2.93837) -- (5.89175,2.70372);
\draw [c,line width=0.9] (6.04,3.17301) -- (6.04,2.70372);
\draw [anchor=base] (6.04,1.82984) node[scale=1.52731, color=c, rotate=0]{$10^{-14}$};
\draw [c,line width=0.9] (7.01534,2.93837) -- (7.01534,2.70372);
\draw [c,line width=0.9] (7.58588,2.93837) -- (7.58588,2.70372);
\draw [c,line width=0.9] (7.99068,2.93837) -- (7.99068,2.70372);
\draw [c,line width=0.9] (8.30466,2.93837) -- (8.30466,2.70372);
\draw [c,line width=0.9] (8.56121,2.93837) -- (8.56121,2.70372);
\draw [c,line width=0.9] (8.77812,2.93837) -- (8.77812,2.70372);
\draw [c,line width=0.9] (8.96601,2.93837) -- (8.96601,2.70372);
\draw [c,line width=0.9] (9.13175,2.93837) -- (9.13175,2.70372);
\draw [c,line width=0.9] (9.28,3.17301) -- (9.28,2.70372);
\draw [anchor=base] (9.28,1.82984) node[scale=1.52731, color=c, rotate=0]{$10^{-13}$};
\draw [c,line width=0.9] (10.2553,2.93837) -- (10.2553,2.70372);
\draw [c,line width=0.9] (10.8259,2.93837) -- (10.8259,2.70372);
\draw [c,line width=0.9] (11.2307,2.93837) -- (11.2307,2.70372);
\draw [c,line width=0.9] (11.5447,2.93837) -- (11.5447,2.70372);
\draw [c,line width=0.9] (11.8012,2.93837) -- (11.8012,2.70372);
\draw [c,line width=0.9] (12.0181,2.93837) -- (12.0181,2.70372);
\draw [c,line width=0.9] (12.206,2.93837) -- (12.206,2.70372);
\draw [c,line width=0.9] (12.3717,2.93837) -- (12.3717,2.70372);
\draw [c,line width=0.9] (12.52,3.17301) -- (12.52,2.70372);
\draw [anchor=base] (12.52,1.82984) node[scale=1.52731, color=c, rotate=0]{$10^{-12}$};
\draw [c,line width=0.9] (13.4953,2.93837) -- (13.4953,2.70372);
\draw [c,line width=0.9] (14.0659,2.93837) -- (14.0659,2.70372);
\draw [c,line width=0.9] (14.4707,2.93837) -- (14.4707,2.70372);
\draw [c,line width=0.9] (14.7847,2.93837) -- (14.7847,2.70372);
\draw [c,line width=0.9] (15.0412,2.93837) -- (15.0412,2.70372);
\draw [c,line width=0.9] (15.2581,2.93837) -- (15.2581,2.70372);
\draw [c,line width=0.9] (15.446,2.93837) -- (15.446,2.70372);
\draw [c,line width=0.9] (15.6117,2.93837) -- (15.6117,2.70372);
\draw [c,line width=0.9] (15.76,3.17301) -- (15.76,2.70372);
\draw [anchor=base] (15.76,1.82984) node[scale=1.52731, color=c, rotate=0]{$10^{-11}$};
\draw [c,line width=0.9] (16.7353,2.93837) -- (16.7353,2.70372);
\draw [c,line width=0.9] (17.3059,2.93837) -- (17.3059,2.70372);
\draw [c,line width=0.9] (17.7107,2.93837) -- (17.7107,2.70372);
\draw [c,line width=0.9] (18.0247,2.93837) -- (18.0247,2.70372);
\draw [c,line width=0.9] (18.2812,2.93837) -- (18.2812,2.70372);
\draw [c,line width=0.9] (18.4981,2.93837) -- (18.4981,2.70372);
\draw [c,line width=0.9] (18.686,2.93837) -- (18.686,2.70372);
\draw [c,line width=0.9] (18.8517,2.93837) -- (18.8517,2.70372);
\draw [c,line width=0.9] (19,3.17301) -- (19,2.70372);
\draw [anchor=base] (19,1.82984) node[scale=1.52731, color=c, rotate=0]{$10^{-10}$};
\draw [c,line width=0.9] (2.8,18.3467) -- (19,18.3467);
\draw [c,line width=0.9] (2.8,17.8774) -- (2.8,18.3467);
\draw [c,line width=0.9] (3.77534,18.1121) -- (3.77534,18.3467);
\draw [c,line width=0.9] (4.34588,18.1121) -- (4.34588,18.3467);
\draw [c,line width=0.9] (4.75068,18.1121) -- (4.75068,18.3467);
\draw [c,line width=0.9] (5.06467,18.1121) -- (5.06467,18.3467);
\draw [c,line width=0.9] (5.32121,18.1121) -- (5.32121,18.3467);
\draw [c,line width=0.9] (5.53812,18.1121) -- (5.53812,18.3467);
\draw [c,line width=0.9] (5.72601,18.1121) -- (5.72601,18.3467);
\draw [c,line width=0.9] (5.89175,18.1121) -- (5.89175,18.3467);
\draw [c,line width=0.9] (6.04,17.8774) -- (6.04,18.3467);
\draw [c,line width=0.9] (7.01534,18.1121) -- (7.01534,18.3467);
\draw [c,line width=0.9] (7.58588,18.1121) -- (7.58588,18.3467);
\draw [c,line width=0.9] (7.99068,18.1121) -- (7.99068,18.3467);
\draw [c,line width=0.9] (8.30466,18.1121) -- (8.30466,18.3467);
\draw [c,line width=0.9] (8.56121,18.1121) -- (8.56121,18.3467);
\draw [c,line width=0.9] (8.77812,18.1121) -- (8.77812,18.3467);
\draw [c,line width=0.9] (8.96601,18.1121) -- (8.96601,18.3467);
\draw [c,line width=0.9] (9.13175,18.1121) -- (9.13175,18.3467);
\draw [c,line width=0.9] (9.28,17.8774) -- (9.28,18.3467);
\draw [c,line width=0.9] (10.2553,18.1121) -- (10.2553,18.3467);
\draw [c,line width=0.9] (10.8259,18.1121) -- (10.8259,18.3467);
\draw [c,line width=0.9] (11.2307,18.1121) -- (11.2307,18.3467);
\draw [c,line width=0.9] (11.5447,18.1121) -- (11.5447,18.3467);
\draw [c,line width=0.9] (11.8012,18.1121) -- (11.8012,18.3467);
\draw [c,line width=0.9] (12.0181,18.1121) -- (12.0181,18.3467);
\draw [c,line width=0.9] (12.206,18.1121) -- (12.206,18.3467);
\draw [c,line width=0.9] (12.3717,18.1121) -- (12.3717,18.3467);
\draw [c,line width=0.9] (12.52,17.8774) -- (12.52,18.3467);
\draw [c,line width=0.9] (13.4953,18.1121) -- (13.4953,18.3467);
\draw [c,line width=0.9] (14.0659,18.1121) -- (14.0659,18.3467);
\draw [c,line width=0.9] (14.4707,18.1121) -- (14.4707,18.3467);
\draw [c,line width=0.9] (14.7847,18.1121) -- (14.7847,18.3467);
\draw [c,line width=0.9] (15.0412,18.1121) -- (15.0412,18.3467);
\draw [c,line width=0.9] (15.2581,18.1121) -- (15.2581,18.3467);
\draw [c,line width=0.9] (15.446,18.1121) -- (15.446,18.3467);
\draw [c,line width=0.9] (15.6117,18.1121) -- (15.6117,18.3467);
\draw [c,line width=0.9] (15.76,17.8774) -- (15.76,18.3467);
\draw [c,line width=0.9] (16.7353,18.1121) -- (16.7353,18.3467);
\draw [c,line width=0.9] (17.3059,18.1121) -- (17.3059,18.3467);
\draw [c,line width=0.9] (17.7107,18.1121) -- (17.7107,18.3467);
\draw [c,line width=0.9] (18.0247,18.1121) -- (18.0247,18.3467);
\draw [c,line width=0.9] (18.2812,18.1121) -- (18.2812,18.3467);
\draw [c,line width=0.9] (18.4981,18.1121) -- (18.4981,18.3467);
\draw [c,line width=0.9] (18.686,18.1121) -- (18.686,18.3467);
\draw [c,line width=0.9] (18.8517,18.1121) -- (18.8517,18.3467);
\draw [c,line width=0.9] (19,17.8774) -- (19,18.3467);
\draw [c,line width=0.9] (2.8,2.70372) -- (2.8,18.3467);
\draw [anchor= east] (0.784,18.3467) node[scale=1.52731, color=c, rotate=90]{$|g_{a\gamma}| (GeV^{-1})$};
\draw [c,line width=0.9] (3.286,2.70373) -- (2.8,2.70373);
\draw [anchor= east] (2.644,2.70373) node[scale=1.52731, color=c, rotate=0]{$10^{-13}$};
\draw [c,line width=0.9] (3.043,3.88098) -- (2.8,3.88098);
\draw [c,line width=0.9] (3.043,4.56963) -- (2.8,4.56963);
\draw [c,line width=0.9] (3.043,5.05823) -- (2.8,5.05823);
\draw [c,line width=0.9] (3.043,5.43722) -- (2.8,5.43722);
\draw [c,line width=0.9] (3.043,5.74688) -- (2.8,5.74688);
\draw [c,line width=0.9] (3.043,6.00869) -- (2.8,6.00869);
\draw [c,line width=0.9] (3.043,6.23548) -- (2.8,6.23548);
\draw [c,line width=0.9] (3.043,6.43553) -- (2.8,6.43553);
\draw [c,line width=0.9] (3.286,6.61447) -- (2.8,6.61447);
\draw [anchor= east] (2.644,6.61447) node[scale=1.52731, color=c, rotate=0]{$10^{-12}$};
\draw [c,line width=0.9] (3.043,7.79172) -- (2.8,7.79172);
\draw [c,line width=0.9] (3.043,8.48037) -- (2.8,8.48037);
\draw [c,line width=0.9] (3.043,8.96898) -- (2.8,8.96898);
\draw [c,line width=0.9] (3.043,9.34797) -- (2.8,9.34797);
\draw [c,line width=0.9] (3.043,9.65762) -- (2.8,9.65762);
\draw [c,line width=0.9] (3.043,9.91943) -- (2.8,9.91943);
\draw [c,line width=0.9] (3.043,10.1462) -- (2.8,10.1462);
\draw [c,line width=0.9] (3.043,10.3463) -- (2.8,10.3463);
\draw [c,line width=0.9] (3.286,10.5252) -- (2.8,10.5252);
\draw [anchor= east] (2.644,10.5252) node[scale=1.52731, color=c, rotate=0]{$10^{-11}$};
\draw [c,line width=0.9] (3.043,11.7025) -- (2.8,11.7025);
\draw [c,line width=0.9] (3.043,12.3911) -- (2.8,12.3911);
\draw [c,line width=0.9] (3.043,12.8797) -- (2.8,12.8797);
\draw [c,line width=0.9] (3.043,13.2587) -- (2.8,13.2587);
\draw [c,line width=0.9] (3.043,13.5684) -- (2.8,13.5684);
\draw [c,line width=0.9] (3.043,13.8302) -- (2.8,13.8302);
\draw [c,line width=0.9] (3.043,14.057) -- (2.8,14.057);
\draw [c,line width=0.9] (3.043,14.257) -- (2.8,14.257);
\draw [c,line width=0.9] (3.286,14.436) -- (2.8,14.436);
\draw [anchor= east] (2.644,14.436) node[scale=1.52731, color=c, rotate=0]{$10^{-10}$};
\draw [c,line width=0.9] (3.043,15.6132) -- (2.8,15.6132);
\draw [c,line width=0.9] (3.043,16.3019) -- (2.8,16.3019);
\draw [c,line width=0.9] (3.043,16.7905) -- (2.8,16.7905);
\draw [c,line width=0.9] (3.043,17.1695) -- (2.8,17.1695);
\draw [c,line width=0.9] (3.043,17.4791) -- (2.8,17.4791);
\draw [c,line width=0.9] (3.043,17.7409) -- (2.8,17.7409);
\draw [c,line width=0.9] (3.043,17.9677) -- (2.8,17.9677);
\draw [c,line width=0.9] (3.043,18.1678) -- (2.8,18.1678);
\draw [c,line width=0.9] (3.286,18.3467) -- (2.8,18.3467);
\draw [anchor= east] (2.644,18.3467) node[scale=1.52731, color=c, rotate=0]{$10^{-9}$};
\draw [c,line width=0.9] (19,2.70372) -- (19,18.3467);
\draw [c,line width=0.9] (18.514,2.70373) -- (19,2.70373);
\draw [c,line width=0.9] (18.757,3.88098) -- (19,3.88098);
\draw [c,line width=0.9] (18.757,4.56963) -- (19,4.56963);
\draw [c,line width=0.9] (18.757,5.05823) -- (19,5.05823);
\draw [c,line width=0.9] (18.757,5.43722) -- (19,5.43722);
\draw [c,line width=0.9] (18.757,5.74688) -- (19,5.74688);
\draw [c,line width=0.9] (18.757,6.00869) -- (19,6.00869);
\draw [c,line width=0.9] (18.757,6.23548) -- (19,6.23548);
\draw [c,line width=0.9] (18.757,6.43553) -- (19,6.43553);
\draw [c,line width=0.9] (18.514,6.61447) -- (19,6.61447);
\draw [c,line width=0.9] (18.757,7.79172) -- (19,7.79172);
\draw [c,line width=0.9] (18.757,8.48037) -- (19,8.48037);
\draw [c,line width=0.9] (18.757,8.96898) -- (19,8.96898);
\draw [c,line width=0.9] (18.757,9.34797) -- (19,9.34797);
\draw [c,line width=0.9] (18.757,9.65762) -- (19,9.65762);
\draw [c,line width=0.9] (18.757,9.91943) -- (19,9.91943);
\draw [c,line width=0.9] (18.757,10.1462) -- (19,10.1462);
\draw [c,line width=0.9] (18.757,10.3463) -- (19,10.3463);
\draw [c,line width=0.9] (18.514,10.5252) -- (19,10.5252);
\draw [c,line width=0.9] (18.757,11.7025) -- (19,11.7025);
\draw [c,line width=0.9] (18.757,12.3911) -- (19,12.3911);
\draw [c,line width=0.9] (18.757,12.8797) -- (19,12.8797);
\draw [c,line width=0.9] (18.757,13.2587) -- (19,13.2587);
\draw [c,line width=0.9] (18.757,13.5684) -- (19,13.5684);
\draw [c,line width=0.9] (18.757,13.8302) -- (19,13.8302);
\draw [c,line width=0.9] (18.757,14.057) -- (19,14.057);
\draw [c,line width=0.9] (18.757,14.257) -- (19,14.257);
\draw [c,line width=0.9] (18.514,14.436) -- (19,14.436);
\draw [c,line width=0.9] (18.757,15.6132) -- (19,15.6132);
\draw [c,line width=0.9] (18.757,16.3019) -- (19,16.3019);
\draw [c,line width=0.9] (18.757,16.7905) -- (19,16.7905);
\draw [c,line width=0.9] (18.757,17.1695) -- (19,17.1695);
\draw [c,line width=0.9] (18.757,17.4791) -- (19,17.4791);
\draw [c,line width=0.9] (18.757,17.7409) -- (19,17.7409);
\draw [c,line width=0.9] (18.757,17.9677) -- (19,17.9677);
\draw [c,line width=0.9] (18.757,18.1678) -- (19,18.1678);
\draw [c,line width=0.9] (18.514,18.3467) -- (19,18.3467);
\definecolor{c}{rgb}{1,0.8,0};
\draw [c, fill=c] (10.155,2.70372) -- (10.1573,6.52302) -- (10.1576,6.7316) -- (10.1579,6.93513) -- (10.159,7.42025) -- (10.1598,7.76502) -- (10.1602,7.90893) -- (10.1605,8.05147) -- (10.1618,8.48071) -- (10.1621,8.59756) -- (10.1624,8.71281) --
 (10.1633,8.98701) -- (10.1635,9.08615) -- (10.1637,9.18307) -- (10.1642,9.37993) -- (10.1635,9.46513) -- (10.1644,9.54755) -- (10.1646,9.62703) -- (10.1647,9.70152) -- (10.1647,9.77481) -- (10.1647,9.84489) -- (10.1647,9.97411) -- (10.1646,10.0366)
 -- (10.1644,10.0957) -- (10.1642,10.2108) -- (10.164,10.2634) -- (10.1637,10.3126) -- (10.1632,10.4203) -- (10.1622,10.4635) -- (10.1626,10.5034) -- (10.1618,10.6081) -- (10.1608,10.6423) -- (10.1605,10.6717) -- (10.1591,10.8043) --
 (10.1593,10.8276) -- (10.1577,10.9345) -- (10.1574,10.952) -- (10.1571,10.9677) -- (10.155,11.0785) -- (10.1545,11.088) -- (10.1546,11.0964) -- (10.1533,11.1521) -- (10.1519,11.2121) -- (10.1518,11.2139) -- (10.1518,11.2154) -- (10.1511,11.2419) --
 (10.1501,11.2735) -- (10.1483,11.3254) -- (10.1482,11.331) -- (10.148,11.3373) -- (10.1466,11.3777) -- (10.1461,11.3867) -- (10.1459,11.397) -- (10.1447,11.4282) -- (10.144,11.4407) -- (10.1437,11.455) -- (10.1428,11.4771) -- (10.1419,11.493) --
 (10.1414,11.5111) -- (10.1408,11.5247) -- (10.1396,11.5436) -- (10.1385,11.5665) -- (10.1372,11.5929) -- (10.1361,11.6145) -- (10.1348,11.6407) -- (10.1332,11.6714) -- (10.1323,11.6872) -- (10.1315,11.7002) -- (10.1297,11.7325) -- (10.1293,11.7415)
 -- (10.1275,11.7707) -- (10.127,11.7767) -- (10.1268,11.7814) -- (10.1243,11.8187) -- (10.1242,11.8197) -- (10.1237,11.8271) -- (10.1214,11.8581) -- (10.1211,11.8616) -- (10.1208,11.866) -- (10.1186,11.8949) -- (10.1178,11.9025) -- (10.1154,11.9305)
 -- (10.1144,11.9425) -- (10.1134,11.9571) -- (10.1127,11.966) -- (10.1109,11.9815) -- (10.1096,12.0004) -- (10.1096,12.0011) -- (10.1074,12.0197) -- (10.1064,12.0337) -- (10.1053,12.0446) -- (10.1037,12.057) -- (10.103,12.0664) -- (10.1009,12.0871)
 -- (10.1,12.0936) -- (10.0997,12.0984) -- (10.0963,12.1287) -- (10.0963,12.1293) -- (10.0962,12.1297) -- (10.0959,12.1329) -- (10.0925,12.1602) -- (10.0919,12.1643) -- (10.0914,12.1699) -- (10.0888,12.1901) -- (10.0877,12.1987) -- (10.0862,12.2102)
 -- (10.085,12.2194) -- (10.0829,12.2322) -- (10.0811,12.2481) -- (10.0809,12.2498) -- (10.0782,12.2652) -- (10.0747,12.2895) -- (10.0726,12.3033) -- (10.0687,12.3294) -- (10.0686,12.3305) -- (10.0674,12.3375) -- (10.0639,12.3567) --
 (10.0633,12.3605) -- (10.0623,12.366) -- (10.0576,12.3911) -- (10.0544,12.4076) -- (10.0518,12.4212) -- (10.0479,12.4412) -- (10.0459,12.4507) -- (10.0449,12.4571) -- (10.0405,12.4777) -- (10.04,12.4797) -- (10.0384,12.4872) -- (10.0344,12.5044) --
 (10.0333,12.5083) -- (10.032,12.5145) -- (10.0288,12.5273) -- (10.0261,12.5364) -- (10.0233,12.5498) -- (10.023,12.5509) -- (10.019,12.564) -- (10.0127,12.5876) -- (10.0111,12.5933) -- (10.0088,12.6018) -- (10.0036,12.6179) -- (9.99791,12.6347) --
 (9.99481,12.6442) -- (9.98908,12.6613) -- (9.98598,12.6702) -- (9.98449,12.6752) -- (9.97855,12.6915) -- (9.97675,12.6957) -- (9.96915,12.7132) -- (9.96591,12.7209) -- (9.95906,12.7366) -- (9.95304,12.7499) -- (9.94761,12.7621) -- (9.94316,12.7701)
 -- (9.93491,12.7845) -- (9.92951,12.7942) -- (9.916,12.8176) -- (9.91583,12.8178) -- (9.91573,12.8179) -- (9.91549,12.8183) -- (9.90492,12.8333) -- (9.89813,12.8414) -- (9.89345,12.8483) -- (9.88364,12.8603) -- (9.8794,12.8645) -- (9.87195,12.8727)
 -- (9.86814,12.8766) -- (9.85568,12.8873) -- (9.85054,12.8922) -- (9.83702,12.9008) -- (9.82341,12.9098) -- (9.81663,12.9141) -- (9.79883,12.9206) -- (9.78188,12.927) -- (9.74765,12.9311) -- (9.74477,12.9311) -- (9.7097,12.9267) -- (9.69147,12.9207)
 -- (9.67223,12.9098) -- (9.6494,12.9007) -- (9.63362,12.891) -- (9.61444,12.8756) -- (9.60456,12.8645) -- (9.58372,12.8473) -- (9.57836,12.8414) -- (9.56212,12.823) -- (9.55754,12.8179) -- (9.5561,12.8163) -- (9.5532,12.8129) -- (9.53967,12.7942) --
 (9.53152,12.7826) -- (9.52477,12.7701) -- (9.50838,12.7473) -- (9.50577,12.7425) -- (9.49374,12.7209) -- (9.48763,12.7095) -- (9.48052,12.6957) -- (9.4679,12.6703) -- (9.46787,12.6702) -- (9.46786,12.6702) -- (9.46785,12.6702) -- (9.45024,12.6284)
 -- (9.44611,12.6179) -- (9.43973,12.601) -- (9.43363,12.585) -- (9.42622,12.564) -- (9.42322,12.555) -- (9.4183,12.5394) -- (9.41618,12.5321) -- (9.40922,12.5083) -- (9.40452,12.4915) -- (9.39738,12.4634) -- (9.39465,12.4507) -- (9.39182,12.4414) --
 (9.38128,12.3938) -- (9.38066,12.3911) -- (9.38023,12.3891) -- (9.37708,12.3727) -- (9.37018,12.3339) -- (9.36948,12.3294) -- (9.36874,12.3238) -- (9.36005,12.2652) -- (9.35382,12.2152) -- (9.35212,12.1987) -- (9.35056,12.1795) -- (9.34565,12.1293)
 -- (9.34215,12.084) -- (9.34063,12.057) -- (9.3395,12.0276) -- (9.33701,11.9815) -- (9.33546,11.9375) -- (9.33481,11.9025) -- (9.33473,11.8662) -- (9.33394,11.8197) -- (9.3339,11.7726) -- (9.33442,11.7325) -- (9.3355,11.693) -- (9.33617,11.6407) --
 (9.33766,11.5845) -- (9.3392,11.5436) -- (9.3411,11.5054) -- (9.34343,11.4407) -- (9.34687,11.3668) -- (9.34884,11.331) -- (9.3509,11.2994) -- (9.35538,11.2139) -- (9.36162,11.1093) -- (9.36301,11.088) -- (9.36434,11.0703) -- (9.3717,10.952) --
 (9.37708,10.8721) -- (9.38119,10.8111) -- (9.38162,10.8043) -- (9.38218,10.7958) -- (9.39116,10.6678) -- (9.39314,10.6423) -- (9.39504,10.6094) -- (9.40175,10.5128) -- (9.40553,10.4635) -- (9.40935,10.3965) -- (9.4129,10.3436) -- (9.4188,10.2634) --
 (9.42465,10.1573) -- (9.42516,10.1489) -- (9.43292,10.0366) -- (9.43743,9.95135) -- (9.44306,9.85227) -- (9.44785,9.77481) -- (9.45105,9.72004) -- (9.46352,9.46513) -- (9.4679,9.37315) -- (9.47862,9.13736) -- (9.48071,9.08615) -- (9.48396,9.00627)
 -- (9.49356,8.75658) -- (9.49969,8.59756) -- (9.50728,8.31876) -- (9.50872,8.26807) -- (9.51891,7.90893) -- (9.53395,7.08459) -- (9.53866,6.7316) -- (9.54166,6.47783) -- (9.5532,4.60658) -- (9.55587,2.70372);
\definecolor{c}{rgb}{1,1,0};
\draw [c,line width=0.9] (10.155,2.70372) -- (10.1573,6.52302);
\draw [c,line width=0.9] (10.1573,6.52302) -- (10.1576,6.7316) -- (10.1579,6.93513) -- (10.159,7.42025) -- (10.1598,7.76502) -- (10.1602,7.90893) -- (10.1605,8.05147) -- (10.1618,8.48071) -- (10.1621,8.59756) -- (10.1624,8.71281) -- (10.1633,8.98701)
 -- (10.1635,9.08615) -- (10.1637,9.18307) -- (10.1642,9.37993) -- (10.1635,9.46513) -- (10.1644,9.54755) -- (10.1646,9.62703) -- (10.1647,9.70152) -- (10.1647,9.77481) -- (10.1647,9.84489) -- (10.1647,9.97411) -- (10.1646,10.0366) --
 (10.1644,10.0957) -- (10.1642,10.2108) -- (10.164,10.2634) -- (10.1637,10.3126) -- (10.1632,10.4203) -- (10.1622,10.4635) -- (10.1626,10.5034) -- (10.1618,10.6081) -- (10.1608,10.6423) -- (10.1605,10.6717) -- (10.1591,10.8043) -- (10.1593,10.8276)
 -- (10.1577,10.9345) -- (10.1574,10.952) -- (10.1571,10.9677) -- (10.155,11.0785) -- (10.1545,11.088) -- (10.1546,11.0964) -- (10.1533,11.1521) -- (10.1519,11.2121) -- (10.1518,11.2139) -- (10.1518,11.2154) -- (10.1511,11.2419) -- (10.1501,11.2735)
 -- (10.1483,11.3254) -- (10.1482,11.331) -- (10.148,11.3373) -- (10.1466,11.3777) -- (10.1461,11.3867) -- (10.1459,11.397) -- (10.1447,11.4282) -- (10.144,11.4407) -- (10.1437,11.455) -- (10.1428,11.4771) -- (10.1419,11.493) -- (10.1414,11.5111) --
 (10.1408,11.5247) -- (10.1396,11.5436) -- (10.1385,11.5665) -- (10.1372,11.5929) -- (10.1361,11.6145) -- (10.1348,11.6407) -- (10.1332,11.6714) -- (10.1323,11.6872) -- (10.1315,11.7002) -- (10.1297,11.7325) -- (10.1293,11.7415) -- (10.1275,11.7707)
 -- (10.127,11.7767) -- (10.1268,11.7814) -- (10.1243,11.8187) -- (10.1242,11.8197) -- (10.1237,11.8271) -- (10.1214,11.8581) -- (10.1211,11.8616) -- (10.1208,11.866) -- (10.1186,11.8949) -- (10.1178,11.9025) -- (10.1154,11.9305) -- (10.1144,11.9425)
 -- (10.1134,11.9571) -- (10.1127,11.966) -- (10.1109,11.9815) -- (10.1096,12.0004) -- (10.1096,12.0011) -- (10.1074,12.0197) -- (10.1064,12.0337) -- (10.1053,12.0446) -- (10.1037,12.057) -- (10.103,12.0664) -- (10.1009,12.0871) -- (10.1,12.0936) --
 (10.0997,12.0984) -- (10.0963,12.1287) -- (10.0963,12.1293) -- (10.0962,12.1297) -- (10.0959,12.1329) -- (10.0925,12.1602) -- (10.0919,12.1643) -- (10.0914,12.1699) -- (10.0888,12.1901) -- (10.0877,12.1987) -- (10.0862,12.2102) -- (10.085,12.2194)
 -- (10.0829,12.2322) -- (10.0811,12.2481) -- (10.0809,12.2498) -- (10.0782,12.2652) -- (10.0747,12.2895) -- (10.0726,12.3033) -- (10.0687,12.3294) -- (10.0686,12.3305) -- (10.0674,12.3375) -- (10.0639,12.3567) -- (10.0633,12.3605) --
 (10.0623,12.366) -- (10.0576,12.3911) -- (10.0544,12.4076) -- (10.0518,12.4212) -- (10.0479,12.4412) -- (10.0459,12.4507) -- (10.0449,12.4571) -- (10.0405,12.4777) -- (10.04,12.4797) -- (10.0384,12.4872) -- (10.0344,12.5044) -- (10.0333,12.5083) --
 (10.032,12.5145) -- (10.0288,12.5273) -- (10.0261,12.5364) -- (10.0233,12.5498) -- (10.023,12.5509) -- (10.019,12.564) -- (10.0127,12.5876) -- (10.0111,12.5933) -- (10.0088,12.6018) -- (10.0036,12.6179) -- (9.99791,12.6347) -- (9.99481,12.6442) --
 (9.98908,12.6613) -- (9.98598,12.6702) -- (9.98449,12.6752) -- (9.97855,12.6915) -- (9.97675,12.6957) -- (9.96915,12.7132) -- (9.96591,12.7209) -- (9.95906,12.7366) -- (9.95304,12.7499) -- (9.94761,12.7621) -- (9.94316,12.7701) -- (9.93491,12.7845)
 -- (9.92951,12.7942) -- (9.916,12.8176) -- (9.91583,12.8178) -- (9.91573,12.8179) -- (9.91549,12.8183) -- (9.90492,12.8333) -- (9.89813,12.8414) -- (9.89345,12.8483) -- (9.88364,12.8603) -- (9.8794,12.8645) -- (9.87195,12.8727) -- (9.86814,12.8766)
 -- (9.85568,12.8873) -- (9.85054,12.8922) -- (9.83702,12.9008) -- (9.82341,12.9098) -- (9.81663,12.9141) -- (9.79883,12.9206) -- (9.78188,12.927) -- (9.74765,12.9311) -- (9.74477,12.9311) -- (9.7097,12.9267) -- (9.69147,12.9207) -- (9.67223,12.9098)
 -- (9.6494,12.9007) -- (9.63362,12.891) -- (9.61444,12.8756) -- (9.60456,12.8645) -- (9.58372,12.8473) -- (9.57836,12.8414) -- (9.56212,12.823) -- (9.55754,12.8179) -- (9.5561,12.8163) -- (9.5532,12.8129) -- (9.53967,12.7942) -- (9.53152,12.7826) --
 (9.52477,12.7701) -- (9.50838,12.7473) -- (9.50577,12.7425) -- (9.49374,12.7209) -- (9.48763,12.7095) -- (9.48052,12.6957) -- (9.4679,12.6703) -- (9.46787,12.6702) -- (9.46786,12.6702) -- (9.46785,12.6702) -- (9.45024,12.6284) -- (9.44611,12.6179)
 -- (9.43973,12.601) -- (9.43363,12.585) -- (9.42622,12.564) -- (9.42322,12.555) -- (9.4183,12.5394) -- (9.41618,12.5321) -- (9.40922,12.5083) -- (9.40452,12.4915) -- (9.39738,12.4634) -- (9.39465,12.4507) -- (9.39182,12.4414) -- (9.38128,12.3938) --
 (9.38066,12.3911) -- (9.38023,12.3891) -- (9.37708,12.3727) -- (9.37018,12.3339) -- (9.36948,12.3294) -- (9.36874,12.3238) -- (9.36005,12.2652) -- (9.35382,12.2152) -- (9.35212,12.1987) -- (9.35056,12.1795) -- (9.34565,12.1293) -- (9.34215,12.084)
 -- (9.34063,12.057) -- (9.3395,12.0276) -- (9.33701,11.9815) -- (9.33546,11.9375) -- (9.33481,11.9025) -- (9.33473,11.8662) -- (9.33394,11.8197) -- (9.3339,11.7726) -- (9.33442,11.7325) -- (9.3355,11.693) -- (9.33617,11.6407) -- (9.33766,11.5845) --
 (9.3392,11.5436) -- (9.3411,11.5054) -- (9.34343,11.4407) -- (9.34687,11.3668) -- (9.34884,11.331) -- (9.3509,11.2994) -- (9.35538,11.2139) -- (9.36162,11.1093) -- (9.36301,11.088) -- (9.36434,11.0703) -- (9.3717,10.952) -- (9.37708,10.8721) --
 (9.38119,10.8111) -- (9.38162,10.8043) -- (9.38218,10.7958) -- (9.39116,10.6678) -- (9.39314,10.6423) -- (9.39504,10.6094) -- (9.40175,10.5128) -- (9.40553,10.4635) -- (9.40935,10.3965) -- (9.4129,10.3436) -- (9.4188,10.2634) -- (9.42465,10.1573) --
 (9.42516,10.1489) -- (9.43292,10.0366) -- (9.43743,9.95135) -- (9.44306,9.85227) -- (9.44785,9.77481) -- (9.45105,9.72004) -- (9.46352,9.46513) -- (9.4679,9.37315) -- (9.47862,9.13736) -- (9.48071,9.08615) -- (9.48396,9.00627) -- (9.49356,8.75658)
 -- (9.49969,8.59756) -- (9.50728,8.31876) -- (9.50872,8.26807) -- (9.51891,7.90893) -- (9.53395,7.08459) -- (9.53866,6.7316) -- (9.54166,6.47783) -- (9.5532,4.60658) -- (9.55587,2.70372);
\definecolor{c}{rgb}{0.4,0.8,1};
\draw [c, fill=c, fill opacity=0.3] (19,2.70372) -- (19,18.3467) -- (2.8,18.3467) -- (2.8,11.2696) -- (3.05655,11.2696) -- (3.31309,11.2696) -- (3.57127,11.2696) -- (3.82374,11.2696) -- (4.08369,11.2696) -- (4.34117,11.2696) -- (4.59458,11.2696) --
 (4.85244,11.2696) -- (5.10898,11.2696) -- (5.36508,11.2696) -- (5.622,11.2696) -- (5.87918,11.2696) -- (6.1352,11.2696) -- (6.38736,11.2696) -- (6.64757,11.2696) -- (6.90564,11.2696) -- (7.16218,11.2586) -- (7.41662,11.2586) -- (7.67228,11.2586) --
 (7.92956,11.2475) -- (8.18734,11.2475) -- (8.44388,11.2364) -- (8.69958,11.2139) -- (8.95719,11.1911) -- (9.21374,11.1563) -- (9.46437,11.1088) -- (9.72297,11.0349) -- (9.98465,10.9445) -- (10.2412,10.8063) -- (10.4942,10.6401) -- (10.7537,10.4255)
 -- (11.0102,10.1736) -- (11.2654,9.89745) -- (11.522,9.60297) -- (11.7799,9.30322) -- (12.0361,8.99844) -- (12.293,8.68795) -- (12.5479,8.38129) -- (12.7998,8.07283) -- (13.0621,7.76605) -- (13.3155,7.45467) -- (13.5773,7.14915) -- (13.8317,6.83701)
 -- (14.0891,6.52377) -- (14.3457,6.21412) -- (14.6016,5.90618) -- (14.8587,5.596) -- (15.1143,5.28449) -- (15.3701,4.97557) -- (15.6273,4.66859) -- (15.8813,4.35893) -- (16.14,4.05056) -- (16.3947,3.73936) -- (16.6557,3.42601) -- (16.9073,3.123) --
 (17.168,2.80269) -- (17.2508,2.70372);
\definecolor{c}{rgb}{0,0,0};
\draw [c,dash pattern=on 4pt off 4pt,line width=1.8] (2.8,18.3467) -- (2.8,11.2696);
\draw [c,dash pattern=on 4pt off 4pt,line width=1.8] (2.8,11.2696) -- (3.05655,11.2696) -- (3.31309,11.2696) -- (3.57127,11.2696) -- (3.82374,11.2696) -- (4.08369,11.2696) -- (4.34117,11.2696) -- (4.59458,11.2696) -- (4.85244,11.2696) --
 (5.10898,11.2696) -- (5.36508,11.2696) -- (5.622,11.2696) -- (5.87918,11.2696) -- (6.1352,11.2696) -- (6.38736,11.2696) -- (6.64757,11.2696) -- (6.90564,11.2696) -- (7.16218,11.2586) -- (7.41662,11.2586) -- (7.67228,11.2586) -- (7.92956,11.2475) --
 (8.18734,11.2475) -- (8.44388,11.2364) -- (8.69958,11.2139) -- (8.95719,11.1911) -- (9.21374,11.1563) -- (9.46437,11.1088) -- (9.72297,11.0349) -- (9.98465,10.9445) -- (10.2412,10.8063) -- (10.4942,10.6401) -- (10.7537,10.4255) -- (11.0102,10.1736)
 -- (11.2654,9.89745) -- (11.522,9.60297) -- (11.7799,9.30322) -- (12.0361,8.99844) -- (12.293,8.68795) -- (12.5479,8.38129) -- (12.7998,8.07283) -- (13.0621,7.76605) -- (13.3155,7.45467) -- (13.5773,7.14915) -- (13.8317,6.83701) -- (14.0891,6.52377)
 -- (14.3457,6.21412) -- (14.6016,5.90618) -- (14.8587,5.596) -- (15.1143,5.28449) -- (15.3701,4.97557) -- (15.6273,4.66859) -- (15.8813,4.35893) -- (16.14,4.05056) -- (16.3947,3.73936) -- (16.6557,3.42601) -- (16.9073,3.123) -- (17.168,2.80269) --
 (17.2508,2.70372);
\definecolor{c}{rgb}{0.4,0.8,1};
\draw [c, fill=c, fill opacity=0.2] (19,2.70372) -- (19,18.3467) -- (2.8,18.3467) -- (2.8,9.11144) -- (3.05655,9.11144) -- (3.31309,9.11144) -- (3.57127,9.11144) -- (3.82374,9.11144) -- (4.08369,9.11144) -- (4.34117,9.11144) -- (4.59458,9.11144) --
 (4.85244,9.10753) -- (5.10898,9.10753) -- (5.36508,9.10753) -- (5.622,9.10361) -- (5.87918,9.09968) -- (6.1352,9.09575) -- (6.38736,9.08785) -- (6.64757,9.07991) -- (6.90564,9.06393) -- (7.16218,9.04373) -- (7.41662,9.01505) -- (7.67228,8.97321) --
 (7.92956,8.91286) -- (8.18734,8.82736) -- (8.44388,8.70787) -- (8.69958,8.55242) -- (8.95719,8.35102) -- (9.21374,8.11547) -- (9.46437,7.84192) -- (9.72297,7.5552) -- (9.98465,7.25721) -- (10.2412,6.9522) -- (10.4942,6.6481) -- (10.7537,6.33845) --
 (11.0102,6.03039) -- (11.2654,5.72121) -- (11.522,5.40982) -- (11.7799,5.10017) -- (12.0361,4.79216) -- (12.293,4.48251) -- (12.5479,4.16927) -- (12.7998,3.86391) -- (13.0621,3.55425) -- (13.3155,3.2384) -- (13.5773,2.9411) -- (13.7676,2.70372);
\definecolor{c}{rgb}{0,0,0};
\draw [c,dash pattern=on 4pt off 4pt,line width=1.8] (2.8,18.3467) -- (2.8,9.11144);
\draw [c,dash pattern=on 4pt off 4pt,line width=1.8] (2.8,9.11144) -- (3.05655,9.11144) -- (3.31309,9.11144) -- (3.57127,9.11144) -- (3.82374,9.11144) -- (4.08369,9.11144) -- (4.34117,9.11144) -- (4.59458,9.11144) -- (4.85244,9.10753) --
 (5.10898,9.10753) -- (5.36508,9.10753) -- (5.622,9.10361) -- (5.87918,9.09968) -- (6.1352,9.09575) -- (6.38736,9.08785) -- (6.64757,9.07991) -- (6.90564,9.06393) -- (7.16218,9.04373) -- (7.41662,9.01505) -- (7.67228,8.97321) -- (7.92956,8.91286) --
 (8.18734,8.82736) -- (8.44388,8.70787) -- (8.69958,8.55242) -- (8.95719,8.35102) -- (9.21374,8.11547) -- (9.46437,7.84192) -- (9.72297,7.5552) -- (9.98465,7.25721) -- (10.2412,6.9522) -- (10.4942,6.6481) -- (10.7537,6.33845) -- (11.0102,6.03039) --
 (11.2654,5.72121) -- (11.522,5.40982) -- (11.7799,5.10017) -- (12.0361,4.79216) -- (12.293,4.48251) -- (12.5479,4.16927) -- (12.7998,3.86391) -- (13.0621,3.55425) -- (13.3155,3.2384) -- (13.5773,2.9411) -- (13.7676,2.70372);
\definecolor{c}{rgb}{0.4,0.8,1};
\draw [c, fill=c, fill opacity=0.3] (19,2.70372) -- (19,18.3467) -- (2.8,18.3467) -- (2.8,8.26968) -- (3.05655,8.26968) -- (3.31309,8.26968) -- (3.57127,8.26968) -- (3.82374,8.26326) -- (4.08369,8.26326) -- (4.34117,8.26326) -- (4.59458,8.26326) --
 (4.85244,8.26326) -- (5.10898,8.25681) -- (5.36508,8.25034) -- (5.622,8.24384) -- (5.87918,8.23732) -- (6.1352,8.22421) -- (6.38736,8.20434) -- (6.64757,8.17071) -- (6.90564,8.12945) -- (7.16218,8.07283) -- (7.41662,7.9842) -- (7.67228,7.86648) --
 (7.92956,7.71352) -- (8.18734,7.5157) -- (8.44388,7.28032) -- (8.69958,7.00699) -- (8.95719,6.71343) -- (9.21374,6.42036) -- (9.46437,6.11678) -- (9.72297,5.81076) -- (9.98465,5.50056) -- (10.2412,5.19287) -- (10.4942,4.88399) -- (10.7537,4.57528)
 -- (11.0102,4.26675) -- (11.2654,3.95574) -- (11.522,3.64445) -- (11.7799,3.33479) -- (12.0361,3.02748) -- (12.293,2.72062) -- (12.3066,2.70372);
\definecolor{c}{rgb}{0,0,0};
\draw [c,dash pattern=on 4pt off 4pt,line width=1.8] (2.8,18.3467) -- (2.8,8.26968);
\draw [c,dash pattern=on 4pt off 4pt,line width=1.8] (2.8,8.26968) -- (3.05655,8.26968) -- (3.31309,8.26968) -- (3.57127,8.26968) -- (3.82374,8.26326) -- (4.08369,8.26326) -- (4.34117,8.26326) -- (4.59458,8.26326) -- (4.85244,8.26326) --
 (5.10898,8.25681) -- (5.36508,8.25034) -- (5.622,8.24384) -- (5.87918,8.23732) -- (6.1352,8.22421) -- (6.38736,8.20434) -- (6.64757,8.17071) -- (6.90564,8.12945) -- (7.16218,8.07283) -- (7.41662,7.9842) -- (7.67228,7.86648) -- (7.92956,7.71352) --
 (8.18734,7.5157) -- (8.44388,7.28032) -- (8.69958,7.00699) -- (8.95719,6.71343) -- (9.21374,6.42036) -- (9.46437,6.11678) -- (9.72297,5.81076) -- (9.98465,5.50056) -- (10.2412,5.19287) -- (10.4942,4.88399) -- (10.7537,4.57528) -- (11.0102,4.26675)
 -- (11.2654,3.95574) -- (11.522,3.64445) -- (11.7799,3.33479) -- (12.0361,3.02748) -- (12.293,2.72062) -- (12.3066,2.70372);
\definecolor{c}{rgb}{0.4,0.6,0.8};
\draw [c, fill=c] (19,18.3467) -- (2.8,18.3467) -- (2.8,13.7302) -- (3.05655,13.7302) -- (3.31309,13.7302) -- (3.57127,13.7302) -- (3.82374,13.7302) -- (4.08369,13.7302) -- (4.34117,13.7302) -- (4.59458,13.7302) -- (4.85244,13.7302) --
 (5.10898,13.7302) -- (5.36508,13.7302) -- (5.622,13.7302) -- (5.87918,13.7302) -- (6.1352,13.7302) -- (6.38736,13.7302) -- (6.64757,13.7302) -- (6.90564,13.7302) -- (7.16218,13.7302) -- (7.41662,13.7302) -- (7.67228,13.7302) -- (7.92956,13.7302) --
 (8.18734,13.7302) -- (8.44388,13.7302) -- (8.69958,13.7302) -- (8.95719,13.7277) -- (9.21374,13.7277) -- (9.46437,13.7277) -- (9.72297,13.7251) -- (9.98465,13.7225) -- (10.2412,13.7199) -- (10.4942,13.7147) -- (10.7537,13.7069) -- (11.0102,13.6965)
 -- (11.2654,13.6833) -- (11.522,13.662) -- (11.7799,13.6323) -- (12.0361,13.5881) -- (12.293,13.5283) -- (12.5479,13.4421) -- (12.7998,13.3221) -- (13.0621,13.1644) -- (13.3155,12.9626) -- (13.5773,12.7195) -- (13.8317,12.4523) -- (14.0891,12.1611)
 -- (14.3457,11.8643) -- (14.6016,11.5609) -- (14.8587,11.2475) -- (15.1143,10.9445) -- (15.3701,10.6401) -- (15.6273,10.3254) -- (15.8813,10.0161) -- (16.14,9.70507) -- (16.3947,9.39487) -- (16.6557,9.08785) -- (16.9073,8.77581) -- (17.168,8.46901)
 -- (17.4228,8.15707) -- (17.6787,7.85015) -- (17.9376,7.53556) -- (18.1916,7.23379) -- (18.449,6.92413) -- (18.7052,6.60937) -- (18.9629,6.30005) -- (19,6.25538);
\definecolor{c}{rgb}{0,0,1};
\draw [c,line width=0.9] (2.8,18.3467) -- (2.8,13.7302);
\draw [c,line width=0.9] (2.8,13.7302) -- (3.05655,13.7302) -- (3.31309,13.7302) -- (3.57127,13.7302) -- (3.82374,13.7302) -- (4.08369,13.7302) -- (4.34117,13.7302) -- (4.59458,13.7302) -- (4.85244,13.7302) -- (5.10898,13.7302) -- (5.36508,13.7302)
 -- (5.622,13.7302) -- (5.87918,13.7302) -- (6.1352,13.7302) -- (6.38736,13.7302) -- (6.64757,13.7302) -- (6.90564,13.7302) -- (7.16218,13.7302) -- (7.41662,13.7302) -- (7.67228,13.7302) -- (7.92956,13.7302) -- (8.18734,13.7302) -- (8.44388,13.7302)
 -- (8.69958,13.7302) -- (8.95719,13.7277) -- (9.21374,13.7277) -- (9.46437,13.7277) -- (9.72297,13.7251) -- (9.98465,13.7225) -- (10.2412,13.7199) -- (10.4942,13.7147) -- (10.7537,13.7069) -- (11.0102,13.6965) -- (11.2654,13.6833) -- (11.522,13.662)
 -- (11.7799,13.6323) -- (12.0361,13.5881) -- (12.293,13.5283) -- (12.5479,13.4421) -- (12.7998,13.3221) -- (13.0621,13.1644) -- (13.3155,12.9626) -- (13.5773,12.7195) -- (13.8317,12.4523) -- (14.0891,12.1611) -- (14.3457,11.8643) --
 (14.6016,11.5609) -- (14.8587,11.2475) -- (15.1143,10.9445) -- (15.3701,10.6401) -- (15.6273,10.3254) -- (15.8813,10.0161) -- (16.14,9.70507) -- (16.3947,9.39487) -- (16.6557,9.08785) -- (16.9073,8.77581) -- (17.168,8.46901) -- (17.4228,8.15707) --
 (17.6787,7.85015) -- (17.9376,7.53556) -- (18.1916,7.23379) -- (18.449,6.92413) -- (18.7052,6.60937) -- (18.9629,6.30005) -- (19,6.25538);
\definecolor{c}{rgb}{1,0,0};
\foreach \P in {(9.88757,10.9708)}{\draw[mark options={color=c,fill=c},mark size=2.402402pt,mark=*] plot coordinates {\P};}
\definecolor{c}{rgb}{0,0,0};
\draw [anchor=base west] (3.77534,14.057) node[scale=1.27276, color=c, rotate=0]{CAST};
\draw [anchor=base west] (3.77534,9.34796) node[scale=1.71823, color=c, rotate=0]{IAXO};
\draw [anchor=base west] (3.77534,11.4264) node[scale=1.27276, color=c, rotate=0]{BabyIAXO};
\draw [anchor=base west] (3.77534,8.48037) node[scale=1.27276, color=c, rotate=0]{IAXO+};
\draw [c,line width=0.9] (2.8,2.70372) -- (19,2.70372);
\draw [c,line width=0.9] (2.8,3.17301) -- (2.8,2.70372);
\draw [c,line width=0.9] (3.77534,2.93837) -- (3.77534,2.70372);
\draw [c,line width=0.9] (4.34588,2.93837) -- (4.34588,2.70372);
\draw [c,line width=0.9] (4.75068,2.93837) -- (4.75068,2.70372);
\draw [c,line width=0.9] (5.06467,2.93837) -- (5.06467,2.70372);
\draw [c,line width=0.9] (5.32121,2.93837) -- (5.32121,2.70372);
\draw [c,line width=0.9] (5.53812,2.93837) -- (5.53812,2.70372);
\draw [c,line width=0.9] (5.72601,2.93837) -- (5.72601,2.70372);
\draw [c,line width=0.9] (5.89175,2.93837) -- (5.89175,2.70372);
\draw [c,line width=0.9] (6.04,3.17301) -- (6.04,2.70372);
\draw [c,line width=0.9] (7.01534,2.93837) -- (7.01534,2.70372);
\draw [c,line width=0.9] (7.58588,2.93837) -- (7.58588,2.70372);
\draw [c,line width=0.9] (7.99068,2.93837) -- (7.99068,2.70372);
\draw [c,line width=0.9] (8.30466,2.93837) -- (8.30466,2.70372);
\draw [c,line width=0.9] (8.56121,2.93837) -- (8.56121,2.70372);
\draw [c,line width=0.9] (8.77812,2.93837) -- (8.77812,2.70372);
\draw [c,line width=0.9] (8.96601,2.93837) -- (8.96601,2.70372);
\draw [c,line width=0.9] (9.13175,2.93837) -- (9.13175,2.70372);
\draw [c,line width=0.9] (9.28,3.17301) -- (9.28,2.70372);
\draw [c,line width=0.9] (10.2553,2.93837) -- (10.2553,2.70372);
\draw [c,line width=0.9] (10.8259,2.93837) -- (10.8259,2.70372);
\draw [c,line width=0.9] (11.2307,2.93837) -- (11.2307,2.70372);
\draw [c,line width=0.9] (11.5447,2.93837) -- (11.5447,2.70372);
\draw [c,line width=0.9] (11.8012,2.93837) -- (11.8012,2.70372);
\draw [c,line width=0.9] (12.0181,2.93837) -- (12.0181,2.70372);
\draw [c,line width=0.9] (12.206,2.93837) -- (12.206,2.70372);
\draw [c,line width=0.9] (12.3717,2.93837) -- (12.3717,2.70372);
\draw [c,line width=0.9] (12.52,3.17301) -- (12.52,2.70372);
\draw [c,line width=0.9] (13.4953,2.93837) -- (13.4953,2.70372);
\draw [c,line width=0.9] (14.0659,2.93837) -- (14.0659,2.70372);
\draw [c,line width=0.9] (14.4707,2.93837) -- (14.4707,2.70372);
\draw [c,line width=0.9] (14.7847,2.93837) -- (14.7847,2.70372);
\draw [c,line width=0.9] (15.0412,2.93837) -- (15.0412,2.70372);
\draw [c,line width=0.9] (15.2581,2.93837) -- (15.2581,2.70372);
\draw [c,line width=0.9] (15.446,2.93837) -- (15.446,2.70372);
\draw [c,line width=0.9] (15.6117,2.93837) -- (15.6117,2.70372);
\draw [c,line width=0.9] (15.76,3.17301) -- (15.76,2.70372);
\draw [c,line width=0.9] (16.7353,2.93837) -- (16.7353,2.70372);
\draw [c,line width=0.9] (17.3059,2.93837) -- (17.3059,2.70372);
\draw [c,line width=0.9] (17.7107,2.93837) -- (17.7107,2.70372);
\draw [c,line width=0.9] (18.0247,2.93837) -- (18.0247,2.70372);
\draw [c,line width=0.9] (18.2812,2.93837) -- (18.2812,2.70372);
\draw [c,line width=0.9] (18.4981,2.93837) -- (18.4981,2.70372);
\draw [c,line width=0.9] (18.686,2.93837) -- (18.686,2.70372);
\draw [c,line width=0.9] (18.8517,2.93837) -- (18.8517,2.70372);
\draw [c,line width=0.9] (19,3.17301) -- (19,2.70372);
\draw [c,line width=0.9] (2.8,18.3467) -- (19,18.3467);
\draw [c,line width=0.9] (2.8,17.8774) -- (2.8,18.3467);
\draw [c,line width=0.9] (3.77534,18.1121) -- (3.77534,18.3467);
\draw [c,line width=0.9] (4.34588,18.1121) -- (4.34588,18.3467);
\draw [c,line width=0.9] (4.75068,18.1121) -- (4.75068,18.3467);
\draw [c,line width=0.9] (5.06467,18.1121) -- (5.06467,18.3467);
\draw [c,line width=0.9] (5.32121,18.1121) -- (5.32121,18.3467);
\draw [c,line width=0.9] (5.53812,18.1121) -- (5.53812,18.3467);
\draw [c,line width=0.9] (5.72601,18.1121) -- (5.72601,18.3467);
\draw [c,line width=0.9] (5.89175,18.1121) -- (5.89175,18.3467);
\draw [c,line width=0.9] (6.04,17.8774) -- (6.04,18.3467);
\draw [c,line width=0.9] (7.01534,18.1121) -- (7.01534,18.3467);
\draw [c,line width=0.9] (7.58588,18.1121) -- (7.58588,18.3467);
\draw [c,line width=0.9] (7.99068,18.1121) -- (7.99068,18.3467);
\draw [c,line width=0.9] (8.30466,18.1121) -- (8.30466,18.3467);
\draw [c,line width=0.9] (8.56121,18.1121) -- (8.56121,18.3467);
\draw [c,line width=0.9] (8.77812,18.1121) -- (8.77812,18.3467);
\draw [c,line width=0.9] (8.96601,18.1121) -- (8.96601,18.3467);
\draw [c,line width=0.9] (9.13175,18.1121) -- (9.13175,18.3467);
\draw [c,line width=0.9] (9.28,17.8774) -- (9.28,18.3467);
\draw [c,line width=0.9] (10.2553,18.1121) -- (10.2553,18.3467);
\draw [c,line width=0.9] (10.8259,18.1121) -- (10.8259,18.3467);
\draw [c,line width=0.9] (11.2307,18.1121) -- (11.2307,18.3467);
\draw [c,line width=0.9] (11.5447,18.1121) -- (11.5447,18.3467);
\draw [c,line width=0.9] (11.8012,18.1121) -- (11.8012,18.3467);
\draw [c,line width=0.9] (12.0181,18.1121) -- (12.0181,18.3467);
\draw [c,line width=0.9] (12.206,18.1121) -- (12.206,18.3467);
\draw [c,line width=0.9] (12.3717,18.1121) -- (12.3717,18.3467);
\draw [c,line width=0.9] (12.52,17.8774) -- (12.52,18.3467);
\draw [c,line width=0.9] (13.4953,18.1121) -- (13.4953,18.3467);
\draw [c,line width=0.9] (14.0659,18.1121) -- (14.0659,18.3467);
\draw [c,line width=0.9] (14.4707,18.1121) -- (14.4707,18.3467);
\draw [c,line width=0.9] (14.7847,18.1121) -- (14.7847,18.3467);
\draw [c,line width=0.9] (15.0412,18.1121) -- (15.0412,18.3467);
\draw [c,line width=0.9] (15.2581,18.1121) -- (15.2581,18.3467);
\draw [c,line width=0.9] (15.446,18.1121) -- (15.446,18.3467);
\draw [c,line width=0.9] (15.6117,18.1121) -- (15.6117,18.3467);
\draw [c,line width=0.9] (15.76,17.8774) -- (15.76,18.3467);
\draw [c,line width=0.9] (16.7353,18.1121) -- (16.7353,18.3467);
\draw [c,line width=0.9] (17.3059,18.1121) -- (17.3059,18.3467);
\draw [c,line width=0.9] (17.7107,18.1121) -- (17.7107,18.3467);
\draw [c,line width=0.9] (18.0247,18.1121) -- (18.0247,18.3467);
\draw [c,line width=0.9] (18.2812,18.1121) -- (18.2812,18.3467);
\draw [c,line width=0.9] (18.4981,18.1121) -- (18.4981,18.3467);
\draw [c,line width=0.9] (18.686,18.1121) -- (18.686,18.3467);
\draw [c,line width=0.9] (18.8517,18.1121) -- (18.8517,18.3467);
\draw [c,line width=0.9] (19,17.8774) -- (19,18.3467);
\draw [c,line width=0.9] (2.8,2.70372) -- (2.8,18.3467);
\draw [c,line width=0.9] (3.286,2.70373) -- (2.8,2.70373);
\draw [c,line width=0.9] (3.043,3.88098) -- (2.8,3.88098);
\draw [c,line width=0.9] (3.043,4.56963) -- (2.8,4.56963);
\draw [c,line width=0.9] (3.043,5.05823) -- (2.8,5.05823);
\draw [c,line width=0.9] (3.043,5.43722) -- (2.8,5.43722);
\draw [c,line width=0.9] (3.043,5.74688) -- (2.8,5.74688);
\draw [c,line width=0.9] (3.043,6.00869) -- (2.8,6.00869);
\draw [c,line width=0.9] (3.043,6.23548) -- (2.8,6.23548);
\draw [c,line width=0.9] (3.043,6.43553) -- (2.8,6.43553);
\draw [c,line width=0.9] (3.286,6.61447) -- (2.8,6.61447);
\draw [c,line width=0.9] (3.043,7.79172) -- (2.8,7.79172);
\draw [c,line width=0.9] (3.043,8.48037) -- (2.8,8.48037);
\draw [c,line width=0.9] (3.043,8.96898) -- (2.8,8.96898);
\draw [c,line width=0.9] (3.043,9.34797) -- (2.8,9.34797);
\draw [c,line width=0.9] (3.043,9.65762) -- (2.8,9.65762);
\draw [c,line width=0.9] (3.043,9.91943) -- (2.8,9.91943);
\draw [c,line width=0.9] (3.043,10.1462) -- (2.8,10.1462);
\draw [c,line width=0.9] (3.043,10.3463) -- (2.8,10.3463);
\draw [c,line width=0.9] (3.286,10.5252) -- (2.8,10.5252);
\draw [c,line width=0.9] (3.043,11.7025) -- (2.8,11.7025);
\draw [c,line width=0.9] (3.043,12.3911) -- (2.8,12.3911);
\draw [c,line width=0.9] (3.043,12.8797) -- (2.8,12.8797);
\draw [c,line width=0.9] (3.043,13.2587) -- (2.8,13.2587);
\draw [c,line width=0.9] (3.043,13.5684) -- (2.8,13.5684);
\draw [c,line width=0.9] (3.043,13.8302) -- (2.8,13.8302);
\draw [c,line width=0.9] (3.043,14.057) -- (2.8,14.057);
\draw [c,line width=0.9] (3.043,14.257) -- (2.8,14.257);
\draw [c,line width=0.9] (3.286,14.436) -- (2.8,14.436);
\draw [c,line width=0.9] (3.043,15.6132) -- (2.8,15.6132);
\draw [c,line width=0.9] (3.043,16.3019) -- (2.8,16.3019);
\draw [c,line width=0.9] (3.043,16.7905) -- (2.8,16.7905);
\draw [c,line width=0.9] (3.043,17.1695) -- (2.8,17.1695);
\draw [c,line width=0.9] (3.043,17.4791) -- (2.8,17.4791);
\draw [c,line width=0.9] (3.043,17.7409) -- (2.8,17.7409);
\draw [c,line width=0.9] (3.043,17.9677) -- (2.8,17.9677);
\draw [c,line width=0.9] (3.043,18.1678) -- (2.8,18.1678);
\draw [c,line width=0.9] (3.286,18.3467) -- (2.8,18.3467);
\draw [c,line width=0.9] (19,2.70372) -- (19,18.3467);
\draw [c,line width=0.9] (18.514,2.70373) -- (19,2.70373);
\draw [c,line width=0.9] (18.757,3.88098) -- (19,3.88098);
\draw [c,line width=0.9] (18.757,4.56963) -- (19,4.56963);
\draw [c,line width=0.9] (18.757,5.05823) -- (19,5.05823);
\draw [c,line width=0.9] (18.757,5.43722) -- (19,5.43722);
\draw [c,line width=0.9] (18.757,5.74688) -- (19,5.74688);
\draw [c,line width=0.9] (18.757,6.00869) -- (19,6.00869);
\draw [c,line width=0.9] (18.757,6.23548) -- (19,6.23548);
\draw [c,line width=0.9] (18.757,6.43553) -- (19,6.43553);
\draw [c,line width=0.9] (18.514,6.61447) -- (19,6.61447);
\draw [c,line width=0.9] (18.757,7.79172) -- (19,7.79172);
\draw [c,line width=0.9] (18.757,8.48037) -- (19,8.48037);
\draw [c,line width=0.9] (18.757,8.96898) -- (19,8.96898);
\draw [c,line width=0.9] (18.757,9.34797) -- (19,9.34797);
\draw [c,line width=0.9] (18.757,9.65762) -- (19,9.65762);
\draw [c,line width=0.9] (18.757,9.91943) -- (19,9.91943);
\draw [c,line width=0.9] (18.757,10.1462) -- (19,10.1462);
\draw [c,line width=0.9] (18.757,10.3463) -- (19,10.3463);
\draw [c,line width=0.9] (18.514,10.5252) -- (19,10.5252);
\draw [c,line width=0.9] (18.757,11.7025) -- (19,11.7025);
\draw [c,line width=0.9] (18.757,12.3911) -- (19,12.3911);
\draw [c,line width=0.9] (18.757,12.8797) -- (19,12.8797);
\draw [c,line width=0.9] (18.757,13.2587) -- (19,13.2587);
\draw [c,line width=0.9] (18.757,13.5684) -- (19,13.5684);
\draw [c,line width=0.9] (18.757,13.8302) -- (19,13.8302);
\draw [c,line width=0.9] (18.757,14.057) -- (19,14.057);
\draw [c,line width=0.9] (18.757,14.257) -- (19,14.257);
\draw [c,line width=0.9] (18.514,14.436) -- (19,14.436);
\draw [c,line width=0.9] (18.757,15.6132) -- (19,15.6132);
\draw [c,line width=0.9] (18.757,16.3019) -- (19,16.3019);
\draw [c,line width=0.9] (18.757,16.7905) -- (19,16.7905);
\draw [c,line width=0.9] (18.757,17.1695) -- (19,17.1695);
\draw [c,line width=0.9] (18.757,17.4791) -- (19,17.4791);
\draw [c,line width=0.9] (18.757,17.7409) -- (19,17.7409);
\draw [c,line width=0.9] (18.757,17.9677) -- (19,17.9677);
\draw [c,line width=0.9] (18.757,18.1678) -- (19,18.1678);
\draw [c,line width=0.9] (18.514,18.3467) -- (19,18.3467);
\end{tikzpicture}

%% file: sections/transparency2.tex

IAXO will be capable to probe the ALP parameter space which is currently only accessible to astrophysical observations.
Astrophysical magnetic fields can range from values below $10^{-9}$\,G in intergalactic space to $\gtrsim10^{14}\,$G on the surface of magnetars (highly magnetized neutron stars) on a large range of spatial scales.
As the photon-ALP conversion probability scales with the product of the field strength $B$ and the spatial extend of the field $L$, in some astrophysical environments a large conversion probability could be possible~\cite{hooper2007:alps,deangelis2007,DeAngelis:2007wiw}.
For a homogeneous magnetic field, the conversion probability becomes maximal and independent of energy
for energies $E_\mathrm{crit} \lesssim E \lesssim E_\mathrm{max}$, with \cite[e.g.][]{hooper2007:alps,bassan2009} (neglecting dispersion due to, e.g., the CMB~\cite{Kartavtsev:2016doq})
\begin{eqnarray}
E_\mathrm{crit} &\approx& \frac{|m_a^2 - \omega_\mathrm{pl}^2|}{2g_{a\gamma}B} \sim 2.5\,\mathrm{GeV}\, |m^2_\mathrm{neV} - 1.4\times10^{-3}\, n_{\mathrm{cm}^{-3}}| \,g_{11}^{-1} B_{\mu\mathrm{G}}^{-1},
\label{eqn:ecrit}\\
E_\mathrm{max} &\approx& \frac{90\pi}{7\alpha}\frac{B_\mathrm{cr}^2\,g_{a\gamma}}{B} \sim 2.12\times10^{6}\,\mathrm{GeV}\,g_{11}B_{\mu\mathrm{G}}^{-1}.
\label{eqn:emax}
\end{eqnarray}
In the above equations, we have introduced the notation $B_X = B / X$, $m_X = m_a / X$, and $g_X = g_{a\gamma} \times 10^X / \mathrm{GeV}^{-1}$.
The plasma frequency is connected to the electron density of the medium through $\omega_\mathrm{pl} \sim 0.037 \sqrt{n_{\mathrm{cm}^{-3}}}\,$neV.
Above $E_\mathrm{max}$, the oscillations are damped due to the QED vacuum polarisation.
The chosen units in the above equations already indicate that for magnetic fields with a strength of $\mu$G and ALP masses around neV, a strong mixing can be expected at gamma-ray energies. Such magnetic fields are commonly observed in the intra-cluster medium of galaxy clusters, in galaxies themselves, and in the lobes of jets launched by active galactic nuclei (AGNs)~\cite[e.g.][]{widrow2002,pudritz2012}.
For lower ALP masses $m_a \lesssim 10^{-12}$eV, $E_\mathrm{crit}$ shifts to X-ray energies.
For X-rays or gamma rays produced in AGN,
Tab.~\ref{tabAstrObj} displays the typical strength of the magnetic field $B$, its coherence length $L$, and the photon-ALP oscillation length $\lambda_a$ of magnetic fields traversed by the photons.


\begin{table}[h]
\begin{center}
\begin{tabular}{c|ccc}
\hline

Astronomical object & \ \  $B$ \ \ & \ \  $L$ \ \ & \ \  $\lambda_a$  \ \ \\
\hline
\hline

AGN jet & $0.1-5 \, \rm G$ \ & none & ${\cal O}(0.01-0.5\, \rm pc)^*$ \\
Radio lobes & $10 \, \rm \mu G$ & $10 \, \rm kpc$ & ${\cal O}(10 \, \rm kpc)$ \\
Spiral galaxies & $7 \, \rm \mu G$ & $10 \, \rm kpc$ & ${\cal O}(10 \, \rm kpc)$  \\
Starburst galaxies & $50 \, \rm \mu G$ & $10 \, \rm kpc$ & ${\cal O}(1 \, \rm kpc)$  \\
Elliptical galaxies & $5 \, \rm \mu G$ & $150  \, \rm pc$ &  ${\cal O}(10 \, \rm kpc)$ \\
Cluster & $5 \, \rm \mu G$ & $10-100\, \rm kpc$ & ${\cal O}(10 \, \rm kpc)$ \\
IGM & $< 1.7 \, \rm nG$ & $0.1-10 \, \rm Mpc$ & ${\cal O}(50 \, \rm Mpc)^{\dag}$ \\
Milky Way &$5 \, \rm \mu G$& $  10 \, \rm kpc$ & ${\cal O}(10 \, \rm kpc)$ \\
\hline
\end{tabular}
\caption{Various astronomical objects with typical values of average magnetic field $B$,  domain length $L$ and corresponding photon-ALP oscillation length $\lambda_a$ traversed by a photon beam originating from AGNs. IGM stands for intergalactic medium. The magnetic field in the AGN jet {\it does not} possess a domain-like structure~\cite{Tavecchio:2014yoa}. The values of $\lambda_a$ are calculated in the strong mixing regime where the plasma and ALP mass effects at low energies, the QED vacuum polarization and CMB photon dispersion effects at high energies can be neglected. In the case where $\lambda_a \lesssim L$ the simple sharp edges model model to describe domain-like magnetic fields gives unphysical results because the photon/ALP beam becomes sensitive to the discontinuities of the magnetic field components: in this case, more sophisticated models must be used~\cite[e.g.][]{Galanti:2018nvl,Kartavtsev:2016doq,Meyer:2014epa}. $^*$In the case of the jet QED effects are quite  important and for energies $E \gtrsim {\cal O}(100 \, \rm GeV)$ they strongly reduce the reported value of $\lambda_a$. $^{**}$A similar fact happens in the extragalactic space because of the ALP mass term for $E \lesssim {\cal O}(100 \, \rm GeV)$ and because of the CMB photon dispersion on the CMB for $E \gtrsim {\cal O}(5 \, \rm TeV)$~\cite{2018JHEAp..20....1G}.}
\label{tabAstrObj}
\end{center}
\end{table}

Below, we summarize current hints, constraints, and future sensitivities of gamma-ray and X-ray observations.
In comparison to future IAXO observations, the X-ray and gamma-ray measurements suffer from the unknown exact conditions of the involved magnetic fields and photon fluxes. In this respect, IAXO offers the unique opportunity to test any hints found in astrophysical observations in a more controlled setting.

\subsection{Conversion between gamma-ray photons and ALPs}
For gamma rays, one expects two possible observables as a result of photon-ALP mixing.
On the one hand, the conversions should lead to oscillatory features around $E_\mathrm{crit}$ and $E_\mathrm{max}$ as some photons will oscillate into ALPs thereby reducing the photon flux.
The shape of these oscillations will depend on the morphology of the magnetic field, the plasma frequency, the ALP parameters, and the gamma-ray energy. Since astrophysical magnetic fields are often turbulent in nature, the spectral features are expected to show a complicated dependence on energy.

On the other hand, photon-ALP oscillations could lead to a boost in the photon flux due to an astrophysical version of the ``light shining through a wall'' experiment,
in which photons convert to axions or ALPs and back on either side of an opaque wall.
For gamma rays emitted by AGNs, the opacity is caused by the interaction of gamma rays with lower-energy photons from infrared (IR) to ultraviolet (UV), which produces electron-positron pairs~\cite{nikishov1962,gould1967}.
In intergalactic space, these lower-energy photons originate from the extragalactic background light (EBL), the background radiation field which encompasses the stellar emission integrated over the age of the Universe, and the emission absorbed and re-emitted by dust~\cite[see][for reviews]{hauser2001,dwek2013}.
Due to strong foreground emission in the solar system in the EBL wavelength range, the EBL is extremely difficult to measure directly~\cite{hauser1998}.
Within the AGN jet, the radiation fields from the dusty torus, the accretion disk, or optical emission from ionized clouds (the so-called broad line region, BLR) can cause the attenuation~\cite[e.g.][]{finke2016}. The presence of the BLR is usually associated with a sub-class of AGNs, so called flat-spectrum radio quasars (FSRQs).
Basically, because of photon-ALP oscillations, the photon acquires a ``split personality'': sometimes it travels as an ordinary photon and gets absorbed by the EBL, but sometimes it travels as an ALP which does not interact with EBL photons. As a consequence, the effective optical depth, $\tau_\mathrm{eff}$ is smaller than the optical depth as evaluated by conventional physics.
Since the photon survival probability is given by $P^\mathrm{ALP}_{\gamma\leftarrow\gamma} = \exp[-\tau_\mathrm{eff}]$, even a small decrease of the effective optical depth can give rise to a large enhancement in photon flux.

The exponential attenuation of the initial gamma-ray flux is
described with the optical depth $\tau$,
which increases monotonically with the primary gamma-ray energy, the source distance (in the case of absorption on the EBL), and the photon density of the radiation field.
ALPs are not absorbed during their propagation
and if a sizeable amount of photons converts into ALPs, which then reconvert back to photons close to Earth, a boost of the expected gamma-ray flux is expected~\cite[e.g.][]{deangelis2007,mirizzi2007,tavecchio2012,2018JHEAp..20....1G}.

\subsection{Hints for anomalies in the gamma-ray opacity?}

Several authors have found evidence that state-of-the-art EBL models over-predict the attenuation of gamma rays using published data points of AGN spectra obtained with imaging air Cherenkov telescopes (IACTs), which measure gamma rays above energies of $\sim 50\,$GeV.
Such an over-prediction would manifest itself through a hardening of the AGN spectra.
For example, if the observed spectrum at low energies can be described with a power law, $dN/dE \propto E^{-\Gamma_\mathrm{low}}$ a spectral hardening would mean that at high energies $dN/dE \propto E^{-\Gamma_\mathrm{high}}$ with $\Gamma_\mathrm{high} < \Gamma_\mathrm{low}$.
Such a behavior is in general not expected from standard gamma-ray emission scenarios (although specific models can produce such spectra~\cite[e.g.][]{lefa2011b}), especially if the hardening $\Delta\Gamma = \Gamma_\mathrm{low} - \Gamma_\mathrm{high}$ correlates with increasing optical depth for several sources.
Such a correlation has indeed been found and ALPs have been proposed as a possible explanation~ \cite{deangelis2007,simet2008,deangelis2009,sanchezconde2009,
dominguez2011alps,deangelis2011,essey2012,
rubtsov2014,galanti2015}.
An over-predicted EBL attenuation should also lead to a correlation between fit residuals and the optical depth when smooth concave, i.e. non-hardening, functions are assumed for the emitted AGN spectra. A $4\,\sigma$ indication for this effect has been found~\cite{horns2012,meyer2012ppa}, and ALP parameters reducing this tension were derived~\cite{meyer2013} (see the ``T-Hint'' labelled region in Fig.~\ref{fig:transpa}).
Furthermore, using recent EBL measurements with the CIBER experiment,  which suggest a larger attenuation than current EBL models, the authors of Ref.~\cite{Kohri:2017ljt} found that ALPs can improve the fits of IACT spectra when again concave intrinsic spectra are assumed (region on top of the ``T-Hint'' region in Fig.~\ref{fig:transpa}).
Instead of ALPs, these evidences have also been interpreted as evidence for particle cascades initiated by ultra-high energy cosmic rays~\cite{essey2010} or a correlation between AGN lines of sight with cosmic voids~\cite{furniss2014}.

However, recent analyses could not confirm the above correlations for a spectral hardening.
Extending the IACT data sample of Ref.~\cite{meyer2013}, the authors of Ref.~\cite{biteau2015} did not find a correlation of fit residuals with the optical depth.
Furthermore, when including systematic uncertainties such as the IACT energy resolution, no spectral hardening between IACT spectra and spectra measured at lower gamma-ray energies with the Large Area Telescope (LAT) on board the \emph{Fermi} satellite could be found~\cite{Sanchez:2013lla}.
Using \emph{Fermi}-LAT data alone,
no spectral hardening as a function of redshift (or equivalently the optical depth) was found in a recent analysis~\cite{dominguez2015}.

It has also been hypothesized that photon-ALP conversions could be responsible for IACT observations of FSRQs above 100\,GeV.
If gamma rays are produced close to the central super massive black hole, their flux should be severely attenuated due to the interaction with the radiation fields within the jet as mentioned above.
Yet, a number of FSRQs has indeed been observed~\cite[e.g.][]{3c279,pks1222,pks1510}\footnote{See also \url{http://tevcat.uchicago.edu/}.} and
it has been shown that the inclusion of ALPs can reproduce the observed spectra for ALP parameters $g_{11}\sim 1$ and $m_\mathrm{neV} \lesssim 10$~\cite{tavecchio2012}.
One has to keep in mind, though, that astrophysical mechanisms could also produce the gamma-ray emission beyond the BLR and thus circumvent the pair production and the attenuation~\cite[e.g.][]{tavecchio2011pks1222,Petropoulou2017}.
But it was the dissatisfaction of the {\it ad hoc} nature of these astrophysical attempted explanations that led to the above ALP-based proposal. In addition for BL Lacs (a sub-class of AGNs), by combining all the magnetic environments crossed by the photon/ALP beam -- namely BL Lac jet, host galaxy, extragalactic space and Milky Way -- it is possible to infer important predictions for BL Lacs spectra which present peculiar observable features induced by photon-ALP oscillations: (i) the oscillatory behavior of the energy spectrum (ii) photon excess above $20 \, \rm TeV$~\cite{Galanti:2018upl}. 

As shown in Fig.~\ref{fig:transpa}, IAXO is sensitive to the entire  parameter space where ALPs have been proposed to alter the gamma-ray transparency.
Therefore, IAXO measurements could give the definite answer on this matter.

\subsubsection{Statistical distributions of X-ray source luminosities}

While this part does not deal specifically with opacity, we mention this effect here as another possible hint for a light ALP ($\sim$ 10 neV) with the same coupling as what is discussed in this section ($10^{-11}\;\rm GeV^{-1}$). In the strong mixing regime, it is easy to show that the average luminosity of a photon beam is 2/3 that without mixing~\cite{Csaki2002} . It means that the luminosity of X-ray sources would be actually higher than what we observe. As all source would be concerned, this effect based on average luminosities would not be observable. However, photon-ALP mixing would also affect higher moments of the luminosity distribution of the sources. By studying these higher moments, the authors of Ref.~\cite{Burrage2009} showed an anomaly compatible with ALPs in the same parameter space region as mentioned in the previous subsection. It has later been noted that this effect could be due to outliers in the used source catalog~\cite{Pettinari2010} so it could be a selection-bias-induced fake signal.

\subsection{Constraints from high-energy astronomy and remaining parameter space}

\begin{figure}[h]
\centering
\tikzsetnextfilename{pics/pics/IAXO_star_anomaly}
\resizebox{\textwidth}{!} {\input{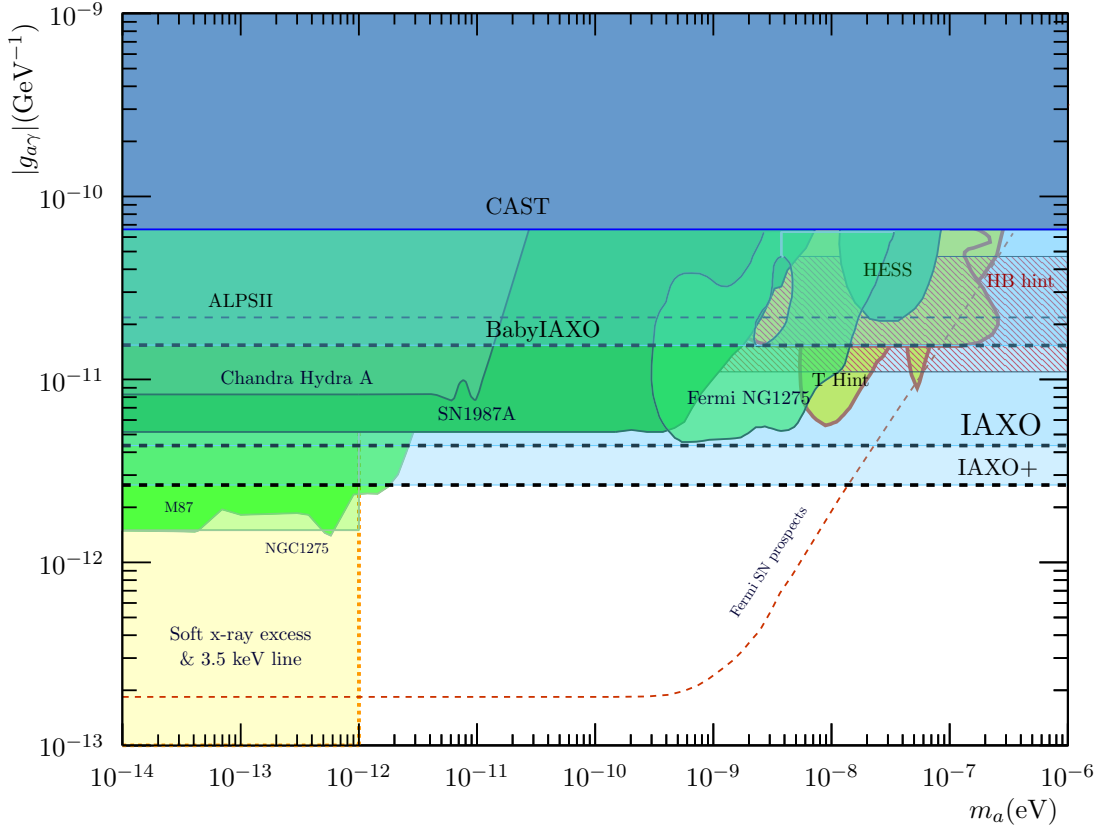}}
 \caption{The IAXO potential in the low ALP mass part of the $\gagamma$ vs. $m_a$ plane. The hashed region indicates the R-parameter hint, discussed in section~\ref{sec:stellarHints_GC}, in the case of ALPs interacting only with photons. In this case, this is known also as the HB-hint~\cite{Ayala:2014pea,Straniero:2015nvc,Giannotti:2016hnk}. 
           The region indicated with "T-hint" is the transparency region, discussed in the section~\ref{sec:transparency}.
                ALPs with parameters in this region have been invoked to address the unexpected transparency of the Universe to very high energy photons~\cite{meyer2013,Kohri:2017ljt} and some anomalous redshift-dependence of AGN gamma-ray spectra~\cite{Galanti:2015rda}.
                Notice that the lower mass section of the hinted region has been excluded by the non-observation of gamma rays from SN~1987A~\cite{Payez:2014xsa} and, more recently, by the search for spectral irregularities in the gamma ray spectrum of NGC~1275~\cite{ajello2016}.
                The region enclosed within the dashed red line, labelled Fermi SN prospects, shows the Fermi LAT potential to probe the ALP parameter space in case of a new nearby (galactic) SN explosion~\cite{Meyer:2016wrm}.
Finally, the figure shows the regions excluded by the analysis of the spectral distortions of X-ray point sources in galaxy clusters: the Chandra observations of the AGN in Hydra A~ \cite{Wouters:2013hua}, the Perseus cluster NGC1275~\cite{berg2016} (indicated in the figure as NGC1275 and to be distinguished from the region labelled Fermi NG1275), and the M87 AGN of the Virgo cluster~\cite{170307354}. Refer to the text for more details.
}
\label{fig:transpa}
\end{figure}

Searches for spectral distortions in gamma-ray spectra have already constrained the parameter space where ALPs could explain a reduced transparency.
Observations with the H.E.S.S. telescopes of one blazar, i.e. an AGN with its jet closely aligned to the line of sight, have led to the exclusions labelled ``H.E.S.S.'' in Fig.~\ref{fig:transpa} under the assumption that the blazar is located in a Galaxy group that harbours a magnetic field of $B = 1\,\mu$G~\cite{hess2013:alps}.
Further constraints were derived using \emph{Fermi}-LAT observations of NGC\,1275, the central AGN of the Perseus galaxy cluster~\cite{ajello2016}.
This galaxy cluster could have a central magnetic field as large as $25\,\mu$G~\cite{taylor2006}.
The constraints are the strongest to date in the mass range between $0.5\lesssim m_\mathrm{neV} \lesssim 20$ (see Fig.~\ref{fig:transpa}).
Similar analyses using X-ray observations of AGN in galaxy clusters have constrained lower mass ALPs (see below).

For lower ALP masses, strong constraints where derived from the non-observation of a gamma-ray burst from the core-collapse supernova (SN) SN1987A~\cite{brockway1996,Grifols:1996id,Payez:2014xsa}.
During the core collapse, gamma rays in the core could convert to ALPs in the electrostatic fields of ions and escape the explosion.
If they convert back into gamma rays in the Galactic magnetic field, a gamma ray burst lasting tens of seconds could be observed in temporal co-incidence with the SN neutrino burst.
Interestingly, if a Galactic SN occurred in the field of view of the  \emph{Fermi}~LAT, a wide range of photon-ALP couplings could be probed for $m_\mathrm{neV} \lesssim 100$~(see the red dashed line in Fig~\ref{fig:transpa}).
These prospects for a detection rely on a simplified picture of the proton-neutron star nuclear medium~\cite{Payez:2014xsa}
and encourage future refinements of the theoretically expected ALP flux from such objects~\cite{2017arXiv171206205B,2018JHEAp..20....1G}.

The next generation of gamma-ray observatories will allow deeper studies of ALP-induced effects and will have enough sensitivity to probe the whole parameter
space relevant for the gamma ray transparency hints.
In particular, analyses of Cherenkov Telescope Array (CTA) data will consist of searches for the spectral irregularities~\cite[see][for preliminary results]{gate2017},
the reduced opacity~\cite{meyer2014cta}, or spatial correlations between blazar spectra with the magnetic field of the Milky Way~\cite{wouters2014}.
A similar correlation study could also be conducted with future gamma-ray observations of the HAWC or LHAASO observatories~\cite{vogel2017}.
However, CTA will be fully operational within a decade, and the necessary accumulation of data could take years.

On the other hand, as shown on Fig.~\ref{fig:transpa}, IAXO will be sensitive to the relevant masses and coupling. In conclusion, light ALPs can have a strong impact on gamma-ray astronomy and IAXO will cover the whole relevant parameter space.

\subsection{Conversion between X-ray photons and ALPs}

Let us now describe further aspects of ALP physics that rely on the interconversion of ALPs and X-ray photons in the magnetic field of galaxy clusters.

\begin{itemize}

\item {\bf Spectral distortions of X-ray point sources in galaxy clusters}


Just as with gamma-ray observations, it is possible to search for ALPs by looking for oscillatory features in the spectra of arriving X-rays from AGN or quasars in or behind galaxy clusters  -- and conversely, the absence of
such features can be used to place bounds on ALP parameter space.

The main disadvantage of this method is that the actual magnetic field along the line of sight is unknown. 
For any one source, it is then never possible to exclude the possibility that the magnetic field configuration along the line of sight is particularly unfavourable for ALP-photon conversion. This motivates considering sufficiently broad energy ranges so that multiple features are expected, and observing multiple X-ray sources within distinct astrophysical environments.

In X-rays, this method has been applied for the AGN in Hydra A \cite{Wouters:2013hua}, the central AGN of the Perseus cluster NGC1275~\cite{berg2016} (see above for gamma rays), the central M87 AGN of the Virgo cluster \cite{170307354}, and a variety of weaker quasars and AGNs in and behind clusters \cite{170405256}.
Of these sources, the two bright local AGNs NGC1275 and M87 are the most constraining, as both are bright AGNs with deep exposures at the heart of large cool-core galaxy clusters (which tend to have the highest magnetic fields). For reasonable and observationally supported values for the magnetic field structure within the Perseus and Virgo cluster, the absence of large spectral modulations was used in \cite{berg2016} and  \cite{170307354} to constrain $g_{a\gamma} \lesssim 1.5 \times 10^{-12} {\rm GeV}^{-1}$ for ALP masses $m_a \lesssim 10^{-12} {\rm eV}$.
These bounds depend on the magnetic field model for the cluster. For magnetic fields weaker or stronger than assumed, the constraint scales inversely with the magnetic field.

\item {\bf Spectral distortions of the continuum thermal bremsstrahlung emission of galaxy clusters}

The advantage of bright point sources is that they are sensitive to the magnetic field along a single line of sight, and so avoid effects of destructive interference
when averaging over many different sightlines. However bright point sources are also rather rare.

Another approach to searching for ALPs is to use instead the continuum emission from galaxy clusters. This arises as thermal bremsstrahlung from the intracluster medium, the hot ($T \sim 2 - 8 {\rm keV}$) ionised plasma where around 90\% of the baryons in a cluster are located. The thermal emission contains both continuum and line emission, and provides an excellent fit to the overall photon spectrum from clusters.

As with point sources, the aim is to observe (or constrain) ALP-induced deviations from the thermal bremsstrahlung fit.
 In the presence of $g_{a\gamma} \sim 10^{11} {\rm GeV}$, this thermal emission will experience significant spectral modulation along
a single line of sight (as for point sources). Averaged over a large region of the cluster, the modulations from individual sightlines will average out.
However, in the presence of significant ALP-photon conversion significant spectral distortions will be present on small scales \cite{Conlon:2015uwa}.

\item {\bf The 3.5 keV Line}

One of the most interesting recent results in particle astrophysics has been the observation of an unexplained line at $E \sim 3.55\, {\rm keV}$ \cite{Bulbul:2014sua, Boyarsky:2014jta}. The line was found originally in observations of stacked samples of galaxy clusters and is not at an energy that corresponds to a known atomic line. There has been considerable interest around the possibility that this could arise from dark matter \cite{Ishida:2014dlp, Finkbeiner:2014sja, Higaki:2014zua, Jaeckel:2014qea, Lee:2014xua, Kong:2014gea, Frandsen:2014lfa, Nakayama:2014ova, Choi:2014tva}, while also various possible astrophysical explanations have been proposed.

If the line arises from dark matter, one difficulty with `standard' interpretations (such as a sterile neutrino) is that the line is much stronger in
clusters than in galaxies, and in particular is much stronger at the centre of the Perseus cluster than at any other location. This implies that the line
is not sensitive only to the dark matter content, but also to some aspect of the astrophysical environment. One way this can arise is in models where the
dark matter decays originally to a relativistic ALP with energy $E = 3.5\, {\rm keV}$, and the observed photon signal comes from conversion of this ALP
in the magnetic field of the cluster \cite{Cicoli:2014bfa}. Depending on the dark matter lifetime, the line signal can be reproduced for axion-photon couplings of order $10^{-15}\,{\rm GeV}^{-1}\lesssim g_{a\gamma}\lesssim 10^{-10}\,{\rm GeV}^{-1}$ for ALP masses $m_a<10^{-12}$ eV. Interestingly, IAXO will be able to test a relatively large region of this parameter space.

The morphology of the 3.5 keV signal can be explained by ALP-photon conversion.
The stronger signal in clusters would then arise from the larger and more extended magnetic fields present in clusters compared to galaxies. As Perseus is a close cool-core cluster, observations of Perseus only cover the
central region with a large magnetic field (as $B \propto n_e(r)^{\frac12}$, the magnetic field is significantly enhanced in the central
high-$n_e$ cool core). The extremely strong signal in the centre of Perseus would then be a consequence of the efficiency of ALP-photon conversion in strong magnetic fields.

If this line is found to arise from new physics, ALPs then offer a way to reproduce the unusual morphology.

\end{itemize}

%% file: sections/iaxosensitivity.tex

\begin{figure}[t] \centering
\includegraphics[width=\textwidth]{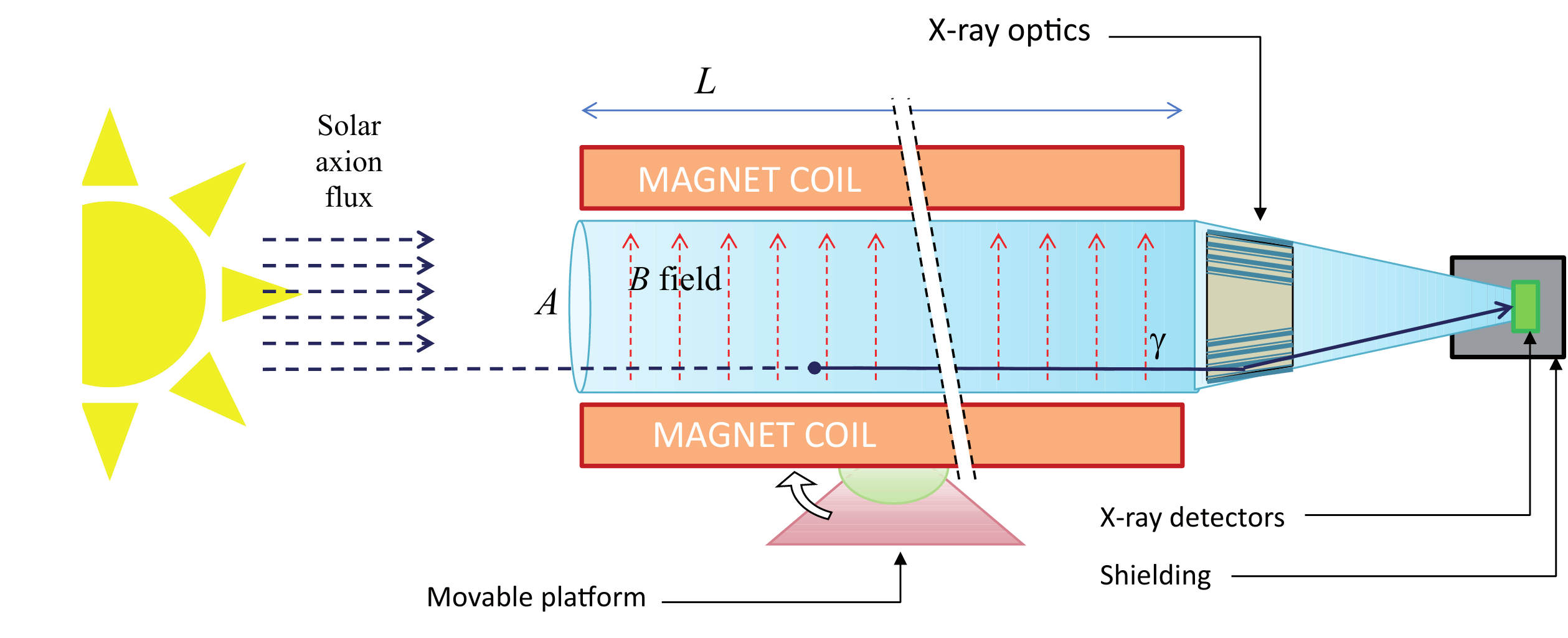}\hspace{2pc}%
\caption{\label{fig:NGAH_sketch} Conceptual arrangement
of an enhanced axion helioscope with X-ray focalization. Solar axions are converted into photons by the transverse magnetic field inside the bore of a powerful magnet. The resulting quasi-parallel beam of photons of cross sectional area $A$ is concentrated by an appropriate X-ray optics onto a small spot area $a$ in a low background detector. The envisaged design for IAXO, shown in Fig.~\ref{fig:IAXO_sketch}, includes eight such magnet bores, with their respective optics and detectors. }
\end{figure}

The International Axion Observatory (IAXO) is a next generation axion helioscope that aims at a substantial step forward, of more than one order of magnitude, in sensitivity to $\gagamma$ with respect to current best limits. The baseline layout of the experiment is based on the enhanced axion helioscope studied in~\cite{Irastorza:2011gs} and sketched in Fig.~\ref{fig:NGAH_sketch}. In this configuration the entire cross sectional area of the magnet is equipped with X-ray focusing optics to increase the signal-to-noise ratio. When the magnet is pointing to the Sun, solar axions are converted into photons, that are  focused and detected by low background X-ray detectors placed at the focal point of the telescopes. In this way a larger magnet aperture $A$ translates directly into the figure of merit of the experiment, as a larger signal is expected while the detector background remains low. This opens the way for new large-volume magnet configurations, like the ones of superconducting \textit{detector} magnets typically developed for high energy physics.

A useful figure of merit (FOM) was introduced in~\cite{Irastorza:2011gs} to easily gauge the relative importance of the various experimental parameters affecting the sensitivity of a helioscope:

\begin{equation}
f \equiv  f_M \: f_{DO} \:f_T
\end{equation}
where we have factored the FOM to explicitly show the contributions from various experimental subsystems: magnet, detectors and optics, and tracking (effective exposure time of
the experiment)
\begin{equation} \label{eq:foms}
f_M  = B^2\:L^2\:A \;\;\;\;\;\; f_{DO}=\frac{\epsilon_d \:
\epsilon_o}{\sqrt{b\:a}} \:\:\:\:\:\: f_T=\sqrt{\epsilon_t\:t} \ ,
\end{equation}


\noindent where $B$, $L$ and $A$ are the magnet field, length and
cross sectional area, respectively. The efficiency $\epsilon =
\epsilon_d \: \epsilon_o \: \epsilon_t$, being $\epsilon_d$ the
detectors' efficiency, $\epsilon_o$ the optics throughput or
focusing efficiency (it is assumed that the optics covers the
entire area $A$), and $\epsilon_t$ the data-taking efficiency,
i.~e.~the fraction of time the magnet tracks the Sun (a parameter
that depends on the extent of the platform movements). Finally,
$b$ is the normalized (in area and time) background of the
detector,  $a$ the total focusing spot area and $t$ the duration
of the data taking campaign.
The expressions in \eqref{eq:foms} assume some simplifications, like that $B$ is constant in all the volume of the magnet, or that $b$ and $\epsilon$ are constant throughout the energy range of interest. Generalizations of the figure of merit for arbitrary distributions of the parameters are straighforward, and the simplified versions quoted here are in any case useful to see the main dependencies.

Following this metric, IAXO thus aims at a  $f$ more than a factor 10$^4$ larger that its predecessor CAST. The conceptual design report (CDR) of the experiment~\cite{Armengaud:2014gea,Irastorza:1567109} demonstrates the technical feasibility of this step in sensitivity, and its main parameters will be reviewed below. Previous implementations of axion helioscopes have relied on existing equipment that were originally built for other experimental purposes. On the contrary, IAXO subsystems (magnet, optics and detectors) are entirely conceived and optimized for solar axion detection. The design prescriptions have been to rely on state-of-the-art technologies scaled up within realistic limits, i.e. no R\&D is needed to reach the stated experimental parameters. The \textit{baseline} sensitivity projections studied below will refer to the experimental parameters anticipated in the CDR. More recently, the realization of an intermediate experimental stage, BabyIAXO, featuring a scaled-down prototype version of the magnet, optics and detectors, is being considered. This new activity brings the opportunity to explore potential improvements over the CDR figure of merit of the full infrastructure, potentially resulting in an \textit{upgraded} sensitivity scenario for IAXO. The sensitivity projections shown below and in the plots all throughout this paper refer to these three experimental scenarios: BabyIAXO, IAXO-baseline and IAXO-upgraded. The experimental parameters corresponding to each of them are listed in table~\ref{tab:scenarios}, and the prescriptions to define them are discussed in the following.


\begin{figure}[!t]
\begin{center}
    \includegraphics[width=0.8\textwidth] {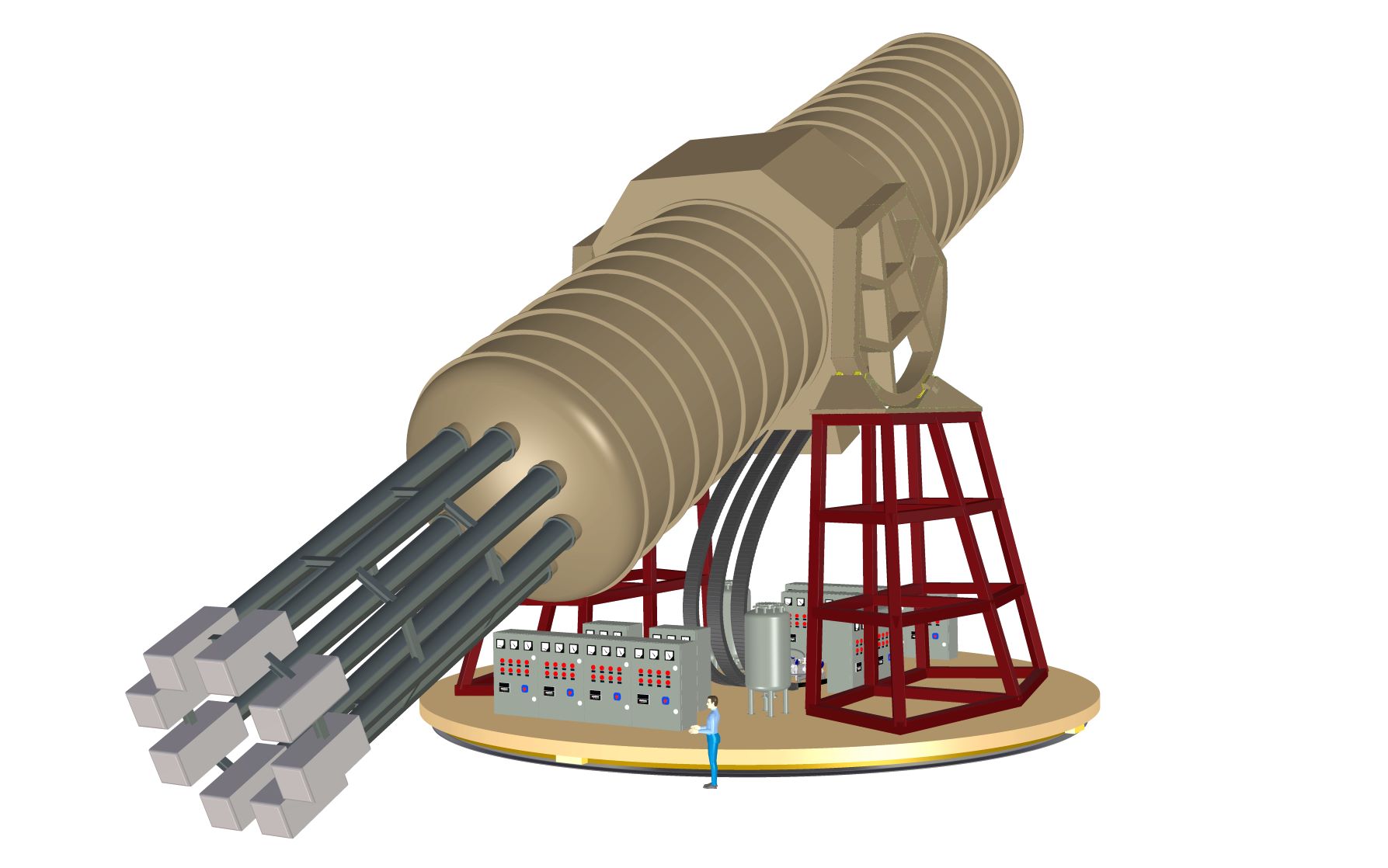}
    \caption{Schematic view of IAXO. Shown are the cryostat, eight telescopes+detector lines, the flexible lines guiding services into the magnet, cryogenics and powering services units, inclination system and the rotating platform for horizontal movement. The dimensions of the system can be appreciated by a comparison to the human figure positioned by the rotating table~\cite{Armengaud:2014gea}.}
    \label{fig:IAXO_sketch}
\end{center}
\end{figure}

The central component of IAXO is therefore a large superconducting magnet. Contrary to previous helioscopes, IAXO's magnet will follow a toroidal multibore configuration~\cite{Shilon:2012te}, to efficiently produce an intense magnetic field over a large volume. The baseline layout of the IAXO magnet is a 25 m long and 5.2 m diameter toroid assembled from 8 coils, and generating effectively 2.5 T average (5 T maximum) in 8 bores of 600 mm diameter.
The toroid's stored energy is 500 MJ. The design is inspired by the ATLAS barrel and end-cap toroids~\cite{tenKate:1158687,tenKate:1169275}, the largest superconducting toroids built and presently in operation at CERN. The superconductor used is a NbTi/Cu based Rutherford cable co-extruded with Aluminum, a successful technology common to most modern detector magnets. Figure~\ref{fig:IAXO_sketch} shows the conceptual design of the overall infrastructure~\cite{Armengaud:2014gea}. IAXO needs to track the Sun for the longest possible period. For the rotation around the two axes to happen, the 250 tons magnet is supported at the centre of mass by a system also used for very large telescopes. The necessary magnet services for vacuum, helium supply, current and controls are rotating along with the magnet.

\begin{figure}[!b]
\begin{center}
    \raisebox{-.5\height}{\includegraphics[width=7cm] {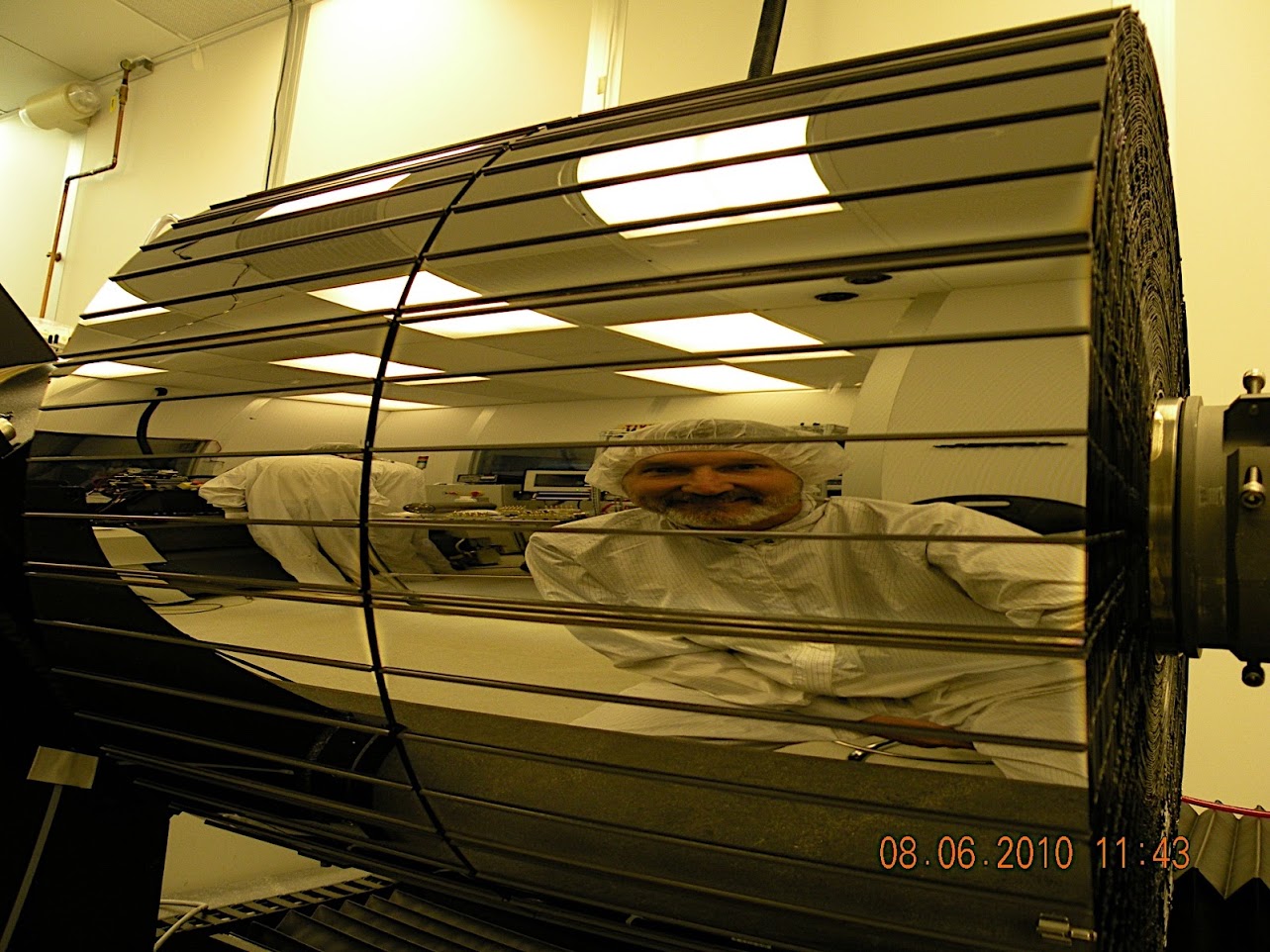}}
    \raisebox{-.5\height}{\includegraphics[width=8cm] {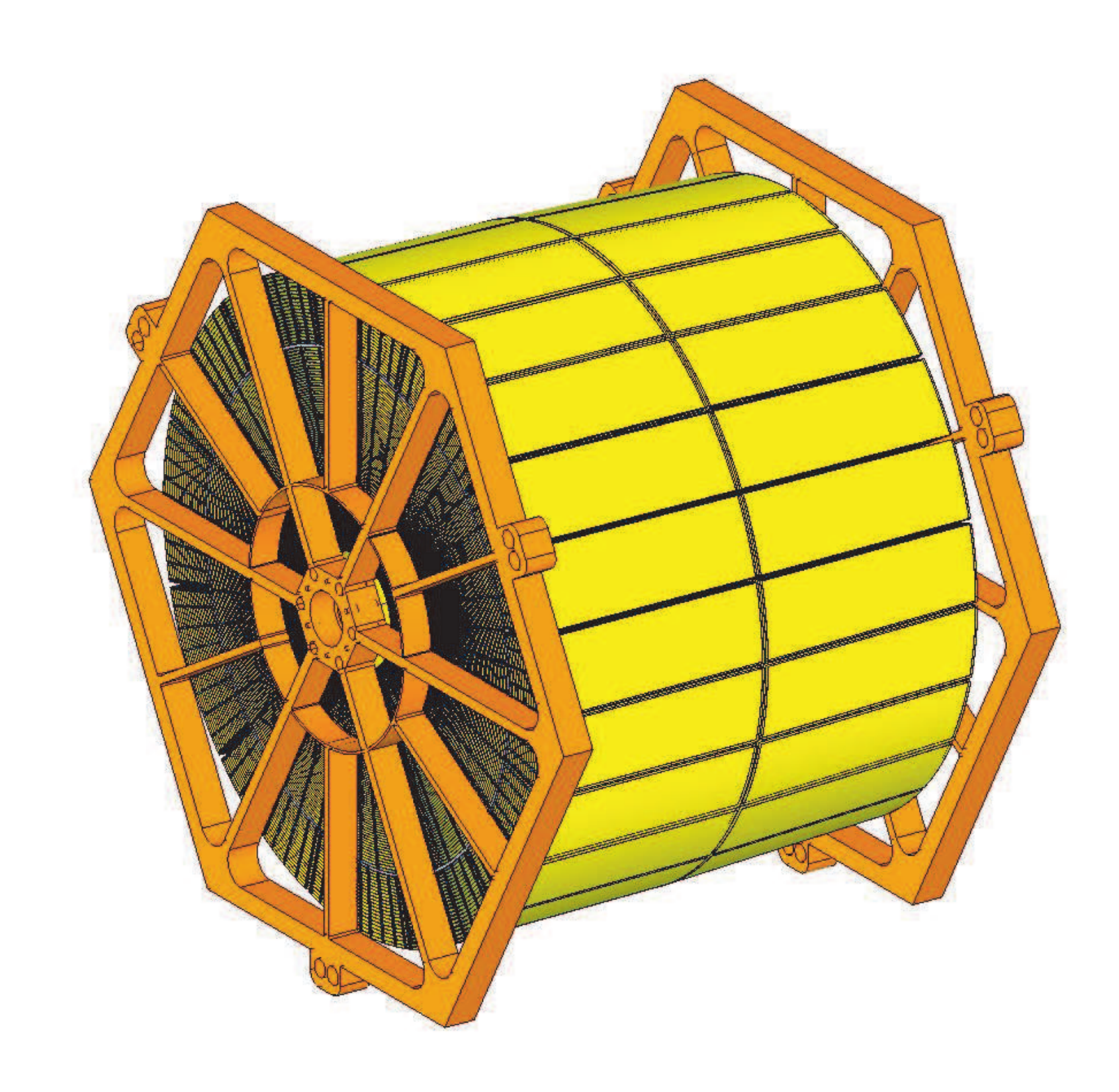}}
    \caption{Left: the NuSTAR X-ray telescope, with optics very similar to that proposed for IAXO.  Right: conceptual design of the X-ray optics needed for IAXO.}
    \label{fig:IAXO_optic}
\end{center}
\end{figure}

Each of the eight magnet bores is equipped with X-ray telescopes that rely on the high X-ray reflectivity on multi-layer surfaces at grazing angles. By means of nesting, that is, placing concentric co-focal X-ray mirrors inside one another, large surface of high-throughput optics can be built. The IAXO collaboration envisions using optics similar to those used on NASA's NuSTAR~\cite{nustar2013}, an X-ray astrophysics satellite with two focusing telescopes that operate in the 3 - 79 keV band.  The NuSTAR's optics, shown in Fig.~\ref{fig:IAXO_optic}, consists of thousands of thermally-formed glass substrates deposited with multilayer coatings to enhance the reflectivity above 10 keV.  For IAXO, the mirror arrangement and coatings are designed to match the solar axion spectrum. The conceptual design of the IAXO telescopes~\cite{doi:10.1117/12.2024476} can be seen on the right of Fig.~\ref{fig:IAXO_optic}. As proven in~\cite{Armengaud:2014gea,Irastorza:1567109,doi:10.1117/12.2024476}, this technology can equip the aperture area of IAXO magnet bores with focusing efficiency of around 0.6 and focal spot areas of about 0.2 cm$^2$.


At the focal plane in each of the optics, IAXO will have low-background X-ray  detectors. The baseline technology for these detectors are small gaseous chambers read by pixelated planes of micro-mesh gas structure (Micromegas)~\cite{Giomataris:1995fq} manufactured with the microbulk technique~\cite{Andriamonje:2010zz}. These detectors have been successfully used and developed in CAST and other low background applications~\cite{Abbon:2007ug}. The latest CAST detectors have achieved background levels of $10^{-6}$ \ckcs\ with prospects for improvement down to $10^{-7}$ or even $10^{-8}$ \ckcs\ ~\cite{Aune:2013nza}. These background levels are achieved by the use of radiopure detector components, appropriate shielding, and offline discrimination-algorithms on the 3D event topology in the gas registered by the pixelised readout. A pathfinder system combining an X-ray optics of the same type as proposed for IAXO and a Micromegas detector has been operated in CAST during 2014 and 2015 with the expected performance~\cite{Aznar:2015iia}. Alternative or additional technologies are being considered to complement the capabilities of Micromegas detectors or to extent them in specific aspects. 
GridPix detectors, similar to Micromegas detectors but built on a small CMOS pixelized readout~\cite{Krieger:2014wxa}, enjoy very low energy threshold down to the tens of eV, and thus are of interest for the search of specific solar axion production channels lying at lower energies, like the ones mediated by the axion-electron coupling. Silicon Drift Detectors (SDD) offer better energy resolution with flexible and cost-effective implementations~\cite{Mertens:2014nha,Dolde:2016wnv}.
Finally, bolometric detectors like Magnetic Metallic Calorimeters (MMC)~\cite{Ranitzsch2012}, or Transition Edge Sensors (TES)~\cite{Gottardi:2016cdx}
, enjoying much lower energy threshold and energy resolution, are also under consideration. An R\&D activity is ongoing to assess the low-background capabilities of all this technologies, and their suitability as detectors for IAXO. 
Low energy threshold and resolution are also of interest to extract model parameters in case of a positive signal (see next section).

\begin{figure}[t]
\begin{center}
    \includegraphics[height=6cm] {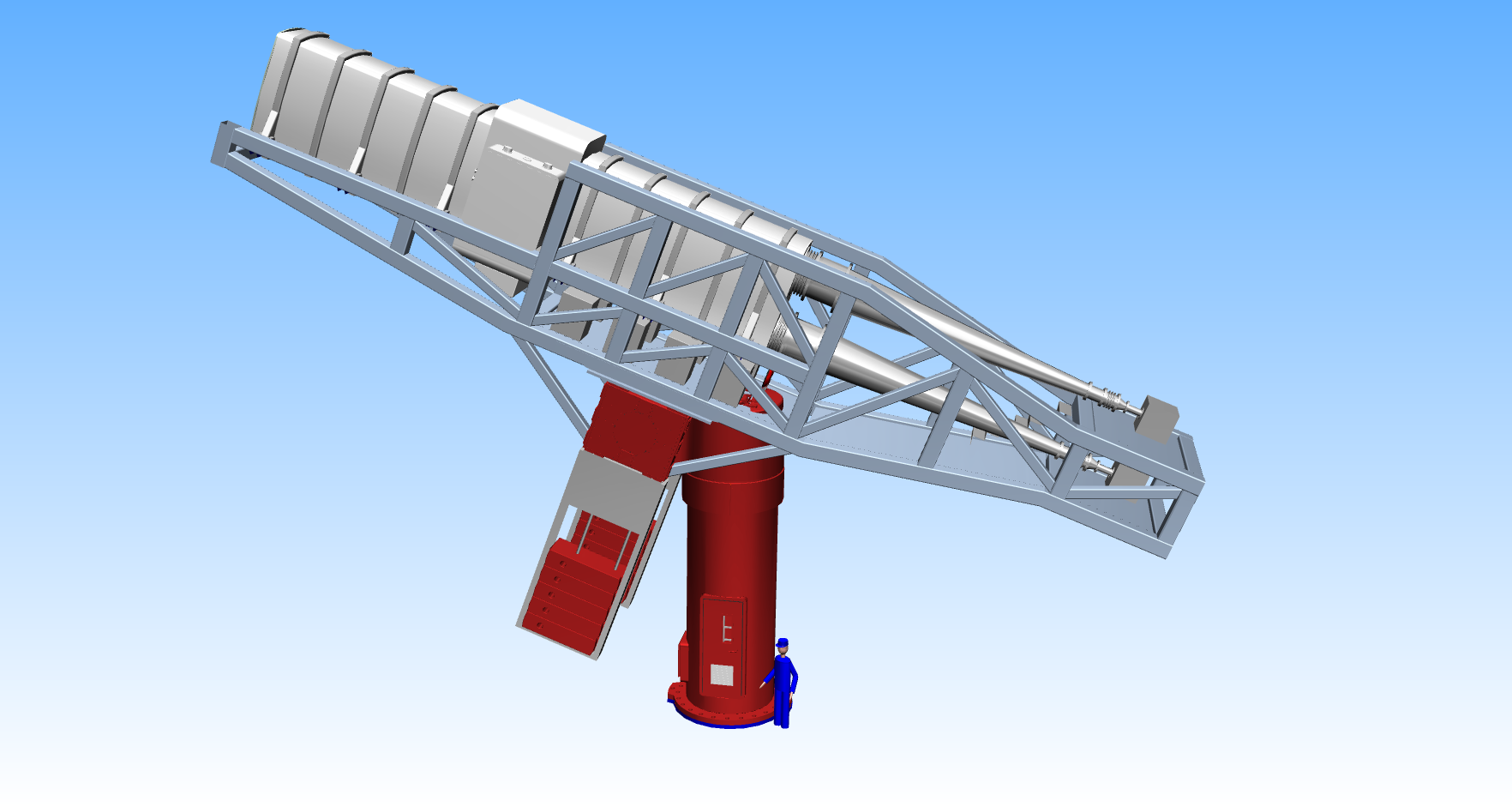}
    \caption{Conceptual design of BabyIAXO.}
 \label{fig:babyiaxo}
\end{center}
\end{figure}

The values considered for the main experimental parameters are listed in table~\ref{tab:scenarios} for three different scenarios. The column labelled \textit{IAXO baseline} shows the set of values anticipated in our CDR for IAXO, and are considered a realistic estimation within current state of the art. We refer to~\cite{Armengaud:2014gea} for a careful justification of those values. However, we now envision, as a first step, the realization of an intermediate stage, called BabyIAXO, featuring a scaled-down prototype magnet, as well as prototype optics and detectors, all representative of the final systems. BabyIAXO will serve as a testbed for all technologies in the full IAXO, but at the same time will deliver competitive physics. The BabyIAXO magnet is conceived as a common-coil dipole magnet, with two bores placed in between the coils. The two superconducting coils are 10-m long, but otherwise they enjoy quite similar engineering parameters than the ones proposed for the final IAXO toroid. The BabyIAXO magnet bores will have a diameter of 70~cm and each one will be equipped with one full detection line, optics and detector, of similar dimensions than the final IAXO systems. In the baseline configuration of the experiment, one of the BabyIAXO lines will host a newly built IAXO optics prototype similar to the one described above (potentially extended to 70~cm diameter), while the second one is expected to host an existing XMM spare optics~\cite{Jansen:2001bi}. A technical description of the BabyIAXO system will be object of another dedicated publication. Figure~\ref{fig:babyiaxo} shows a conceptual design of the BabyIAXO setup, whose expected experimental parameters are listed in the first column of table~\ref{tab:scenarios}. 

In addition, the experience with BabyIAXO is expected to test enhanced design choices that eventually lead to improved IAXO FOM values, especially $f_M$, beyond the ones anticipated in our CDR. The column labelled \textit{IAXO upgraded} represent this possible improved scenario. Collectively they constitute a factor $\sim$10 better $f$ (of which a factor 4 better $f_M$) than the baseline scenario. However, the extent to which those possible improvements may get eventually realized is tentative, and this scenario must be considered as a desirable target whose feasibility will be studied as part of the BabyIAXO stage.

\begin{table}[!b]
\footnotesize
\centering
\begin{tabular}{ccccc}
\hline  \textbf{Parameter} & \textbf{Units} & \textbf{ BabyIAXO} & \textbf{IAXO baseline }  & \textbf{IAXO upgraded }\\
\hline \\
 $B$           & T         & $\sim$2       & $\sim$2.5  & $\sim$3.5        \\
 $L$           & m         & 10       & 20    & 22    \\
 $A$           & m$^2$    & 0.77  & 2.3  & 3.9  \\
                                                                    \\
\hline
 $f_M$         &   T$^2$m$^4$        & $\sim$230              & $\sim$6000     & $\sim$24000 \\
                                                    \\

 $b$             & ${\rm keV^{-1}\, cm^{-2}\, s^{-1}}$ &  $1\times10^{-7}$ & $10^{-8}$& $10^{-9}$\\
 $\epsilon_d$  &              & 0.7      & 0.8 & 0.8   \\
 $\epsilon_o$  &              & 0.35      & 0.7 & 0.7   \\
 $a$             &  cm$^{2}$    &  2 $\times$ 0.3   & 8 $\times$ 0.15  & 8 $\times$ 0.15  \\
                                                                        \\
 $\epsilon_t$  &                     & 0.5   & 0.5  & 0.5  \\
 $t$             & year                    & 1.5     & 3 & 5    \\
 \hline \hline
\end{tabular}
\caption{Indicative values of the relevant experimental
parameters representative of BabyIAXO as well as IAXO, both the \textit{baseline} and \textit{upgraded} scenarios, based on the considerations explained in the text.}\label{tab:scenarios}
\end{table}

\begin{figure}[!b]
\begin{center}
\tikzsetnextfilename{pics/ExclusionPlot}
 \resizebox{0.8\textwidth}{!} {\input{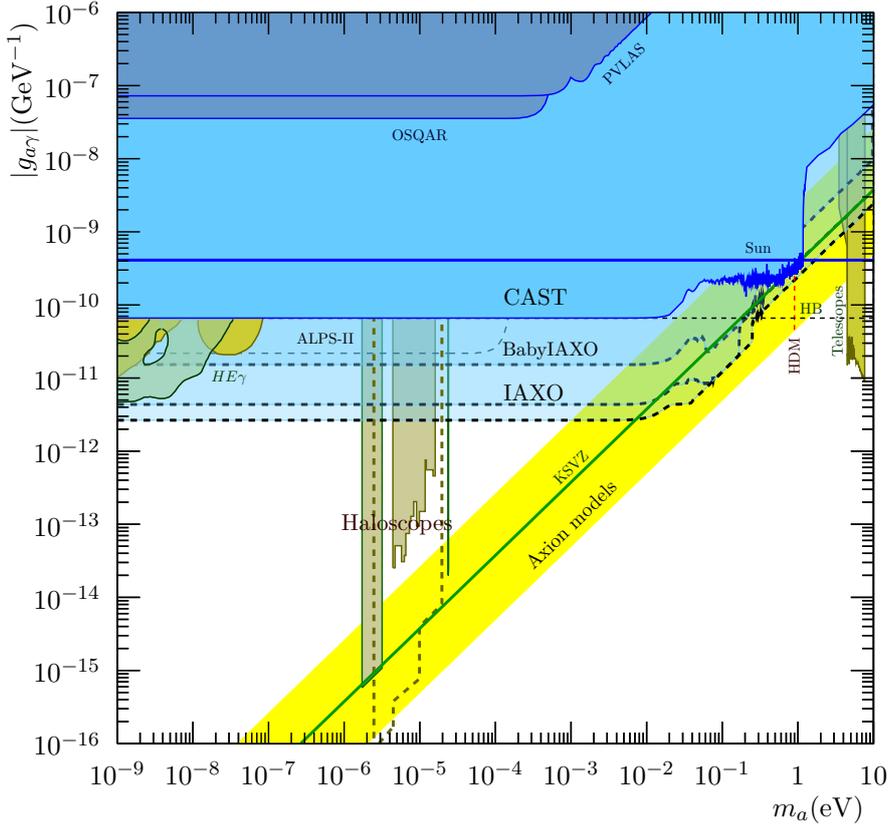}}
    \caption{Sensitivity prospects of BabyIAXO and IAXO (semitransparent regions) in the overall context of other experimental and observational bounds. We refer to~\citep{Irastorza:2018dyq} for details on the latter.}
    \label{fig:excluion_iaxo}
\end{center}
\end{figure}

The IAXO sensitivity projections shown in Fig.~\ref{fig:excluion_iaxo} as well as in all plots throughout this paper refer to the three scenarios of table~\ref{tab:scenarios}. They have been computed  by means of Monte Carlo simulation of the expected background counts in the optics spot area, computation of the likelihood function and subsequent derivation of the 95\% upper limit on the $\gagamma$ assuming no detected signal. The calculation is repeated for a range of $m_a$ values in order to build full sensitivity lines in the $(\gagamma,m_a)-$plane. For the purpose of this analysis, and following similar prescriptions as in~\cite{Irastorza:1567109}, detector background and efficiency are assumed flat with energy down to arbitrarily low energies. The axion-photon conversion in the magnet is approximated to the conversion in an homogeneous field, and the focusing effect is reduced to the equivalent effect of enhancing the signal-to-noise ratio due to the fact of confining the signal counts into the spot area. For all scenarios, an additional buffer gas data taking phase is considered, giving rise to the extended step-wise sensitivity line at high masses (0.01-0.25 eV). This data taking phase is composed by a number of overlapping gas density steps spanning the desired mass range. While in previous projections~\cite{Irastorza:2011gs} an equal exposure time is assigned to each step, resulting in a exclusion line more or less horizontal in $\gagamma$, in this case the exposure time is different for every step and adjusted to obtain a sensitivity down to the DFSZ $\gagamma$ (KSVZ for the BabyIAXO case) for every $m_a$ value. Of course different prescriptions to distribute the total exposure time among the steps are possible, depending on the motivation, e.g. to go for larger $m_a$ or lower $\gagamma$. The total exposure of this second phase is the same as the vacuum phase ($t$ in Table~\ref{tab:scenarios}).

\subsection{Measuring axion parameters}\label{sec:measaxipara}

In the case of a positive signal, and depending on the axion parameters, IAXO will be able to extract information on its mass $m_a$ and relative coupling with electrons and photons, potentially providing invaluable information on the underlying theoretical model. If the axion mass is above around 0.02 eV, axion-photon oscillations destroy the coherence of the conversion along the magnet length. This coherence can be restored if the conversion takes place in a buffer gas with density matching the axion mass, this being the rationale of the gas phase of both IAXO and BabyIAXO. If a positive detection happens during the gas scanning phase, the gas density provides the axion mass. But also in the vacuum phase and for lower mass values the onset of these spectral oscillations can be observed and used to determine the axion mass. As studied in~\cite{Dafni:2018tvj}, provided the signal is measured with sufficient statistics, $m_a$ values as low as $3\times10^{-3}$~eV could be determined by this method. 

Moreover, if the axion signal is composed by significant fractions of Primakoff and ABC solar axions, the combined spectral fitting can provide independent estimations of $\gagamma$ and $g_{ae}$~\cite{Jaeckel:2018mbn}. This combined determination works in areas of parameter space particularly well motivated by the stellar cooling anomalies. To exploit these capabilities high-resolution and low-threshold detectors are preferred, because part of the ABC solar axion spectrum lies at lower ($<1$~keV) energies, and also because it features several high-resolution peaks~\cite{Redondo:2013wwa}. Devices like the bolometric detectors described above, under consideration for IAXO, could play a major role in a post-discovery high-precision measurement campaign.


\subsection{Direct search for DM axions with IAXO and BabyIAXO}

Although the focus of this paper has been the physics potential of IAXO in its baseline configuration as a helioscope, the (Baby)IAXO magnet constitutes a remarkable infrastructure to implement additional setups to search for axions/ALPs in alternative ways. A most appealing option is to implement DM axion detectors that could exploit the particular features of the IAXO magnet. The basic idea of
the axion-haloscope-technique proposed in ~\cite{Sikivie:1983ip} is to
combine a high-$Q$ microwave cavity
inside a magnetic field to trigger the conversion of axions of the DM
halo into photons. A straightforward implementation of this concept in one bore of the IAXO  magnet gives competitive sensitivity~\cite{redondo_patras_2014} even using relatively conservative values for detection parameters, thanks to the large $B^2 V$ available, in the approximate mass range of 0.6 -- 2 $\mu$eV. Another recent concept proposes the use of a carefully placed pick-up coil to sense the small oscillating magnetic field that could be produced by the DM axion interacting in a large magnetic volume~\cite{Sikivie:2013laa}. This technique is best suited for even lower $m_a$ and is better implemented in toroidal magnets~\cite{Kahn:2016aff,Silva-Feaver:2016qhh}. The large size and toroidal geometry of IAXO suggest that it could host a very competitive version of this detection technique.

However, as can be seen from Fig.\ref{fig:DM_predictions}, there is a strong motivation to explore the mass range for QCD axion DM well above $10^{-5}$~eV.
The difficulty in searching in the ``high-mass'' range can be understood from
the fact that the figure of merit of scanning with haloscopes for axion DM scales
with the square of the cavity volume times the $Q$ factor of the cavity.
Most existing setups use solenoidal magnets and cylindrical
cavities. The diameter of the cylinder sets the frequency scale of the resonance
and thus the axion mass scale which the experiment is sensitive to. Thus going to
high mass means going to small diameters. In addition, the cavity quality factor $Q$ typically decreases for smaller cavities.

To tackle higher masses, different strategies are being developed in the community~\citep{Irastorza:2018dyq}. For instance, to compensate the loss in $V$ one can go to very strong magnetic fields and/or use superconducting cavities to keep $Q$ large. Another avenue is to decouple $V$ from the resonant frequency and go for large $V$ structures resonating at high frequency. Several strategies in this direction are being explored. One of them, currently being tested
in exploratory set-ups at the CAST experiment at CERN~\cite{Melcon:2018dba,Fischer:2289074}, employ long rectangular cavities~\cite{Baker:2011na}.
The particular advantage is that the volume can be kept very large (long cavity), while the resonance frequency
can be rather high, through the usage of relatively thin cavities. Given the size of the (Baby)IAXO magnet, it could host a multitude of rectangular cavities.

The implementation of one those concepts in IAXO is under consideration, but is out of the scope of the present paper. It clearly deserves serious study and will be the object of one or more future publications.